**Title:** Complex Patterns of Local Adaptation in Teosinte

**Short title:** Local Adaptation in Teosinte

**Authors and affiliations:** Tanja Pyhäjärvi[1], Matthew B. Hufford[1], Sofiane Mezmouk[1] and Jeffrey Ross-Ibarra[1,2,3]

[1] Department of Plant Sciences, University of California, Davis, CA, USA.

[2] Center for Population Biology, University of California, Davis, CA, USA.

[3] The Genome Center, University of California, Davis, CA, USA.

**Corresponding author:** Jeffrey Ross-Ibarra

## Abstract

Populations of widely distributed species often encounter and adapt to specific environmental conditions. However, comprehensive characterization of the genetic basis of adaptation is demanding, requiring genome-wide genotype data, multiple sampled populations, and a good understanding of population structure. We have used environmental and high-density genotype data to describe the genetic basis of local adaptation in 21 populations of teosinte, the wild ancestor of maize. We found that altitude, dispersal events and admixture among subspecies formed a complex hierarchical genetic structure within teosinte. Patterns of linkage disequilibrium revealed four mega-base scale inversions that segregated among populations and had altitudinal clines. Based on patterns of differentiation and correlation with environmental variation, inversions and nongenic regions play an important role in local adaptation of teosinte. Further, we note that strongly differentiated individual populations can bias the identification of adaptive loci. The role of inversions in local adaptation has been predicted by theory and requires attention as genome-wide data become available for additional plant species. These results also suggest a potentially important role for noncoding variation, especially in large plant genomes in which the gene space represents a fraction of the entire genome.

## Author Summary

Understanding how organisms adapt to different environments is a question of fundamental importance in evolution. Here, we investigated local adaptation in teosinte, the wild relative of domesticated maize. We find that the subspecies of teosinte studied here exhibit complex population structure, patterned by altitude, geography, dispersal events and admixture. We used dense genome-wide SNP genotyping to identify genetic variants likely associated with local adaptation. Our results point to a significant adaptive role for inversion polymorphisms, and reveal that the majority of variants differentiated among populations are found in noncoding regions of the genome. The size and complexity of most plant genomes suggests that our results may be indicative of an important role for structural and noncoding variants in the evolution of many plant species.

# Introduction

One of the enduring goals of evolutionary genetics is to understand the genomic and geographic extent of adaptation in natural populations. While many examples of adaptation in nature are known [1-4], our theoretical understanding [5-7] of the genomic signatures of adaptation has so far outpaced natural studies, and a number of outstanding questions remain. What are the prevailing forms (i.e. clinal, local) of adaptation in nature? What parts of the genome and what kinds of loci are involved? At what geographic scales can genome-wide signatures of adaptation be detected? The advent of high-throughput genotyping, however, is beginning to allow genomic analysis of populations across their natural range. For example, recent studies have shown geographically localized fitness effects [8], enrichment for nonsynonymous variants at adaptive loci [9] and adaptation across a number of environmental variables (e.g., soil type, temperature and precipitation) [8,10,11].

The wild relatives of domesticated maize offer an excellent opportunity to investigate these questions across a number of natural populations. In addition to the well-known domesticated maize, *Zea mays* ssp. *mays*, the species *Zea mays* includes three wild subspecies, collectively known as teosinte. Two of these, *Zea mays* ssp. *parviglumis* (hereafter *parviglumis*) and *Zea mays* ssp. *mexicana* (hereafter *mexicana*) are widespread taxa found in natural populations in central and southwest Mexico (Figure 1). Their combined range extends across varying environments, from the warmer low elevations of the Balsas River Valley to the cooler high elevations of Mexico's Central Plateau, across two-fold differences in precipitation and several soil types. Ecological niche modeling suggests that both taxa have moved little in response to climate change during the Holocene (unpublished data), and their annual life history and outcrossing mating system have likely provided both the time and the genetic basis for adaptation to local conditions.

Previous work has investigated the genetic architecture of morphological differences in the teosintes [12-16]. Overall patterns of genetic diversity and structure in these taxa have also been detailed using various types of markers [17-20], but our understanding of genome-wide patterns of variation has been limited to a handful of accessions or comprehensive analyses of individual populations [21]. Even less is known regarding the genetics of local adaptation, which has only been studied at a few loci in the context of plant immunity [22].

Here we present a detailed population genetic examination of local adaptation across 21 populations of *mexicana* and *parviglumis* (Figure 1), based on high-density genotype data from 250 sampled individuals. We observed hierarchical population structure that correlated strongly with elevation. While patterns of haplotype sharing suggested relatively little gene flow between subspecies even in areas of sympatry, we nonetheless identified a highly admixed population at intermediate elevation. We then scanned the genome for potentially adaptive variants by looking for alleles associated with excess differentiation or correlation with environmental gradients. We identified several inversion polymorphisms and showed that these variants are enriched for signals of adaptation. Finally, we discuss the impacts of different selection pressures, population structure, and geography on the ability to identify potentially adaptive variants using population genetic methods.

## Results

### Genomic diversity in teosinte

We sampled 21 populations of the wild subspecies *Zea mays* ssp. *parviglumis* and *Zea mays* ssp. *mexicana* from across their native ranges in central and southern Mexico (Figure 1). Ten to twelve individuals from each population were genotyped using the Illumina MaizeSNP50 chip. After quality control, the full dataset consisted of 248 samples genotyped at 36,719 SNPs.

Both *parviglumis* and *mexicana* showed generally high levels of heterozygosity and little evidence of inbreeding, and the relatively minor differences in the overall distribution of heterozygosity between the subspecies suggest a relatively limited impact of SNP ascertainment on comparisons of diversity (Figure S1). Considerable variation was evident among populations in both taxa, however (Figure 2, Table S1). The *parviglumis* population of Los Guajes, for example, showed the signs expected of a large outcrossing population, including high diversity, an inbreeding coefficient ($F_{IS}$) close to zero, no deviation from Hardy-Weinberg equilibrium, and relatively brief runs of homozygosity (ROH, Figure S2) across the genome. Other populations, however, showed evidence of past demographic change: diversity in San Lorenzo *parviglumis* was only half that of Los Guajes, and *mexicana* individuals from Xochimilco were characterized by extremely long ROH and extensive within-population haplotype sharing, consistent with a recent population bottleneck.

Patterns of linkage disequilibrium (LD) varied greatly along the genome, with numerous discrete blocks of elevated LD observed in multiple populations. We identified four large (>10Mb) regions of high LD ($r^2 \geq 0.2$), which we interpreted here as inversion polymorphisms (Figure 3, Figure S3). Three of these regions appeared to correspond to inversions previously described cytologically (*Inv9e*; [23]), in mapping populations (*Inv4m*, [24]), or by population genetic analysis (*Inv1n*, [25,26]). Clear haplotype structure was observed for three of the putative inversions, and simple genetic distance-based clustering including *Tripsacum* and maize suggested that the non-maize haplotype was likely the derived state (Figures S4-S7).

**Population structure**

Signs of population structure in teosinte were evident at multiple hierarchical levels. The strongest signal was between subspecies, shown in the principle component analysis (PCA), where the first PC separated the two subspecies and explained ~11.5 % of the total variance (Figure 4A). Differentiation between subspecies was also evident in STRUCTURE results (Figure 4B), haplotype sharing (Figures S8-S9), and in higher pairwise $F_{ST}$ for inter-subspecific (0.33) than intra-subspecific (*parviglumis* 0.24; *mexicana* 0.23) comparisons. Inversion polymorphisms also separated the subspecies, as *Inv9d* and *Inv9e* were found only in *mexicana*, while the derived haplotypes at *Inv1n* and *Inv4m* showed strong frequency differences (Figure S10) among subspecies.

Additional levels of structure were observed within subspecies as well (Figure S11). STRUCTURE and PCA analysis of populations of *parviglumis* identified groups of related populations consistent with previous groupings [17]. Although substructure within *parviglumis* identified a grouping of populations in Jalisco known to be near cultivated maize, [16], a joint STRUCTURE analysis including a diverse sampling of 279 maize inbred lines [27] failed to find evidence that admixture explains the observed population structure (Figure S12). In all, 20 significant principle components identified 21 clusters that generally corresponded to sampling locales (Figure 4A, Figure S13). Three populations of *parviglumis* did not follow this trend, however, as the two Ejutla populations formed a single genetic cluster and the Ahuacatitlan population split into two clusters.

Both geodetic distance (Mantel test p-value: 0.006, Mantel statistic r = 0.40) and altitude (p-value: 0.0008, r = 0.37) appeared to correlate with population differentiation in the combined dataset (Figure S14A; partial Mantel tests for geodetic distance p-value: 0.0005, r = 0.39; altitude p-value: 0.004, r = 0.41). Although the proportion of *mexicana* ancestry appeared to increase with altitude in STRUCTURE results (Figure 4B), differentiation within each subspecies was only significantly correlated with geodetic distance (p-value: 0.004, r = 0.55 in *parviglumis*; p-value: 0.046, r = 0.48 in *mexicana;* Figure S14 B and C). In both subspecies the correlation with geodetic distance appeared to be primarily due to the peripheral Nabogame and Oaxaca populations and was no longer significant when these were removed from the analysis.

Finally, both STRUCTURE and PCA results suggested that the Ahuacatitlan population of *parviglumis* may be admixed (Figure 4). Consistent with this, analysis of haplotype sharing showed that the Ahuacatitlan population shared fewer long haplotypes with *parviglumis* and more long haplotypes with *mexicana* than any other *parviglumis* population (Figure S8 and S9). The average length of shared haplotypes between Ahuacatitlan and *mexicana* (mean 4.8 cM), however, was shorter than the average length of haplotypes shared among *mexicana* populations (mean 30.7 cM), arguing against extensive recent admixture. To explicitly test the position of Ahuacatitlan with respect to the two subspecies, we applied the D-test of derived allele configuration devised by Green et al. [28]. The observed distribution of D was inconsistent with models in which Ahuacatitlan is ancestral to both *parviglumis* and *mexicana* or a sister group to *mexicana* ($p < 0.05$ for both; Figure S15), while the model in which Ahuacatitlan was sister to *parviglumis* could not be rejected ($p < 0.2$).

### Candidate SNP identification
We applied three approaches to identify SNPs underlying local adaptation, and candidate SNPs and genes from each of these approaches are listed in Table S2. We first applied the environmental association method implemented in BAYENV [29] to a set of 76 climate and soil variables. We summarized these variables with 6 principle components which represent 95% of the variation among populations. We calculated environmental PCs and estimated Bayes factors for the joint data set and for each subspecies separately (Table 1, Figure S16-S18), and retained SNPs consistently in the top 5% of Bayes factors and in the top 1% of average Bayes factors across 5 runs of the BAYENV MCMC. In the joint data set, in total 1598 SNPs were associated with at least one PC, while 54 SNPs were associated with more than one PC. The number of candidates was lower within subspecies, where 370 SNPs in

*parviglumis* and 533 SNPs in *mexicana* were associated with at least one of the four PCs representing 95% of within subspecies environmental variation.

Our second approach to identify locally adapted SNPs used an $F_{ST}$ outlier method [7,30] to scan for SNPs showing an excess of differentiation between subspecies ($F_{CT}$) or among populations ($F_{ST}$). Hierarchical analysis yielded 731 $F_{CT}$ (2.0%) and 1363 $F_{ST}$ (3.7%) outliers in the 1% tail of simulated values. A striking peak in differentiation was observed within inversion *Inv4m*, which showed 12-fold and 6-fold enrichment of outlier SNPs for $F_{CT}$ and $F_{ST}$, respectively. Both sets of outliers (Figure 5A, Figure S19, Table S2) differed substantially compared to SNPs identified by BAYENV (Table S3). The strongest correlation (Spearman's rho -0.20, p << 0.001) was seen with Bayes factors for PC2 in the joint data set and p-values of $F_{ST}$ (Figure 5A).

Finally, we performed a scan for SNPs highly differentiated in a single focal population, which we called an $F_{FT}$ outlier approach (see Materials and Methods). Mean $F_{FT}$ across populations was lower than 0.1, suggesting that genome-wide only a small proportion of the total genetic structure was due to drift in individual populations. The 1% tails (354-358 SNPs in each population) of the $F_{FT}$ distribution did not overlap considerably with $F_{ST}$ (Figure 5B) and Spearman's correlation coefficient was not significant for any of the $F_{FT}$ in relation to $F_{ST}$ or $F_{CT}$ values. Populations at range extremes, such as El Rodeo and Nabogame, tended to have higher $F_{FT}$ (1% cutoff at 0.79 and 0.72), and high $F_{FT}$ SNPs from these populations were also overrepresented among $F_{ST}$ (Figure S20; Wilcoxon rank sum test p-values << 0.001) and BAYENV outliers. The most striking case was the effect of the

Nabogame population on association results for PC2 (Figure S21), where SNPs with high $F_{FT}$ in Nabogame had significantly higher Bayes factors than $F_{FT}$ outliers of other populations (Wilcoxon rank sum test p-value << 0.001). Nabogame $F_{FT}$ outliers constituted ~9% of all PC2 outliers for BAYENV, and PC2 Bayes factors were positively correlated with $F_{FT}$ in Nabogame (Spearman's rank correlation p-value <<0.001, rho=0.21).

All four inversions contained an excess of SNPs with high $F_{ST}$ or were enriched for SNPs associated with environmental variables (Figure 6, Table 2). Inversions were 2-fold enriched for candidate SNPs (p-value <0.0001), containing 5.6.% of all SNPs but 11 % of SNPs in the 99th percentile of the maximum rank distribution of our candidate lists. Enrichment was most notable for PC1, which predominantly reflected variation in altitude and annual temperature. As reported in Fang et al. [26], the inverted arrangement of *Inv1n* showed an altitudinal cline, occurring at intermediate frequency in most *parviglumis* populations and at low frequency in *mexicana* (Figure S4 and S10A). The trend was less clear in our data, however, as the derived arrangement of *Inv1n* was absent from the low elevation Crucero Lagunitas population of *parviglumis*. Two additional inversions showed similar altitudinal patterns: the derived arrangement in *Inv4m* was most abundant in high elevation *mexicana* populations and restricted to the highest elevation *parviglumis* population, (Ahuacatitlan; Figure S10B), whereas the non-maize arrangement of *Inv9d* was found only in the highest elevation *mexicana* populations (Figure S10C, Figure S22A).

**Functional evaluation of candidate SNPs**

We tested candidate SNPs for enrichment in various functional categories (genic vs. non-genic, synonymous vs. nonsynonymous) as well as for associations with maize phenotypic traits [31,32] and regions highlighted as putative adaptive introgressions from *mexicana* into maize (unpublished data). Among genes that showed evidence of introgression from *mexicana* into maize both $F_{ST}$ (p-value <0.0001) and $F_{CT}$ (p-value 0.0002) candidates as well as SNPs associated with PC2 within *mexicana* (p-value 0.0061) were more common than expected by chance. $F_{ST}$ outliers were strongly enriched for non-genic SNPs (permutation p-values = 0.001) (Table S4), a result that held even after excluding SNPs in the four identified inversions (data not shown). $F_{ST}$ outliers were also enriched for SNPs associated with male inflorescence architecture and stand count (plant survival and density) in a maize panel (Table S5). Finally, SNPs associated with flowering time were not over-represented under an association model that takes into account structure (both K + Q; see Materials and Methods), but were significantly over-represented among both $F_{CT}$ and $F_{ST}$ outliers when correction for population structure was not applied (Table S6). The collinearity between the structure and flowering time adaptation makes it difficult to control the structure without increasing the rate of false negatives, whereas without a control of the structure all true positives are kept but the rate of false positives is high. The latter is conservative when testing for an enrichment for the adaptive loci.

The functional annotations of nearby or containing genes were investigated for six SNPs that were extreme outliers for $F_{ST}$ (Figure S19). Of these, PZE-104028461 is a synonymous SNP in the maize filtered gene GRMZM2G000471, whose ortholog AT4G10380, *NIP5;1,* encodes a boron channel [33]. The high $F_{ST}$ in this gene is caused by an allele fixed in the Nabogame population that is found only in two heterozygotes in the rest of the sample. The SNP is also significantly correlated with PC2 within *mexicana*. In *Arabidopsis thaliana*, SNPs associated with *in planta* boron concentration were enriched among SNPs showing strong signals of selection [34], and AT4G10380 was also identified as a candidate gene for adaptation to different soil types in *Arabidopsis lyrata* [35].

## Discussion

### Complex population structure

Our results indicate that population structure in teosinte is complex and affected by multiple factors. Across all populations, the effects of altitude on population structure are significant, even after correcting for geodetic distance (Figure S14A). The *mexicana* populations of Santa Clara and Opopeo, for example, show higher $F_{ST}$ with the La Cadena *parviglumis* population 60 km away than with distant *parviglumis* populations at the same elevation (Figure S23). These results are in agreement with earlier studies that have noted the important role of altitude on genetic distance, genome size and morphological characters in *Z. mays* subspecies [36-39]. There is also clear hierarchical population structure (Figure 4, Figure S11), with divisions between subspecies and among previously identified groupings within subspecies [17]. Within subspecies, the effect of altitude is no longer statistically significant, but isolation by distance continues to explain a meaningful portion of the genetic structure observed in both taxa (Figure S14B and S14C). In addition to distance and altitude, stochastic founding events have likely played a role in patterning diversity. This is most evident in the extensive haplotype sharing within the geographically dispersed La Mesa/El Sauz/San Lorenzo group (Figure S8). The importance of founder events is consistent with previous analyses by Fukunaga et al. [17] and Buckler et al. [37] who found that genetic distance was more correlated with specific dispersal routes than with geodetic distance.

While hybridization between *mexicana*, *parviglumis* and cultivated maize is a well-known phenomenon [17], the extent of admixture between the two wild subspecies has not been well documented. Our data suggest that the high elevation *parviglumis* population is extensively admixed, with mean assignment probabilities to *mexicana* ~50% for all individuals in the population. Patterns of haplotype sharing (Figure S8-S9) and derived allele counts (Figure S15), however, support a model of continual gene flow with *parviglumis* over a relatively long period of time. Ross-Ibarra et al. [40] proposed a demographic model including continuous gene flow between *mexicana* and *parviglumis* apparently inconsistent with the general lack of admixture and strong differentiation we observe here for most populations. We note, however, that several of the *parviglumis* individuals sampled by Ross-Ibarra et al. are from localities quite near Ahuacatitlan, including two collected only ~4 km away. Fukunaga et al. [17] also identified a number of putative *mexicana* – *parviglumis* hybrids from localities in the Eastern Balsas near to Ahuacatitlan. Together with our results showing no admixture in *parviglumis* populations such as La Cadena and Los Guajes that are in geographic proximity to *mexicana*, these data indicate the presence of a geographically restricted hybrid zone between *mexicana* and *parviglumis* occurring at mid elevations in the Eastern Balsas region.

Population structure can bias estimates of demographic history and the inference of selection through its effect on the allele frequency spectrum [41]. In addition, Eckert et al. [11] point out that when environmental and population structure gradients coincide, environmental correlation methods may suffer from over-correction of structure. To resolve biased demographic inference caused by sampling, several authors have shown that sampling single individuals from each of many populations largely ameliorates this bias under simple island [42] or stepping stone [43] models. However, such a sampling scheme does not take into account uneven or hierarchical population structure, [44]. In teosinte, hierarchical population structure is observed within individual subspecies and is patterned by both geography and the environment. We have also shown that populations at the edge of the range, like Nabogame and El Rodeo, are unusually distinctive. Because most plants likely show complex patterns of structure, it is clear that careful population level sampling and consideration of population structure is important in studies of plant adaptation and evolutionary history.

### The genetic basis of local adaptation in teosinte

In contrast to the results of similar analyses in *A. thaliana* and human data [9,45], enrichment of neither genic nor nonsynonymous sites was observed among candidate SNPs (Table S4). While the SNP chip developed for maize is less dense than the human or *Arabidopsis* data, it nonetheless samples the gene space of maize quite well, suggesting that SNP density or the makeup of the chip does not explain the difference in nonsynonymous sites. And though the positive correlation between Bayes factors and heterozygosity may lead to biases among SNP categories, the average expected heterozygosity of nongenic SNPs in our data was slightly lower than for genic SNPs (0.275 vs. 0.281). Instead, it seems likely that the complexity of the maize genome, which is more than 85% noncoding sequence [46], may provide greater opportunity for the evolution of functional noncoding elements. The causal quantitative trait loci (QTL) for several well-studied genes has been shown to reside in noncoding regions near genes [e.g., 47,48], and noncoding SNPs have recently been shown to explain the majority of associations for morphological traits [49]. The functional role of candidate SNPs is supported by their enrichment among relevant phenotypic effects (Table S5 and S6). SNPs associated with tassel morphology, that is a major phenotypic difference among *parviglumis* and *mexicana* [50], are enriched among $F_{CT}$ candidates. Further, SNPs associated with flowering time—a common adaptation among plant populations [e.g., 51,52]—are also highly differentiated among populations and subspecies.

In spite of the lack of genic enrichment, we do identify a number of likely candidate genes for local adaptation (Table S2). A SNP in the gene *b1*, for example, was found correlated with PC1—which is largely made up of temperature and altitude—among *mexicana* populations. *b1* is a gene in the anthocyanin synthesis pathway that has been identified as a QTL for sheath color differences among *mexicana* and *parviglumis* by Lauter et al. [12], and pigmentation has been suggested to be an important plant adaptive trait as a response to lower temperatures and changing light conditions [53,54]. Two SNPs in the 3'UTR of the well-known domestication gene *tb1* show unusually strong patterns of differentiation and association with PCs related to temperature range and soil types

(Table S2). *tb1* is a transcription factor that had a significant role in morphological changes during maize domestication [47,55], and SNPs in the 3' UTR have previously been found to be significantly associated with lateral length and tassel branching in *parviglumis* [14,15]. Other candidate genes that have a well known function in maize are, for example, *abph1* [56] and *sh1* [57].

In addition to the rich nucleotide diversity of teosinte, cytological studies of both *mexicana* and *parviglumis* have identified a number of inversion polymorphisms [13,23,58]. In our high density genotyping data we observe four blocks of LD that we interpret as putative inversions. *Inv1n* was described by Fang et al. [26]. *Inv9e* was identified cytologically by Ting [23], who found the inverted arrangement at high frequency in both of the *mexicana* populations in which we see the derived haplotype. Similarly, the LD block we identify as *Inv4m* is in a similar physical position as inversions identified cytologically in maize and teosinte [13,23], and by marker map order in the progeny of a self-fertilized *Zea nicaraguensis* [24]. Synteny maps identify the proximal breakpoint of *Inv4m* as the junction of a chromosomal fusion with the distal end of the telomere of chromosome 5 [Figure S15 in 46], and the breakpoint is common to several distinct inversions in maize [59], suggesting that the region may be prone to structural rearrangements.

All four observed inversions appear to play a role in local adaptation. All show significant enrichment for SNPs with high Bayes factors for PC1 (Table 2), reflecting environmental differences in altitude and temperature. Inversion *Inv1n* is enriched for SNPs with high Bayes factors for PC1, PC3 and PC4, and the clinal patterns observed are consistent with those of Fang et al. [26] with the exception of the Crucero Lagunitas population which was not polymorphic for the inverted allele. Extreme patterns of precipitation in Crucero Lagunitas and shared ancestry with the distant Oaxacan population of El Rodeo suggest that unusual history or distinct selection pressures may help explain the surprising lack of *Inv1n* in this population. *Inv4m* is associated with a striking peak in differentiation among subspecies (Figure 5). This is consistent with an adaptive role for *Inv4m* at higher altitudes, which was also suggested by the evidence of introgression from *mexicana* to sympatric maize populations at *Inv4m* and its rarity in maize outside of the Mexican highlands (unpublished data). Interestingly, QTL for differences between *parviglumis* and *mexicana* for pigment intensity and macrohair count, traits thought to be adaptive at high elevations, co-localize to this region as well [12]. *Inv9d* was only polymorphic in *mexicana* and showed a strong enrichment of SNPs associated with PC1 (Figure 5, Figure S22B). Finally, though not strongly associated with altitude, inversion *Inv9e* showed 8-fold enrichment of SNPs associating with PC4 (top soil) within *mexicana* (Table 1, Figure S18D).

Our evidence suggests that all four inversions carry important adaptive variation, but further experiments such as association mapping and reciprocal transplant experiments will be required to directly measure their effects on phenotype and fitness. The accumulation of locally adaptive loci in inversions has long been predicted by theory [60], and is consistent with observations in a number of plant and animal taxa [61-65]. One of the major effects of chromosomal inversions is to suppress recombination, because inversions that capture two or more locally beneficial alleles may be favored by selection [60,66]. The number and size of the polymorphic inversions observed here underscores the potential importance of inversions in plant local adaptation [66]. These inversions cover ~5% of the *Zea mays* reference genome, and their role in patterning both molecular and phenotypic variation [26] among populations is significant. As even denser marker datasets become available, continued sampling of additional populations will undoubtedly reveal new, smaller, and geographically restricted inversions segregating in plant populations.

## Multiple approaches to understand local adaptation

Different approaches to identify loci conferring local adaptation yielded very different sets of candidates. Rather than indicating a lack of evidence for local adaptation, we argue that the weak correspondence between methods is reflective of biological factors including the kind of natural selection and the effects of population structure. For example, SNPs showing environmental association are only weakly correlated with $F_{ST}$ and $F_{CT}$ p-values (Table S3). The same observation was made for trees by Eckert et al. [67] and Keller et al. [68] who compared $F_{ST}$ outliers to SNPs correlating with environment. Because $F_{ST}$ outlier methods detect excess divergence among populations regardless of the distribution of environmental variation, they are likely more powerful for the identification of loci related to factors that do not correlate with measured environmental variables. All of our measured environmental variables are abiotic, and it seems likely that additional important biotic factors (e.g., herbivory, competition) would not be perfectly reflected in our environmental PCs. For example the presence of *Dalbulus maidis* leafhoppers specialized to *Zea* is thought to depend on the local abundance of maize and proximity to bodies of water [69]. Also, Moeller and Tiffin [22] found signs of local adaptation in a single population and a locus in a study on plant immunity genes of *parviglumis*. They suggested that this was caused by localized selection pressure in herbivory. In humans, too, there is evidence that pathogen diversity may play a larger role in local adaptation than climate [70].

Population structure had a considerable effect on outlier detection in all our methods. *Parviglumis* and *mexicana* do not conform to a simple two-level hierarchical island model where gene flow between populations within subspecies would be equal. As a result, some strongly differentiated populations had undue impact on $F_{ST}$ outliers, and our $F_{FT}$ statistic proved to be informative about such population-specific effects. For example, SNPs specific to the isolated El Rodeo and Nabogame populations have significantly smaller $F_{ST}$ p-values than $F_{FT}$ outliers from other populations (Figure S20). Similarly, because of its extreme temperature and precipitation, SNPs that are strongly differentiated in Nabogame ($F_{FT}$ outliers) have significantly higher Bayes factors for PC2 than SNPs from other *mexicana* populations (Wilcoxon rank sum test, p<<0.001, Figure S21). Such geographically restricted environmental variation has been shown to cause false positives in association studies even when population structure is taken into account [71], and may be one explanation for the observation that new advantageous mutations with narrow geographic distributions were enriched among climate-associated loci in *A. thaliana* [9].

# Material and Methods

## Sampling

Ten to twelve individuals were sampled from each of eleven populations of *parviglumis* and ten populations of *mexicana* (Figure 1, Table S7). Seeds from four populations of *parviglumis* (Ejutla A and B, San Lorenzo and La Mesa) were sampled by MBH in 2007, while the remaining 7 populations were obtained from non-regenerated bulked seed from accessions in the USDA germplasm collections. All *mexicana* populations were collected in 2008 by Pesach Lubinsky and provided by Norman Ellstrand. A *Tripsacum dactyloides* sample provided by Sherry Flint-Garcia was used as an outgroup. Geodetic distances among populations vary from 3 km between Santa Clara and Opopeo to 1503 km between El Rodeo and Nabogame, and the populations cover an altitudinal range from 590 to 2609 meters above sea level. An initial PCA that included 279 maize inbred lines [27] revealed that two *mexicana* individuals appeared to be recent hybrids with maize (Figure S24), and these were removed from subsequent analyses.

## SNP genotyping

DNA was extracted from leaf tips of seedlings at the five-leaf stage, and tissue was stored at 80°C overnight and lyophilized for 48 h. DNA was extracted from homogenized tissue following a modified CTAB protocol [72] and quantified using a NanoDrop spectrophotometer (NanoDrop Technologies, Inc., Wilmington, DE, USA) and Wallac VICTOR2 fluorescence plate reader (Perkin-Elmer Life and Analytical Sciences, Torrance, CA) with Quant-iT™ PicoGreen® dsDNA Assay Kit (Invitrogen, Grand Island, NY).

Single nucleotide polymorphism (SNP) genotyping was conducted at the UC Davis Genome Center using the MaizeSNP50 BeadChip and Infinium® HD Assay (Illumina, San Diego, CA). SNP genotypes were called using GenomeStudio V2009.1 (Illumina). After dropping SNPs with more than 10% missing data in either subspecies and manual adjustment of clustering, average call rates (the proportion of successfully called genotypes) per individual were 99% for both *parviglumis* and *mexicana*. Genotyping error estimated from three technical replicates of *parviglumis* was ~0.0015% after mismatches caused by missing data were excluded.

In total, 43,701 SNPs were called in *parviglumis* and 43,694 in *mexicana*. SNPs within a region on chromosome 8 that shows discrepancies between the physical and genetic maps were removed [73]. After removing monomorphic and duplicate SNPs and SNPs that did not map or mapped multiply to the maize reference genome (release 5b.60), the final data set consisted of 36,719 SNPs genotyped in both subspecies. SNP data, map positions, and annotations are available at www.panzea.org and www.rilab.org.

### Diversity and Population structure

Observed ($H_O$) and expected ($H_E$) heterozygosity and deviation from Hardy-Weinberg equilibrium were calculated separately for each population using the 'genetics' package in R [74]. The inbreeding coefficient $F_{IS}$ was calculated as ($H_E$-$H_O$)/$H_E$.

The prcomp function in R was used to perform principal component analysis (PCA) of SNP genotypes. The number of significant principal components was estimated based on the Tracy-Widom distribution [75]. Individuals were assigned to groups based on significant PCs using Ward clustering via the hclust function of R.

Pairwise $F_{ST}$ [76] was calculated for each pair of genetic groups using all polymorphic SNPs. The relationships between genetic, geodetic and altitudinal distance were evaluated using Mantel tests and partial Mantel tests in the R package 'vegan' [77]. The significance of correlations between pairwise genetic distances measured as $F_{ST}/(1-F_{ST})$ and matrices of both geodetic distance and altitude differences among populations were estimated using 9999 permutations of rows and columns of the distance matrices. The partial Mantel test was used to test the effect of altitude and geodetic distance independent of each other.

Admixture and population structure were estimated using the software package STRUCTURE [78] based on genotypes at 10,000 random SNPs. Initial analyses included 279 maize inbred lines [27]. STRUCTURE was run under the admixture and correlated allele frequency model for a burn-in of 2500 iterations. Two independent runs of 20,000 MCMC iterations were performed for each value of K from 2 to 6 for analyses including maize, and for values of K from 2 to 9 for analyses including only teosinte samples. Results were inspected with STRUCTURE HARVESTER v0.6.8 [79] and visualized using DISTRUCT [80].

To test the placement of the Ahuacatitlan population relative to *mexicana* and *parviglumis*, distributions of D values [28] for three different divergence models were obtained. For each chromosome, a haplotype was sampled from each of three groups (Ahuacatitlan, *parviglumis* excluding Ahuacatitlan and *mexicana*) and the D value was calculated based on all informative sites. Sampling was repeated 1000 times to obtain the distribution of D. Homozygous SNPs from the sister genus *Tripsacum* were used to determine the ancestral state.

### Haplotype sharing and linkage disequilibrium

Haplotypes were inferred with fastPHASE [81] using known haplotypes of 18 teosinte inbred lines [49] and default parameters. Sites with residual heterozygosity were excluded. Segments of identity by state (IBS) and runs of homozygosity were identified using GERMLINE [82], with a seed segment size of 50 SNPs and allowing zero heterozygous and homozygous mismatches. Genetic map coordinates for each SNP were obtained from a modified map of the maize IBM mapping population [73]. Markers for which the genetic and physical map disagreed were omitted from haplotype based analyses.

The software TASSEL [83] was used to calculate linkage disequilibrium ($r^2$ and p-value) using phased data from all individuals and pairs of sites with minimum allele frequency >0.1.

### Candidate SNP identification

BAYENV [29] was used to evaluate the correlation between environmental variables and allele frequencies of individual SNPs. A random set of 10,000 SNPs was used to make three covariance matrices: all populations and separately for each subspecies without the Ahuacatitlan *parviglumis* population. For each covariance matrix, two independent runs of 50,000 iterations were compared to control for convergence. Maximum differences between the two independent estimates of covariance matrices were always less than ~10% of the smallest estimated covariance, indicating good convergence between runs.

A number of environmental variables were analyzed, including 8 soil variables, 19 bioclimatic variables, monthly precipitation and monthly mean, maximum and minimum temperature and altitude (Table S8). For climatic variables, 30 arc-second (~1 km) resolution climate data was downloaded from www.worldclim.org, and DIVA-GIS [84] was used to extract climate data for the population locations. Data for three key soil qualities (rooting conditions, oxygen availability to roots and workability) that varied among populations were downloaded from the Harmonized World Soil Database and data on 5 varying (cracking clays, volcanic, top soil clay, top soil sand, top soil loam) key modifier layers [85] were downloaded from www.harvestchoice.org. All variables were standardized to a mean of 0 and standard deviation of 1. The dimensionality of environmental data was reduced using principal component analysis (prcomp in R). The PCs that captured 95% of the variance in environments among populations were used as environmental variables for BAYENV (Tables S9-S11).

Five independent BAYENV runs with 1,000,000 iterations were used to identify associated SNPs with these PCs. SNPs were considered as candidates if they showed average Bayes factors across runs in the 99th percentile and were consistently in the 95th percentile of each run. A gene was considered as a candidate when one of the candidate SNPs was in the transcribed part of the gene.

To obtain the expected distribution of heterozygosity and the hierarchical differentiation statistics $F_{CT}$ (among subspecies) and $F_{ST}$ (among 20 populations, excluding Ahuacatitlan), 100,000 coalescent simulations were conducted under a hierarchical island model of two groups of 100 demes using Arlequin [86]. The maximum expected heterozygosities for simulations were 0.5.

To identify SNPs underlying differentiation at the level of individual populations, F statistics were calculated in a hierarchical framework using the R package 'hierfstat' [87,88]. Variance components were calculated for three levels within each subspecies: population, focal, and individual. The focal component was calculated for each population and locus, yielding $F_{FT} = \frac{\hat{\sigma}_F^2}{\hat{\sigma}_F^2 + \hat{\sigma}_S^2 + \hat{\sigma}_I^2 + \hat{\sigma}_E^2}$, the proportion of total variance due to differentiation between a focal population and all other populations, where $\hat{\sigma}_F^2$, $\hat{\sigma}_S^2$, $\hat{\sigma}_I^2$ and $\hat{\sigma}_E^2$ are the estimates of variance between focal and other populations, among populations, among individuals and error, respectively. Negative variance components were set to zero. For each population, SNPs with $F_{FT}$ values above the 99th percentile were considered candidates.

**Trait association analysis**

Association mapping tests were carried out on 278 inbred maize lines (Text S1) from the association panel described in [31], which have been genotyped on the Illumina MaizeSNP50 array [27] and phenotyped for 36 traits (127 trait/environment combinations) [32]. The 51,253 SNPs with minor allele frequency >0.05 were tested against at least one of the phenotypes. Associations were tested using the model

$$\hat{G} = \mathbf{1}\mu + M\theta + S\beta + Zu + \varepsilon$$

where $\hat{G}$ is the vector of the estimated genetic values of a given trait, $\mu$ is the trait mean, $M$ is the tested SNP, $\theta$ is the SNP effect, $S$ is the matrix of the panel structure as estimated by [31], $\beta$ is the vector of structure effects, $Z$ is an incidence matrix, $u$ is a vector of random effects assumed to follow a distribution $N(0, \sigma_g^2 K)$ where $K$ is the genetic variance-covariance matrix modeled by a shared allele matrix, and $\varepsilon$ is the model residual assumed to follow $N(0, \sigma_\varepsilon^2 I_n)$. Only SNPs significantly associated to a given trait with p-values < 0.05 were retained for further analyses. To test whether enrichment for flowering time loci was affected by a loss of power due to correction for population structure, we also conducted association analyses for flowering time traits using a simplified model without population structure.

### Enrichment analyses

Enrichment of SNPs within inversions was inspected by ranking SNPs in each candidate list and calculating the proportion of SNPs within inversions in each candidate list. The joint effect of all inversions across all candidate lists was determined by assigning each SNP a maximum rank (more significant ranks are higher) across all candidate lists, and calculating the proportion of SNPs from inversions above the 99th quantile. Bootstrapping was used to obtain the statistical significance.

## Acknowledgements
We thank Jeff Glaubitz for SNP annotations and mapping, Lauren Sagara for technical assistance, Justin Gerke for modifying the IBM map, and Norm Ellstrand,

## Figure legends

**Figure 1. Map of sampled *Zea mays* ssp. *parviglumis* and ssp. *mexicana* populations.**

**Figure 2. Diversity statistics.** A. Proportion of SNPs deviating from Hardy-Weinberg Equilibrium (HWE), proportion of polymorphic SNPs, and mean inbreeding coefficient $F_{IS}$. B. Length and number of runs of homoyzygosity (ROH) and average pairwise length of genomic segments identical by state (IBS).

**Figure 3. Linkage disequilibrium reveals structural rearrangements in teosinte.** Shown are linkage disequilibrium ($r^2$, red) and permutation p-value (black) for pairs of SNPs across chromosomes 1, 4 and 9. Dashed black lines delineate the likely boundaries of structural variants discussed in the text.

**Figure 4. Population structure in teosinte.** A. Principal component analysis of all individuals, labeled according to the sampled population. B. STRUCTURE results for all individuals. Individuals are grouped by population, and populations ordered by increasing altitude.

**Figure 5. Overlap among methods for detecting local adaptation.** Shown is the joint distribution of heterozygosity and $F_{ST}$ under a hierarchical island model (grey dots) The red line indicates the 1 % tail of $F_{ST}$ based on simulations. A. SNPs that have Bayes factors higher than 400 for PC2 are shown in black. B. SNPs that have a significant $F_{FT}$ value in each population (populations are colored as in Figure 3.).

**Figure 6. Association and differentiation in teosinte.** Values are plotted across all 10 chromosomes, with each chromosome plotted in a different color. Black dots represent SNPs considered to be outliers, and black horizontal bars below the plots indicate the positions of inversions in each graph. A. Bayes factors for PC1 in *mexicana*. B. $F_{ST}$ across 20 populations C. $F_{CT}$ between subpscies.

## Supporting information captions

**Figure S1.** Distribution of total expected heterozygosity within *parviglumis* and *mexicana* across all loci.

**Figure S2.** Runs of homozygosity across all 10 chromosomes. Each individual is one horizontal line in the graph. Populations are separated by colors.

**Figure S3.** LD in chromosome 9 among *mexicana* populations based on SNPs with minor allele frequency >0.1

**Figure S4.** Neighbor-joining tree for *Inv1n* based on percentage of SNPs differing among haplotypes. SNPs with missing data were excluded from pairwise comparisons. The tree was rooted based on *Tripsacum*. Red: *parviglumis*, orange, *mexicana*, green: maize, purple: *Tripsacum*. Black branches indicate the inverted haplotype.

**Figure S5.** Neighbor-joining tree for *Inv4m* based on percentage of SNPs differing among haplotypes. SNPs with missing data were excluded from pairwise comparisons. The tree was rooted based on *Tripsacum*. Red: *parviglumis*, orange, *mexicana*, green: maize, purple: *Tripsacum*. Black branches indicate the inverted haplotype.

**Figure S6.** Neighbor-joining tree for *Inv9d* based on percentage of SNPs differing among haplotypes. SNPs with missing data were excluded from pairwise comparisons. The tree was rooted based on *Tripsacum*. Red: *parviglumis*, orange, *mexicana*, green: maize, purple: *Tripsacum*. Black branches indicate the inverted haplotype.

**Figure S7.** Neighbor-joining tree for *Inv9e* based on percentage of SNPs differing among haplotypes. SNPs with missing data were excluded from pairwise comparisons. The tree was rooted based on *Tripsacum*. Red: *parviglumis*, orange, *mexicana*, green: maize, purple: *Tripsacum*. Black branches indicate the inverted haplotype. Black dashed outer lines indicate parts of the tree that are putatively part of the inverted group.

**Figure S8.** The proportion of individuals sharing IBD segments with *parviglumis* (red) or *mexicana* (orange) for each *parviglumis* population along the genome, scaled in genetic distance. Within population and within individual comparisons are excluded. Numbers on the x-axis indicate chromosome midpoints. Total length of the genetic map is ~3700 cM.

**Figure S9**. The proportion of individuals sharing IBD segments with *parviglumis* (red) or *mexicana* (orange) for each *mexicana* population along the genome scaled in genetic distance. Within population and within individual comparisons are excluded. Numbers on the x-axis indicate chromosome midpoints. Total length of the genome is ~3700 cM.

**Figure S10.** Altitudinal clines of three inversions presented as a relationship between altitude and haplotype distance within each inversion. Distance (as a number of SNPs for which they differ) of each haplotype from the most distal haplotype in the main low diversity haplotype group is in the y-axis. Colors indicate populations. A. *Inv1n*, B. *Inv4m* and C. *Inv9d*.

**Figure S11.** Third and fourth principal component from PCA on *parviglumis* and *mexicana*.

**Figure S12.** Structure results for joint analysis of 130 *parviglumis*, 120 *mexicana* and 279 maize lines for K values 2-6, based on 20 000 iterations with a 2500 step burn-in. Colors represent each individual's proportion of ancestry in each genetic group.

**Figure S13.** Dendrogram based on Ward clustering of individuals using the first 20 PCs. Red dashed line indicates the segment that divides individuals into 21 groups.

**Figure S14.** Isolation by geodetic and altitudinal distance among all populations and within both subspecies. Geodetic or altitudinal distance on y-axis and genetic differentiation on x-axis. P-values are from Mantel (A-C) and partial Mantel tests (A),

where correlation with both geodetic and altitude was conditioned on the other variable.

**Figure S15.** Distribution of D values for three possible divergence scenarios of Ahuacatitlan population, *parviglumis* and *mexicana*. The center of expected distribution (0) is marked with vertical dashed line.

**Figure S16.** Natural logarithm of Bayes factors for each of six PCs along chromosomes for the joint analysis of 20 *parviglumis* and *mexicana* populations. Inversions are marked as horizontal segments.

**Figure S17**. Natural logarithm of Bayes factors for each of four PCs along chromosomes for the analysis of 10 *parviglumis* populations. Inversions are marked as horizontal segments.

**Figure S18.** Natural logarithm of Bayes factors for each four PCs along chromosomes for the analysis of 10 *mexicana* populations. Inversions are marked as horizontal segments.

**Figure S19.** Joint distribution of heterozygosity with $F_{CT}$ (A) and $F_{ST}$ (B) under a hierarchical model. Black lines indicate 1% cutoff based on simulations. The six SNPs investigated in detail are shown in red.

**Figure S20.** Boxplot of p-value distribution for $F_{ST}$ outliers for all SNPs (All) and for SNPs that are $F_{FT}$ outliers in each population.

**Figure S21** Boxplot of the logarithm of Bayes factors for PC2 among *mexicana* populations for all SNPs (All) and for SNPs that are $F_{FT}$ outliers in each *mexicana* population.

**Figure S22.** Diversity and environmental association along *mexicana* chromosome 9. A) Per SNP expected heterozygosity $H_E$. Darker lines indicate higher elevation populations. B) Natural logarithm of Bayes factors for PC1. Candidate SNPs for PC1 are indicated as black circles.

**Figure S23.** Neighbour-joining tree based on pairwise $F_{ST}$ among groups identified by PCA.

**Figure S24.** Principal component analysis of 130 *parviglumis*, 120 *mexicana* and 279 maize lines based on 37,021 SNPs. Two outlier *mexicana* individuals are likely recent hybrids between maize and *mexicana*.

**Table S1.** Summary statistics of genetic diversity for each population.

**Table S2.** Candidate status for each 36,719 SNPs in filtered dataset. Marker: marker name at panzea.org, Locus: Original SNP name, Chr: chromosome, Pos: B73

RefGen_v2 position (bp), Type: annotation for SNPs in coding regions. Column names indicate different candidate sets. TRUE: the SNP is a candidate, FALSE: the SNP is not a candidate, NA: the SNP was not included in the analysis.

**Table S3.** Spearman's correlation between Bayes factors and p-values for $F_{CT}$ and $F_{ST}$.

**Table S4.** Enrichment of functional properties among candidate SNPs and genes.

**Table S5.** Enrichment of SNPs that are associated with a maize phenotypic trait for each list of adaptation candidates. GWAS was done using a mixed linear model that takes both population structure and relatedness into account.

**Table S6.** Enrichment of SNPs that are associated with a maize phenotypic trait for each list of adaptation candidates. GWAS was done using a simple model not taking population structure into account.

**Table S7.** Information about sampled populations ordered by ascending altitude.

**Table S8.** Environmental variables and abbreviations used in the study.

**Table S9.** Variable loadings/rotations for each of 6 PCs that were used in BAYENV for the joint dataset of 20 *parviglumis* and *mexicana* populations.

**Table S10.** Variable loadings/rotations for each of 4 PCs that were used in BAYENV for 10 *parviglumis* populations

**Table S11.** Variable loadings/rotations for each of 4 PCs that were used in BAYENV for 10 *mexicana* populations

**Text S1.** List of 278 maize inbred lines used in the association analysis

# Tables

## Table 1. Summary of environmental correlation and differentiation outlier results.

| Analysis | Variable | Major loadings | BF, 99th[a] | No. cand SNPs | No. cand genes |
|---|---|---|---|---|---|
| Both | PC1 | altitude, temperature | 116.8 | 262 | 162 |
|  | PC2 | temperature seasonality, soil quality and precipitation | 400.0 | 359 | 222 |
|  | PC3 | precipitation | 40.4 | 308 | 201 |
|  | PC4 | topsoil variables and precipitation seasonality | 39.4 | 229 | 145 |
|  | PC5 | mean diurnal range of temperature plus some soil variables | 72.9 | 291 | 184 |
|  | PC6 | mean diurnal range of temperature plus some soil variables | 10.2 | 225 | 151 |
| *parviglumis* | PC1 | altitude, temperature | 2.2 | 60 | 36 |
|  | PC2 | temperature range and seasonality | 6.4 | 173 | 118 |
|  | PC3 | soil type and precipitation | 2.0 | 102 | 69 |
|  | PC4 | monthly precipitation and mean diurnal range of temperature | 4.8 | 42 | 24 |
| *mexicana* | PC1 | altitude, temperature | 6.9 | 108 | 69 |
|  | PC2 | temperature seasonality and range and precipitation. | 52.3 | 309 | 204 |
|  | PC3 | precipitation and volcanic soil | 31.1 | 27 | 16 |
|  | PC4 | top soil variables and temperature variability | 3.8 | 93 | 58 |
| $F_{CT}$ |  |  |  | 731 | 728 |
| $F_{ST}$ |  |  |  | 1363 | 411 |

[a] Bayes factor at 99th percentile of distribution

**Table 2. Enrichment of sets of candidate SNPs within four inversions.**

| Analysis | Variable | p-value Inv1n | Inv4m | Inv9d | Inv9e | All inversions |
|---|---|---|---|---|---|---|
| Both | PC1 | <0.001 (4.5) | 0.003 (3.7) | <0.001 (7.6) | 0.027 (2.2) | 0.0001 (3.3.) |
|  | PC2 | 0.032 (1.7) | 0.824 | <0.001 (3.9) | 0.331 | 0.0016 (1.8) |
|  | PC3 | <0.001 (5.7) | 1.000 | 0.819 | 0.186 | 0.0001 (2.2) |
|  | PC4 | 0.002 (2.7) | 0.574 | 0.001 (3.1) | <0.001 (4.0) | 0.0001 (2.5) |
|  | PC5 | 0.440 | 1.000 | 1.000 | 0.938 | 0.9942 |
|  | PC6 | 0.672 | 1.000 | 0.859 | 0.687 | 0.9745 |
| *parviglumis* | PC1 | 0.450 | 1.000 | 0.290 | 1.000 | 0.0181 (1.5) |
|  | PC2 | 0.616 | 0.016 (3.2) | 0.045 (2.3) | 0.728 | 0.1555 |
|  | PC3 | 0.261 | 1.000 | 1.000 | 0.812 | 0.3764 |
|  | PC4 | 0.271 | 1.000 | 1.000 | 0.474 | 0.4543 |
| *mexicana* | PC1 | 0.001 (3.8) | 1.000 | <0.001 (33.0) | 0.095 | 0.0001 (3.2) |
|  | PC2 | 0.748 | 0.694 | <0.001 (4.3) | 0.027 (2.0) | 0.0007 (1.8) |
|  | PC3 | 0.136 | 1.000 | 1.000 | 1.000 | 0.799 |
|  | PC4 | 0.002 (3.9) | 1.000 | 1.000 | <0.001 (8.2) | 0.0001 (2.0) |
| $F_{CT}$ |  | <0.001 (2.9) | <0.001 (11.7) | 0.049 (1.6) | 0.705 | 0.0001 (3.4) |
| $F_{ST}$ |  | <0.001 (2.7) | <0.001 (6.3) | <0.001 (1.8) | 0.866 | 0.0001 (3.2) |

For each p-value < 0.05 the fold enrichment is reported in parenthesis. P-values are based on bootstrapping.

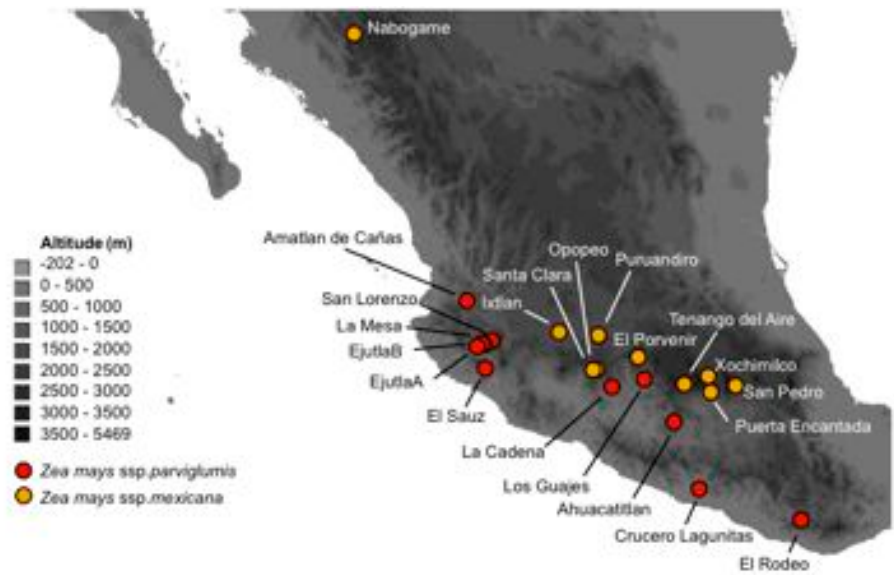

Figure 1

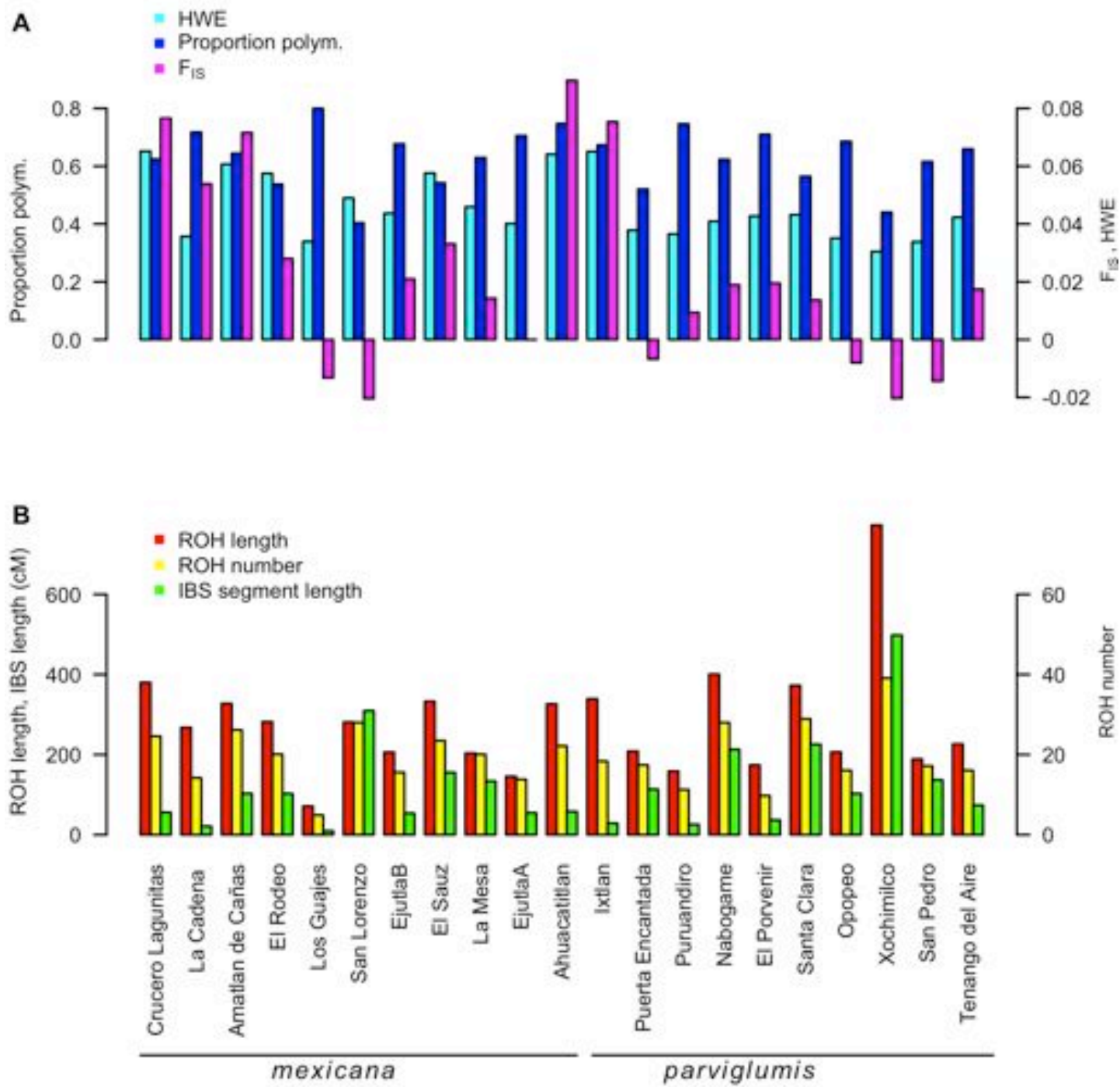

Figure 2

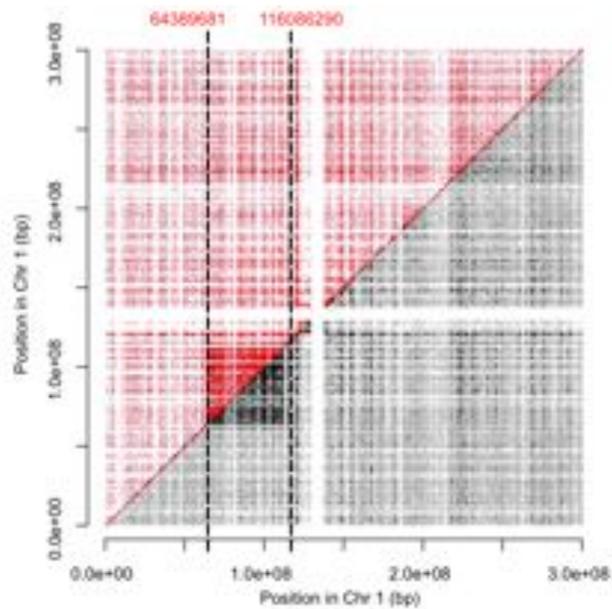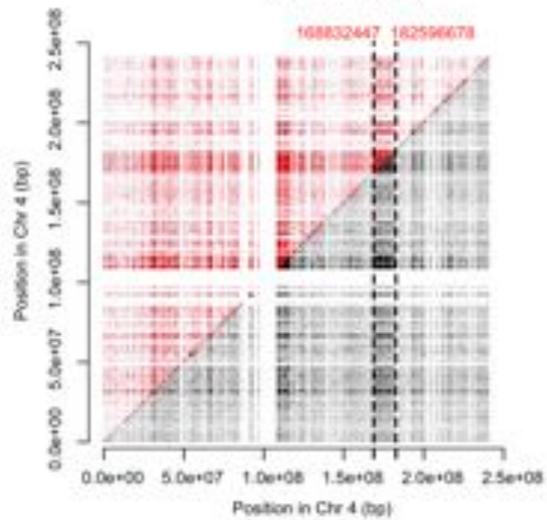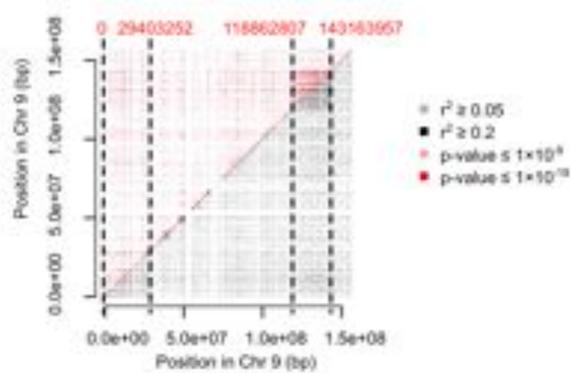

Figure 3

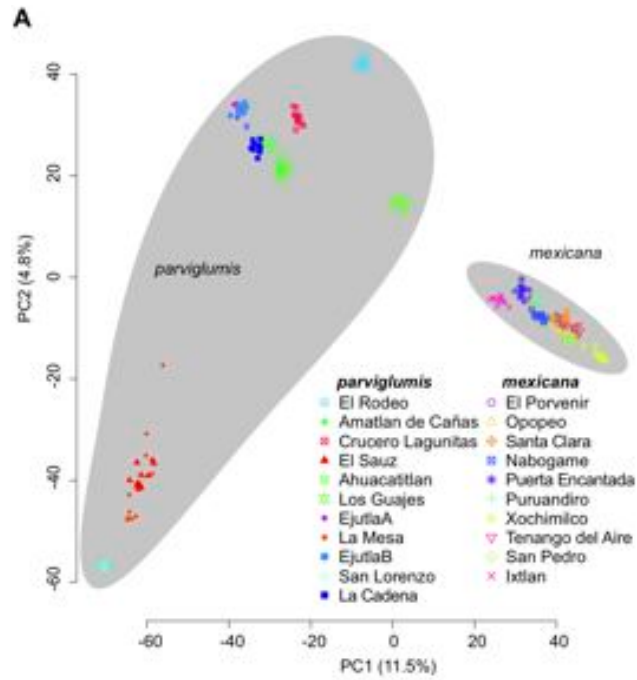
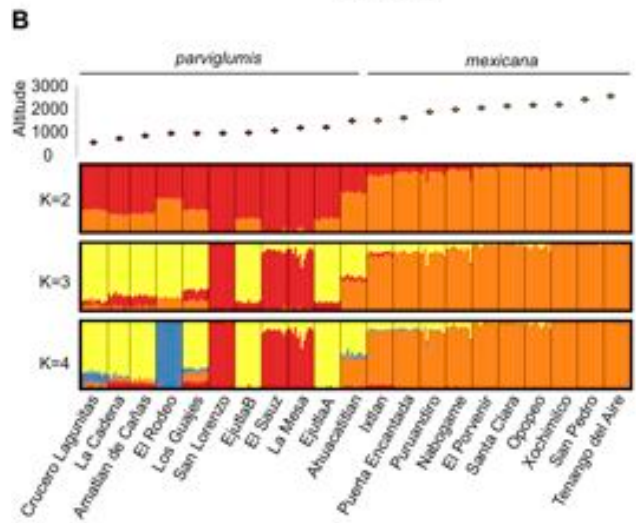

Figure 4

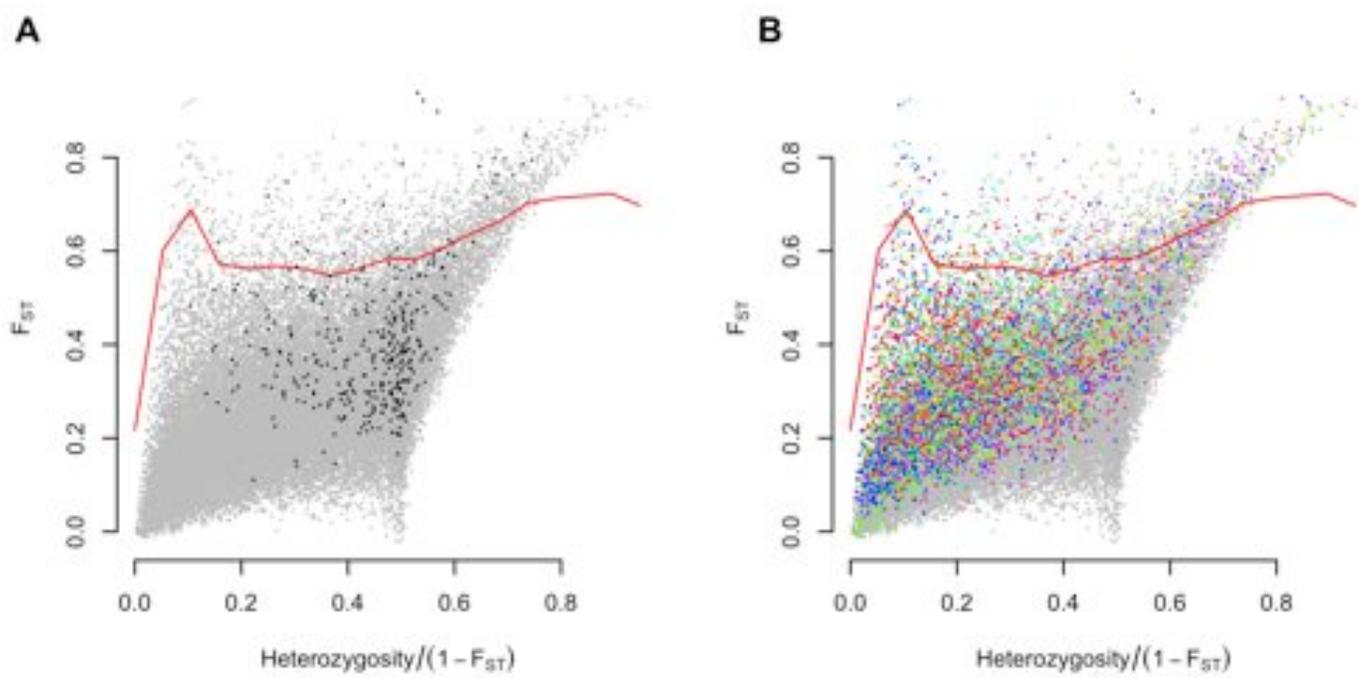

Figure 5

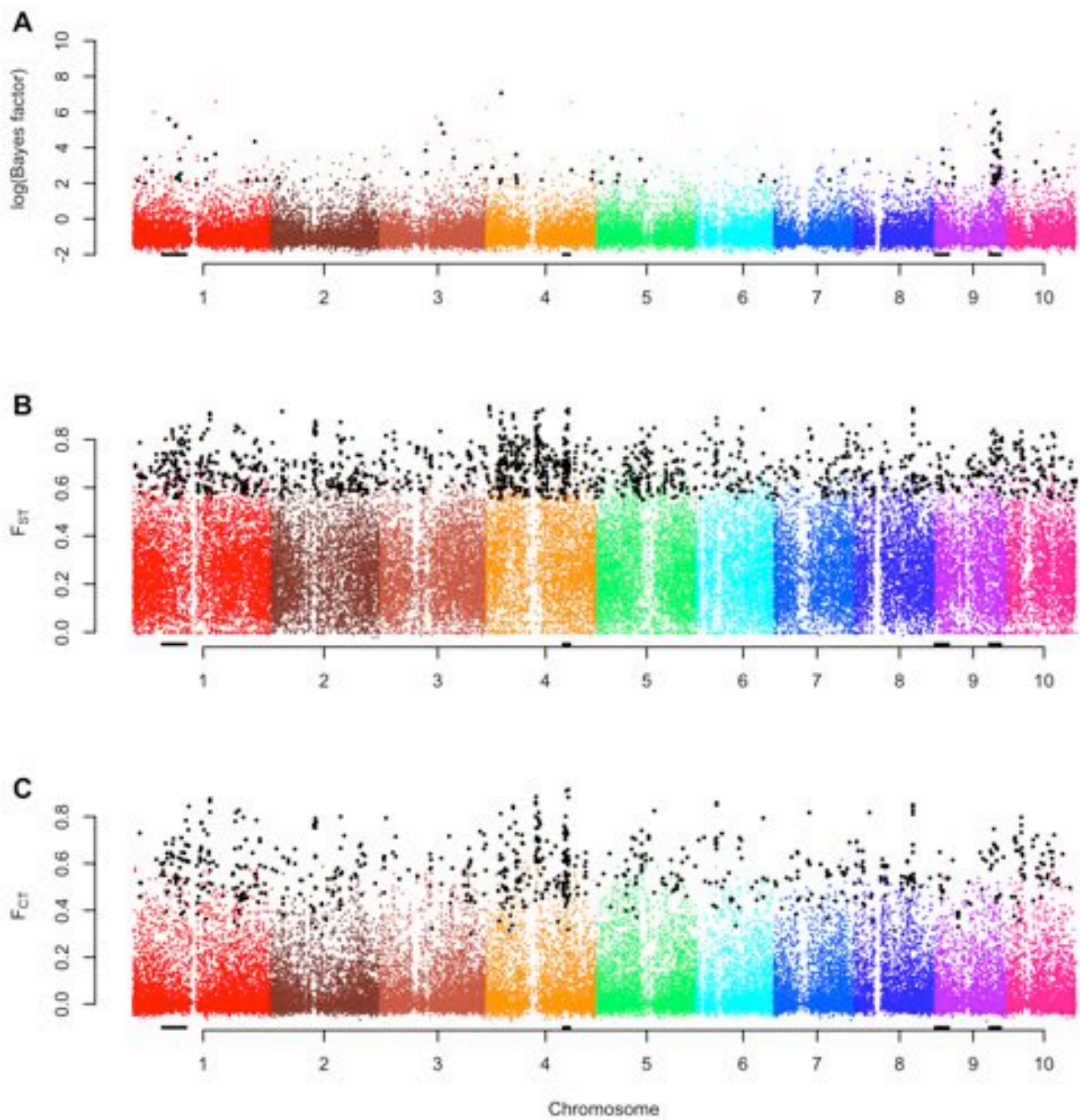

Figure 6

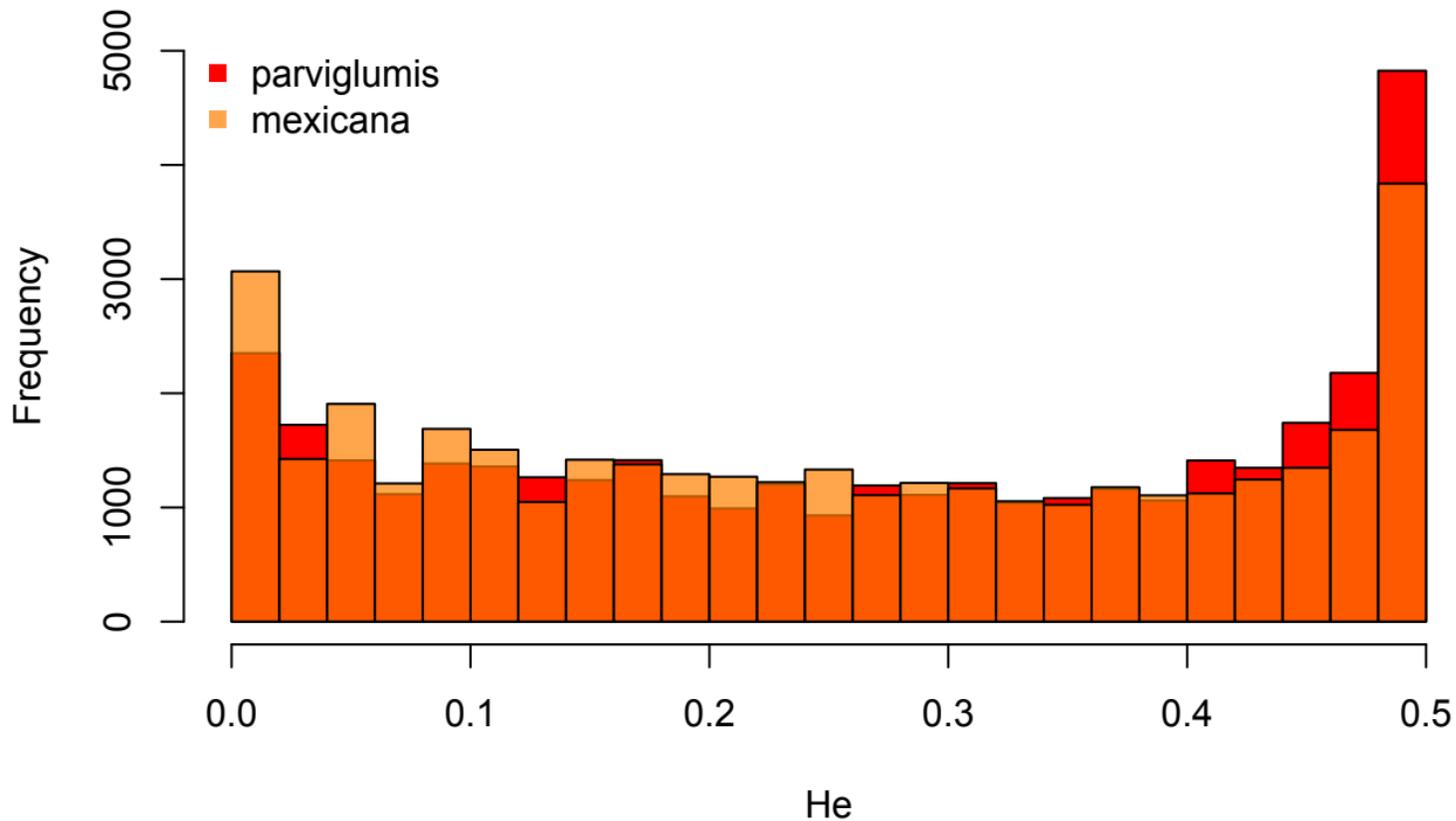

Figure S1

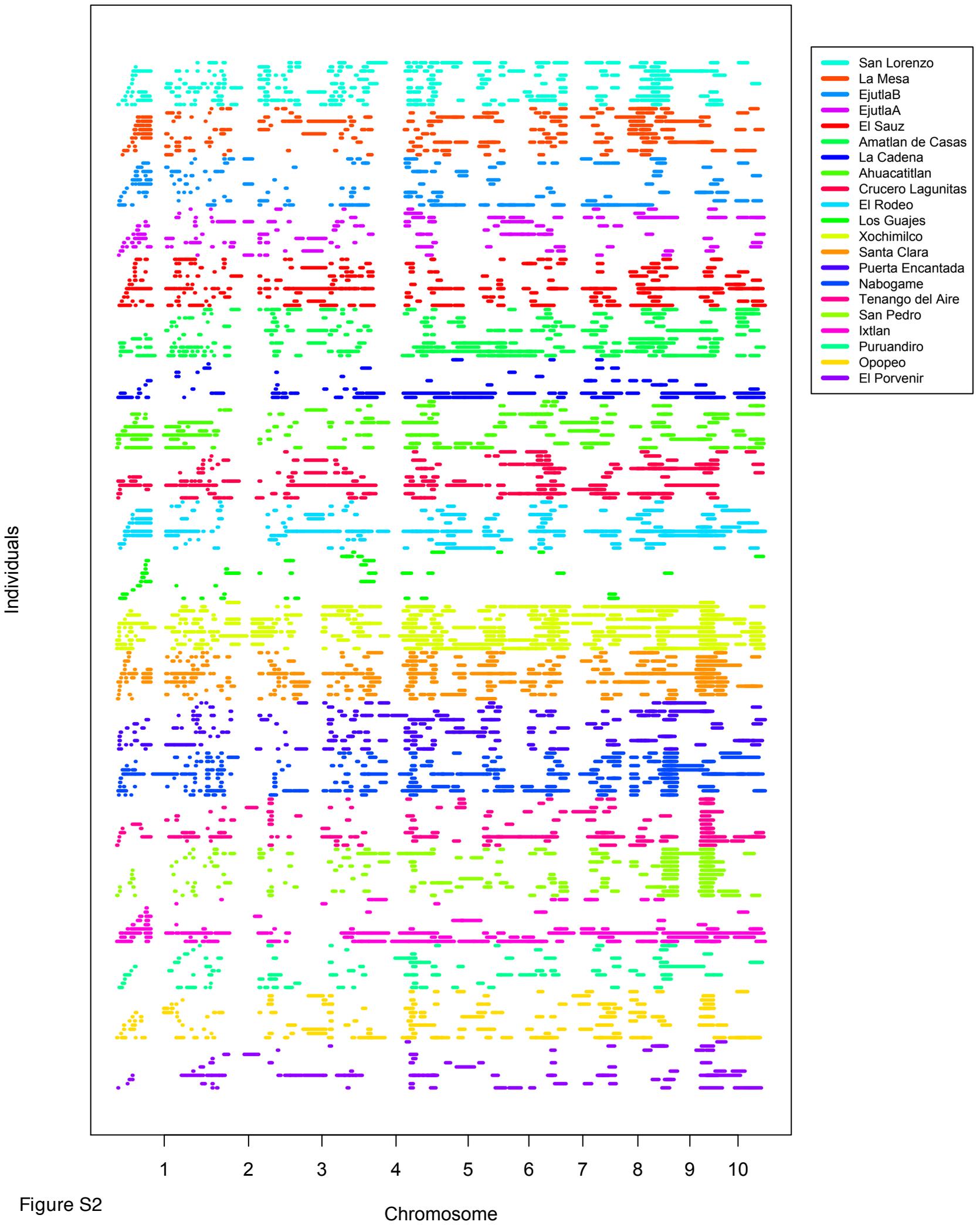
Figure S2

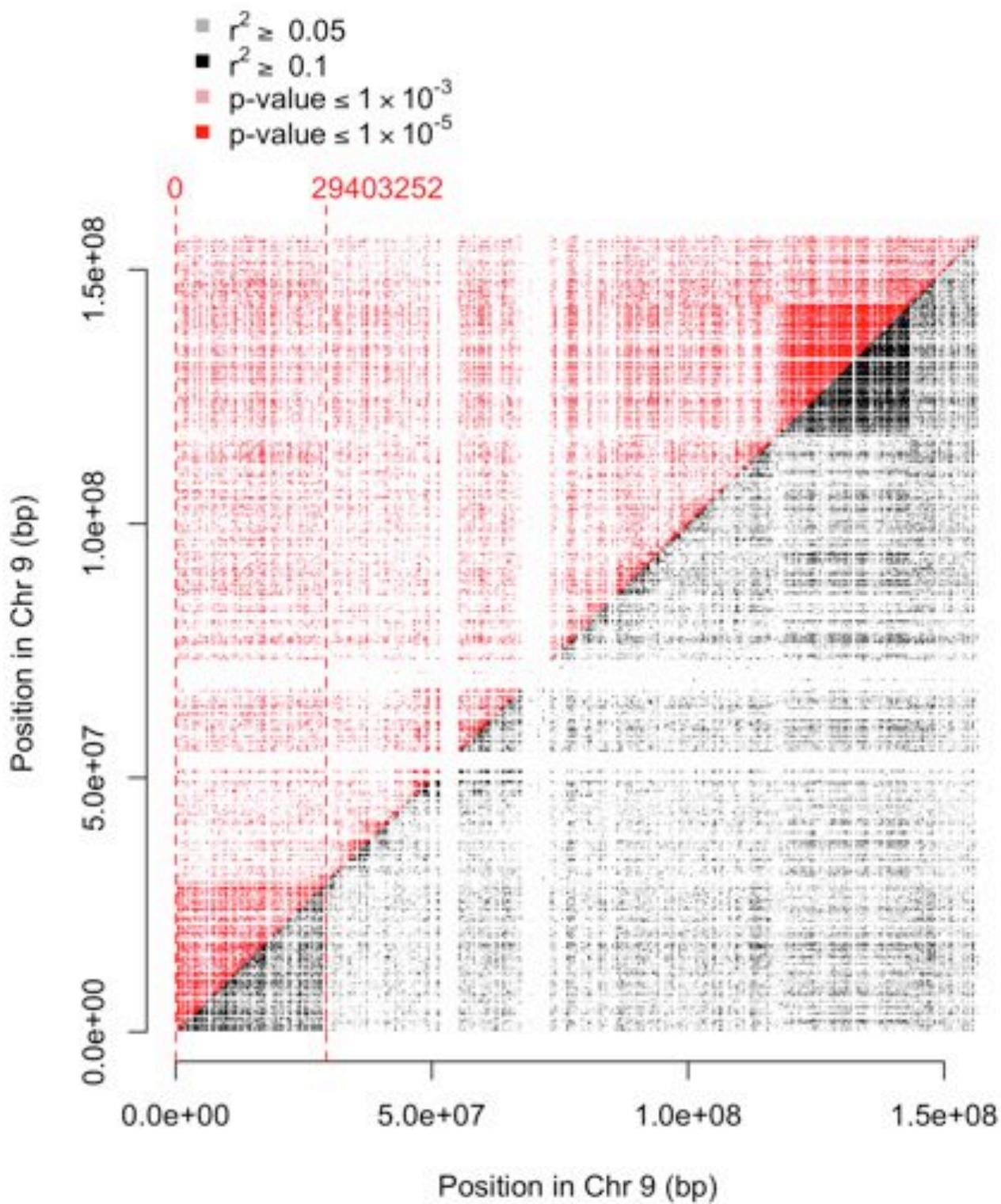

Figure S3

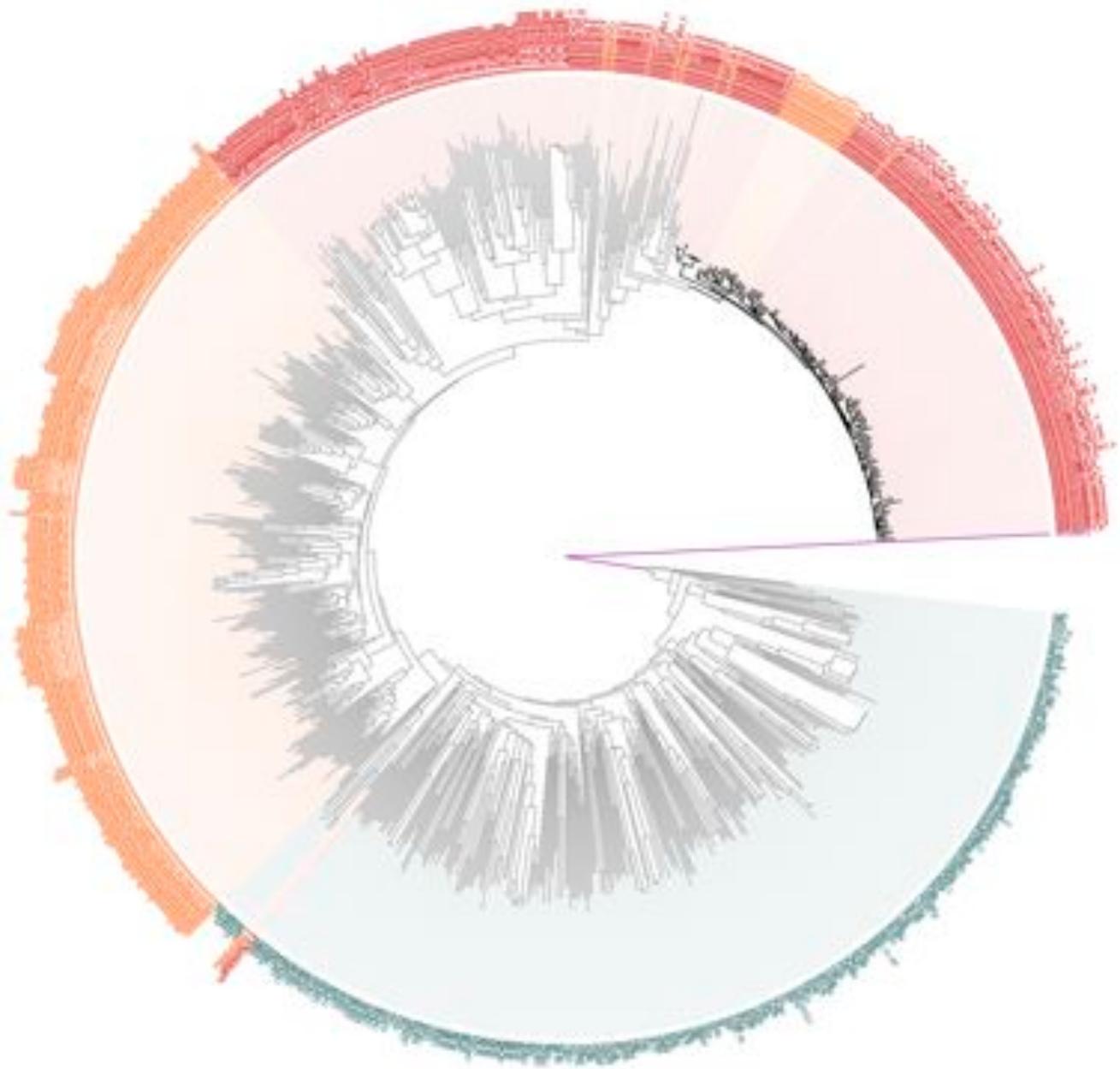

Figure S4

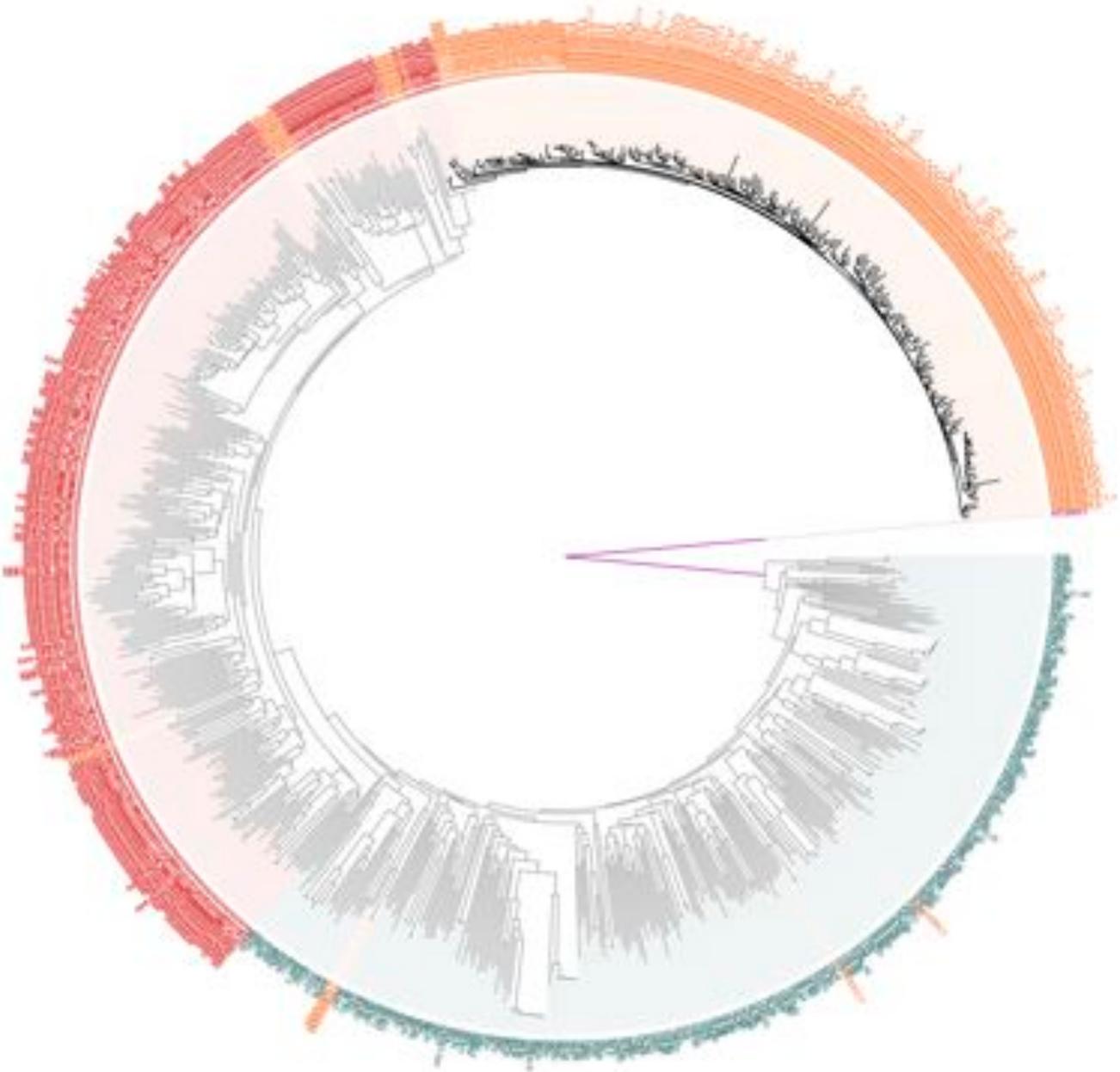

Figure S5

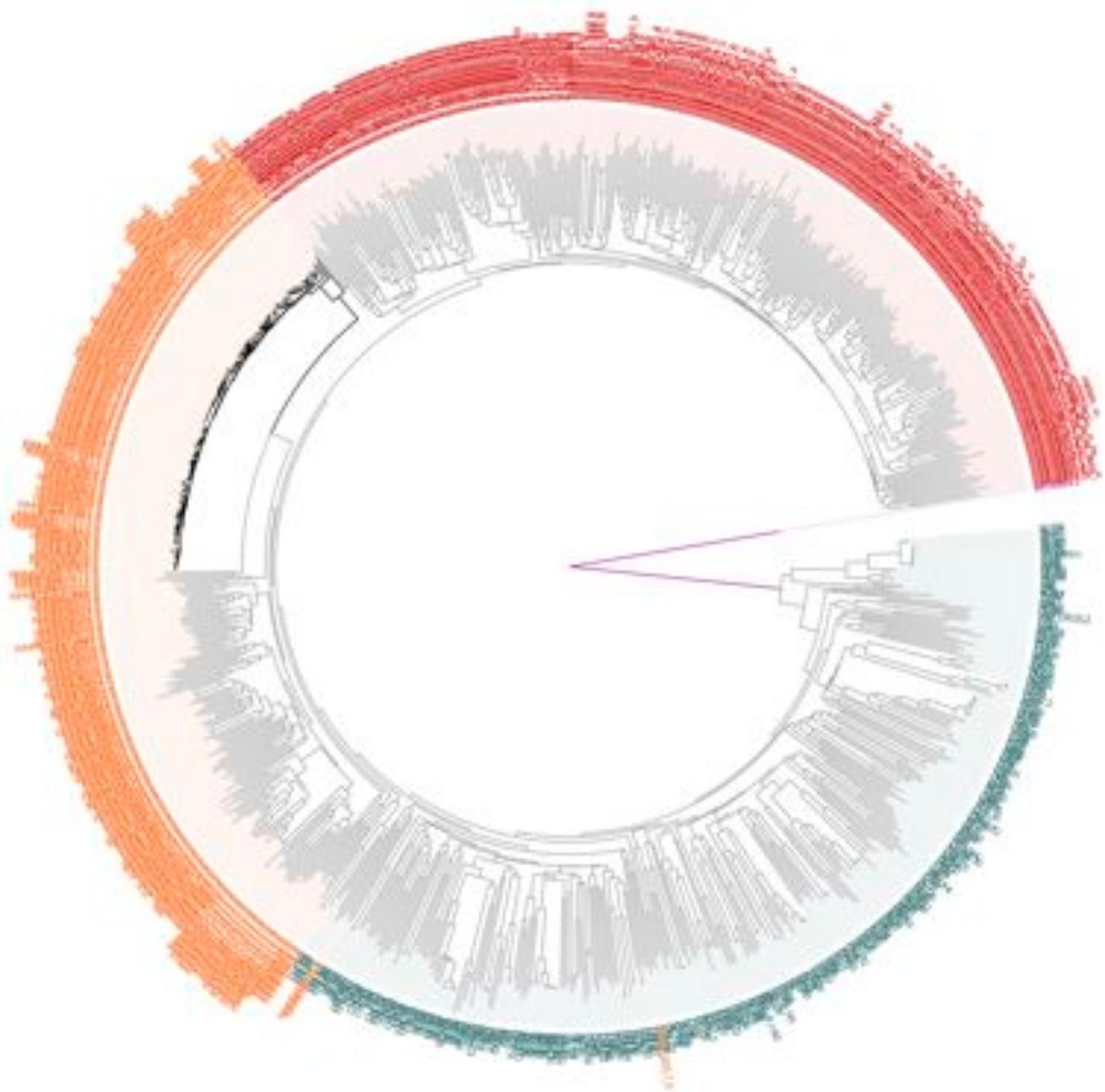

Figure S6

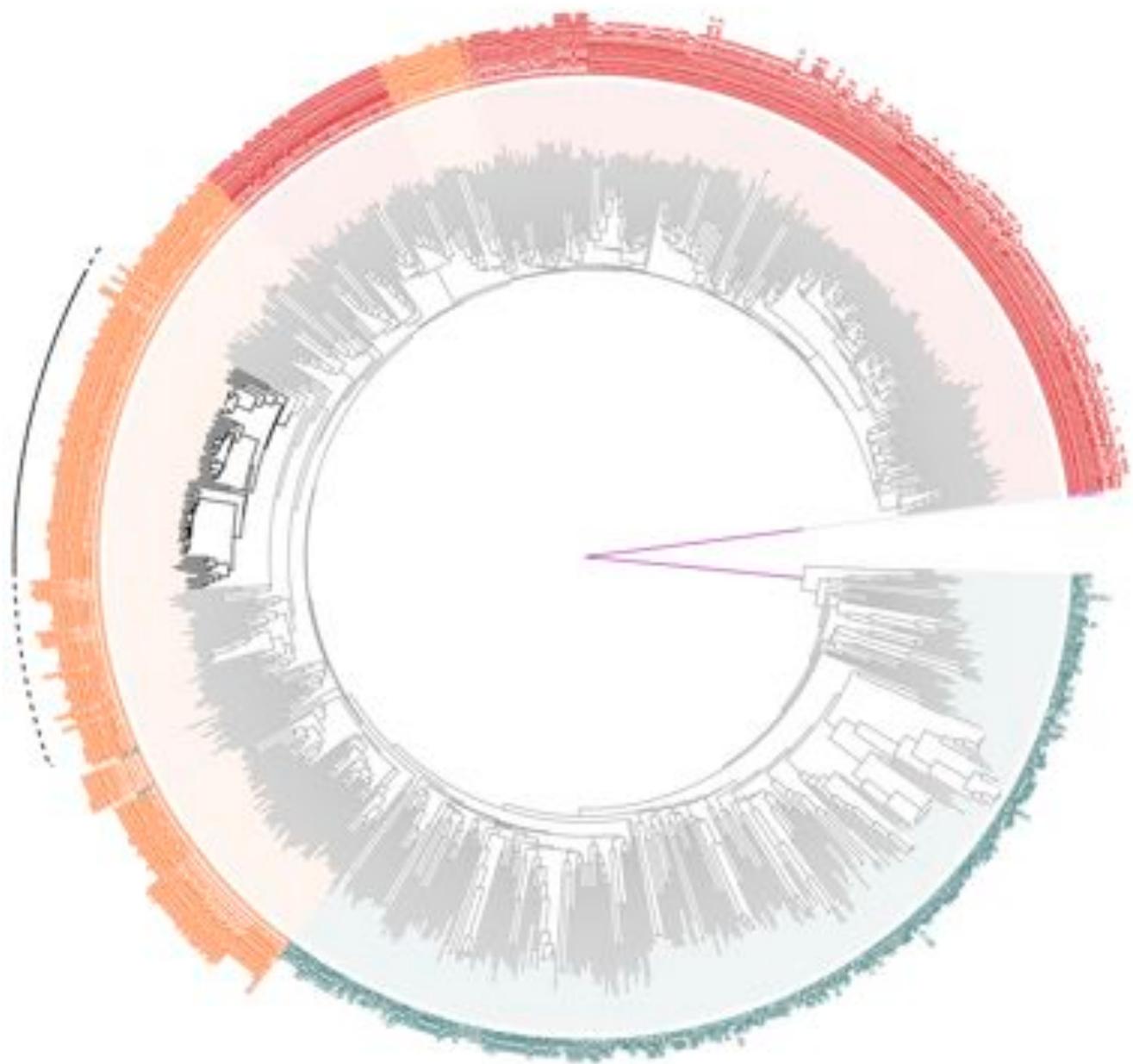

Figure S7

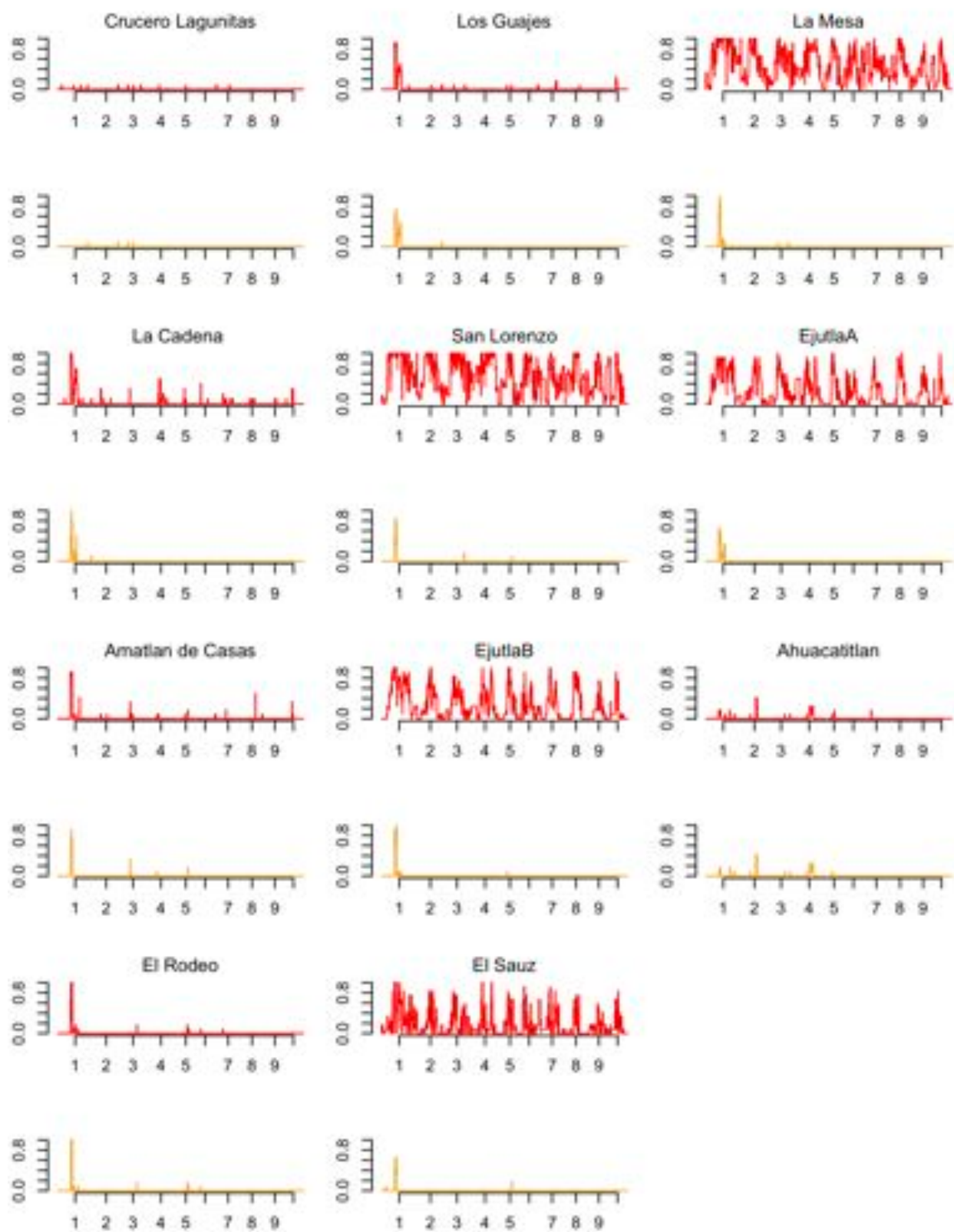

Figure S8

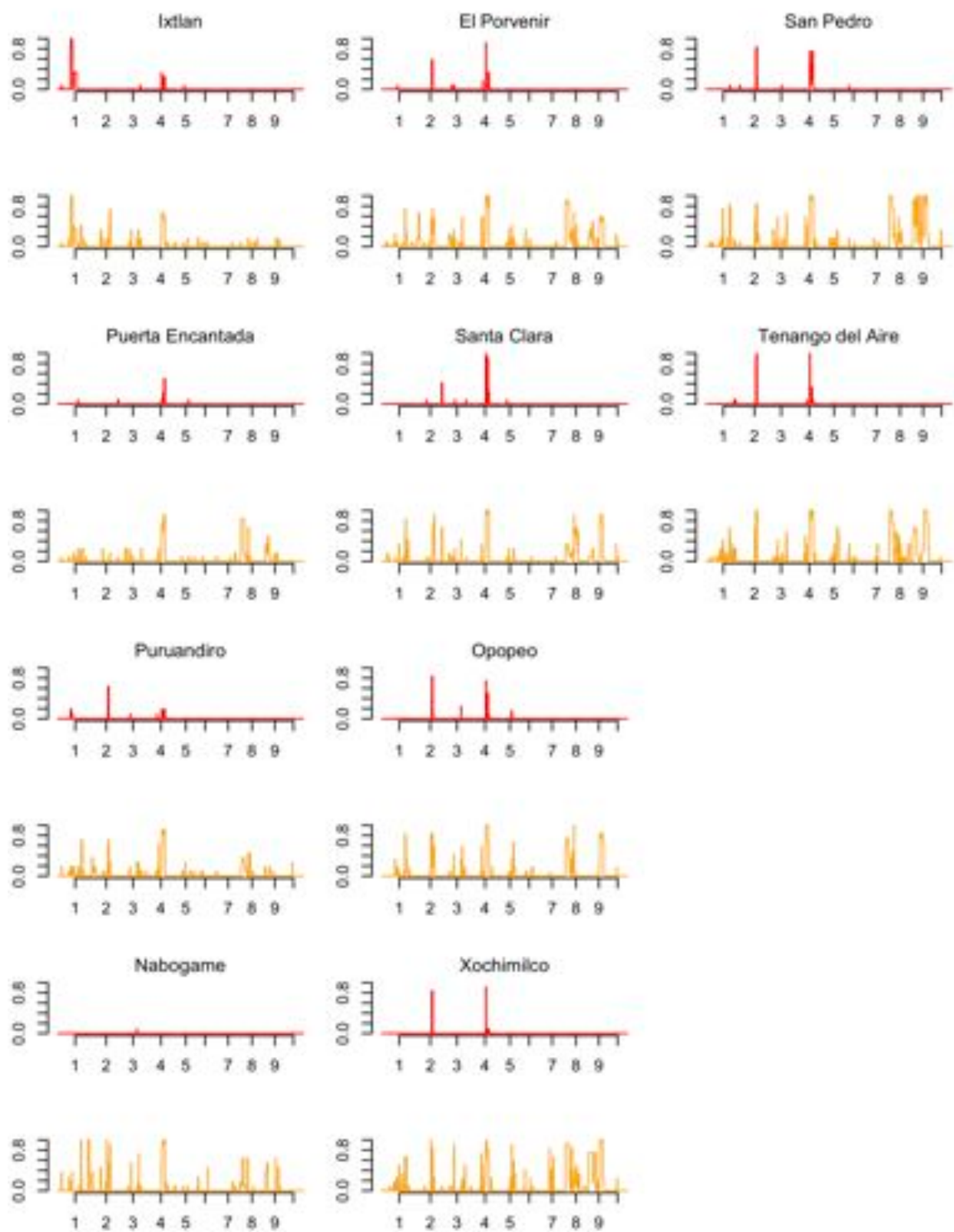

Figure S9

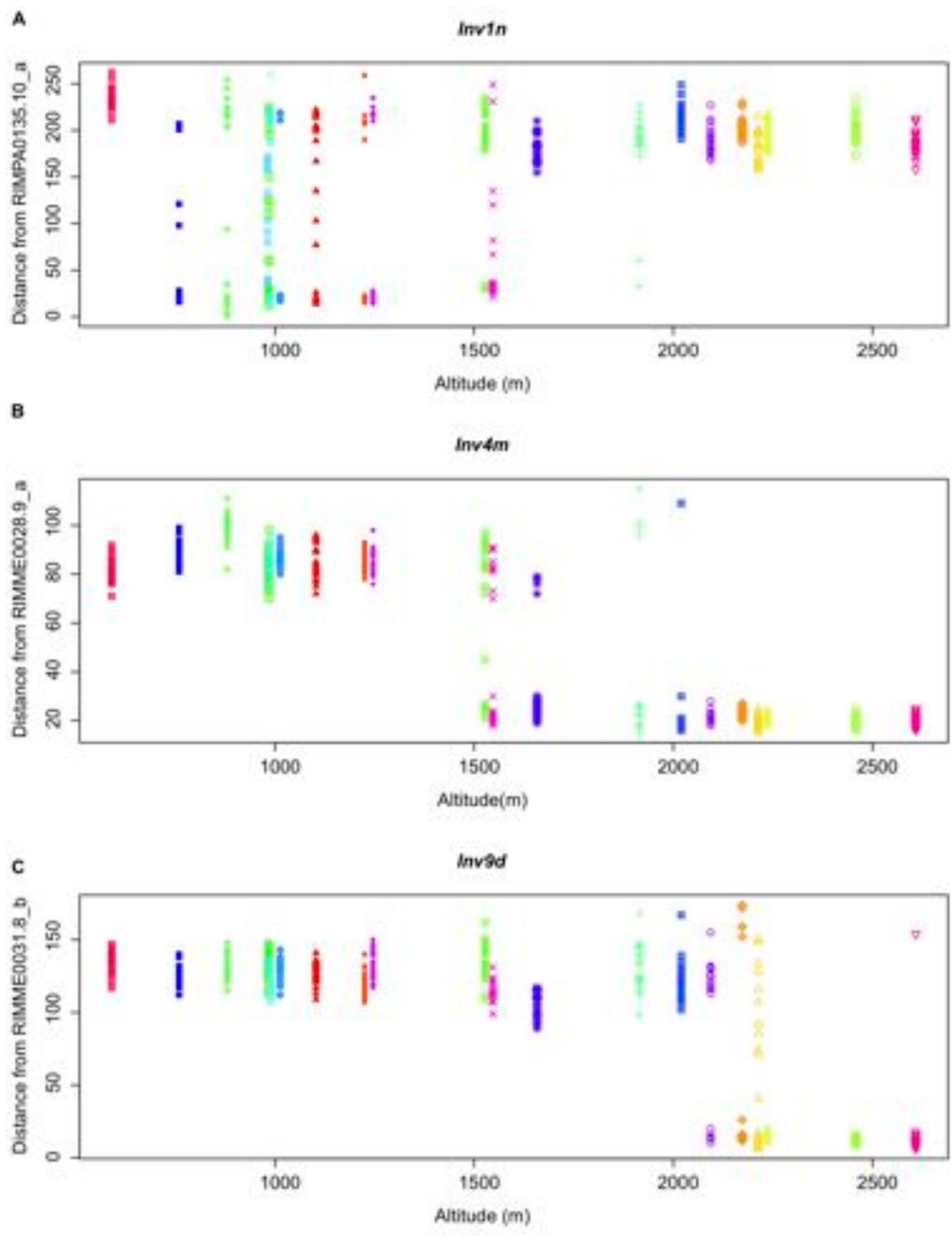

Figure S10

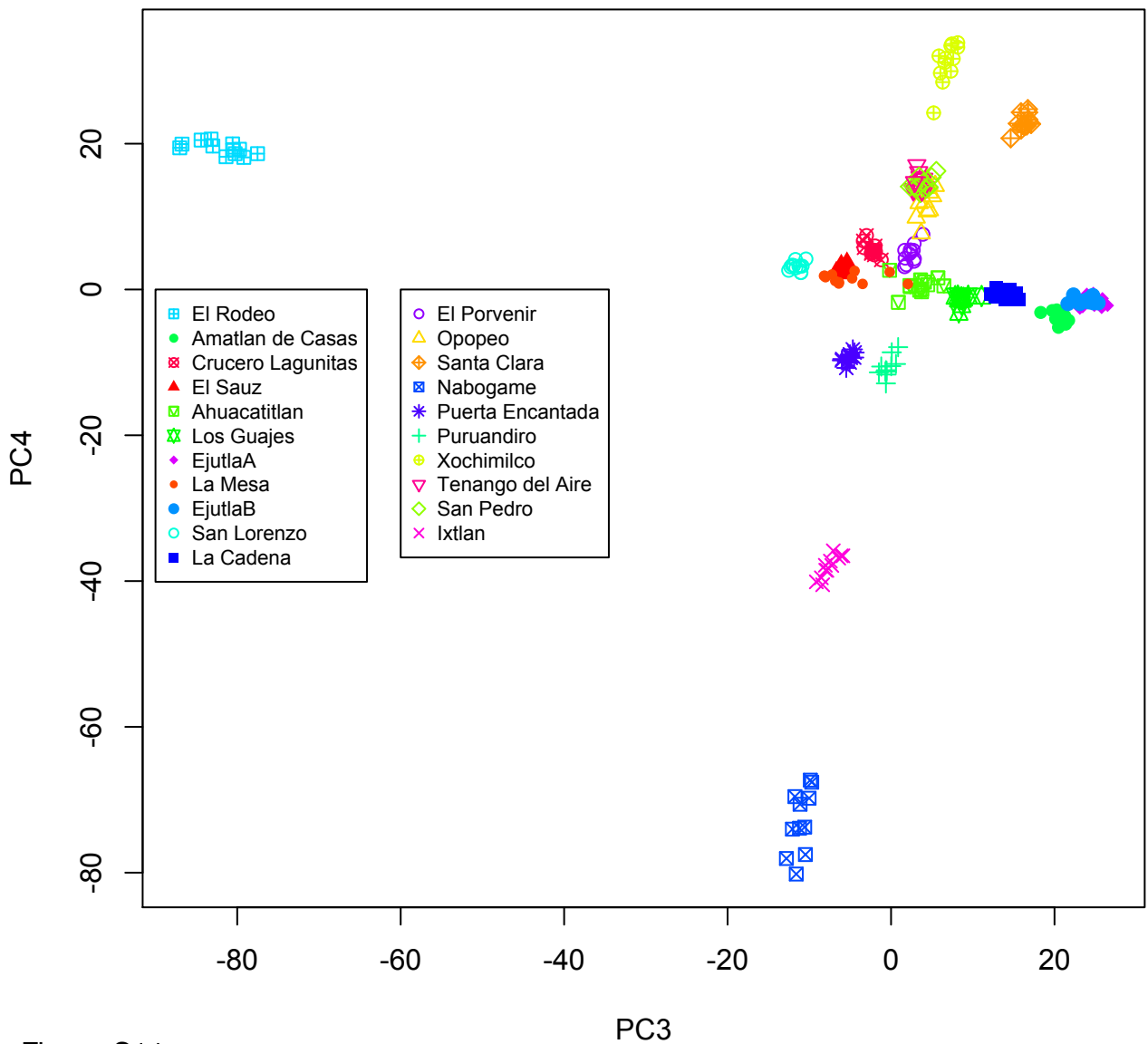

Figure S11

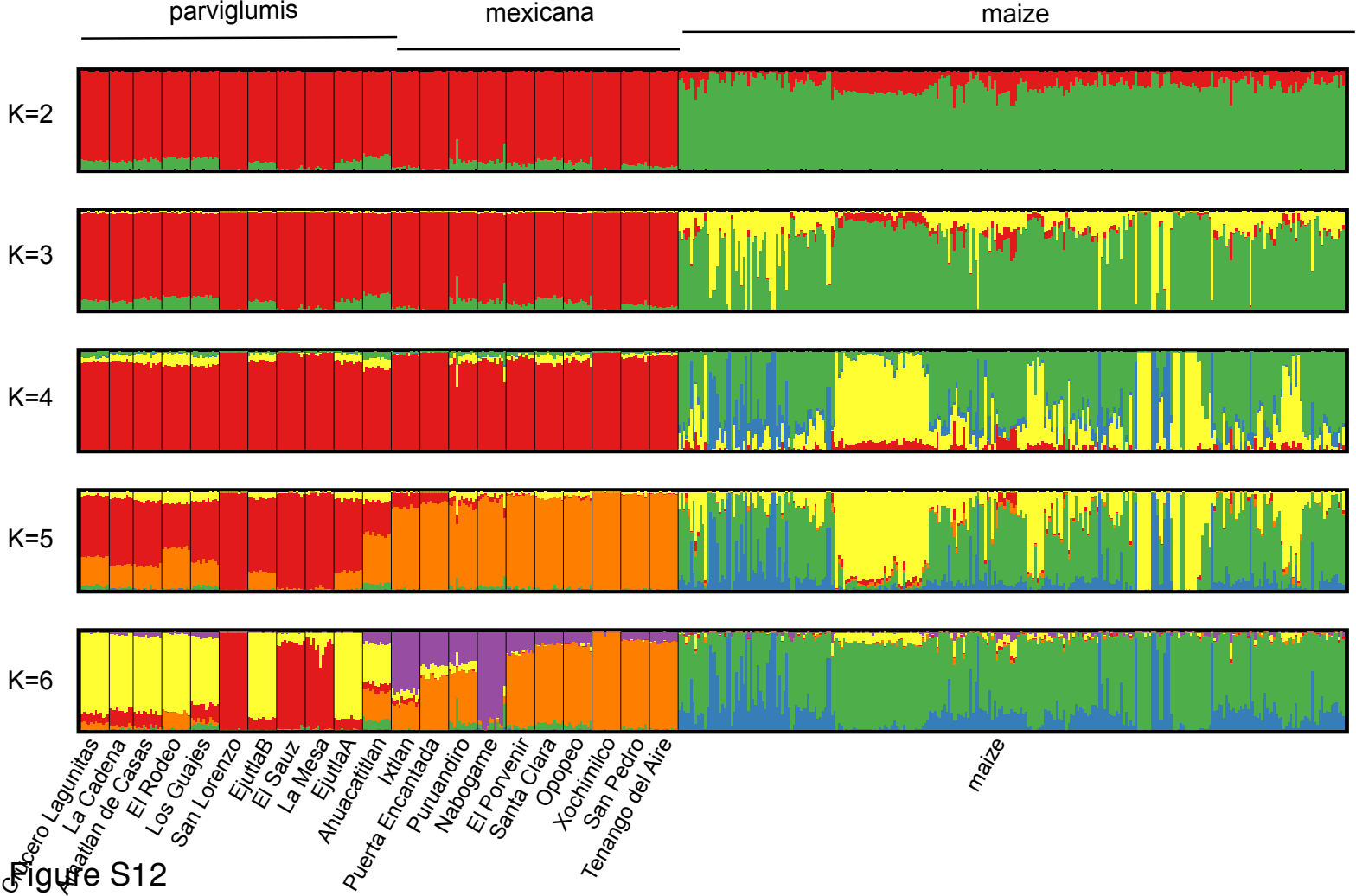

Figure S12

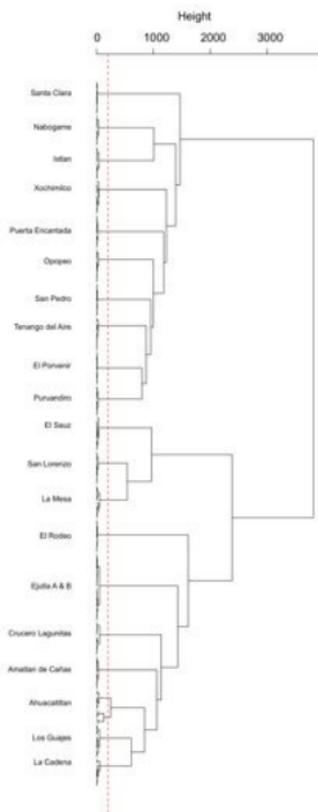

Figure S13

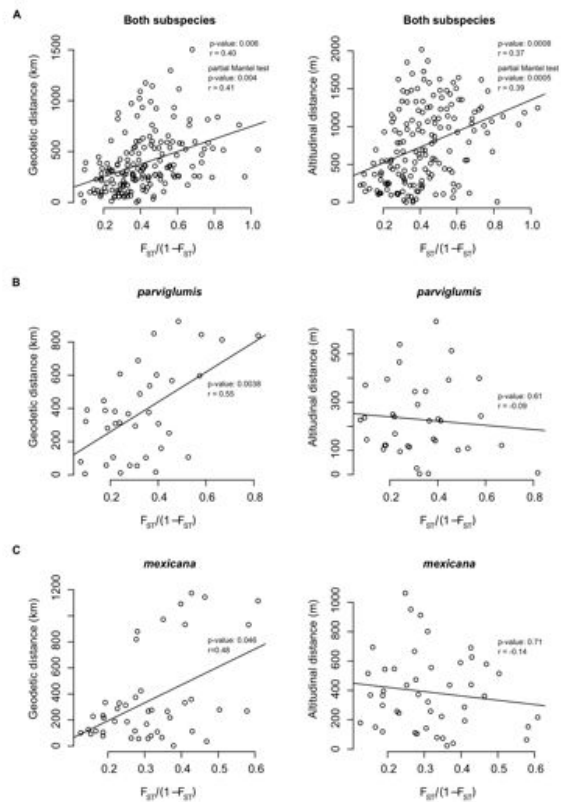

Figure S14

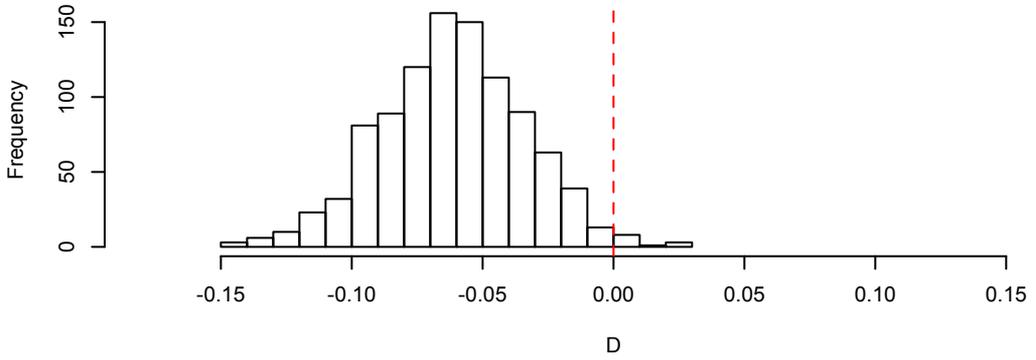
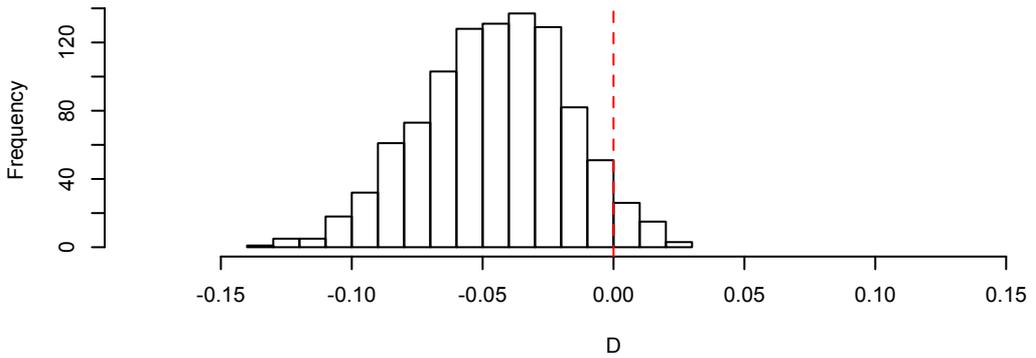
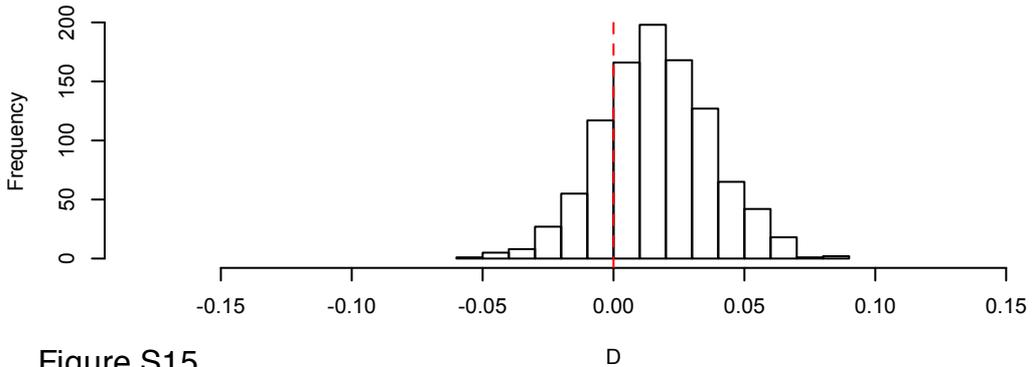

Figure S15

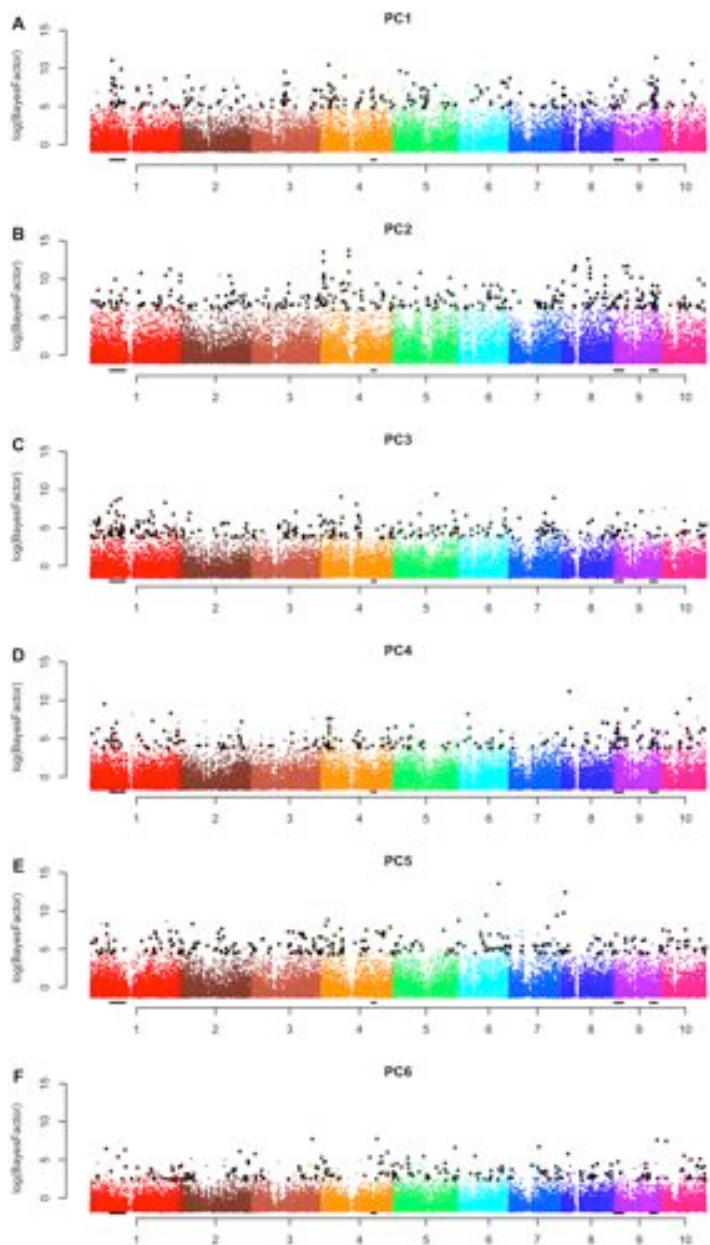

Figure S16

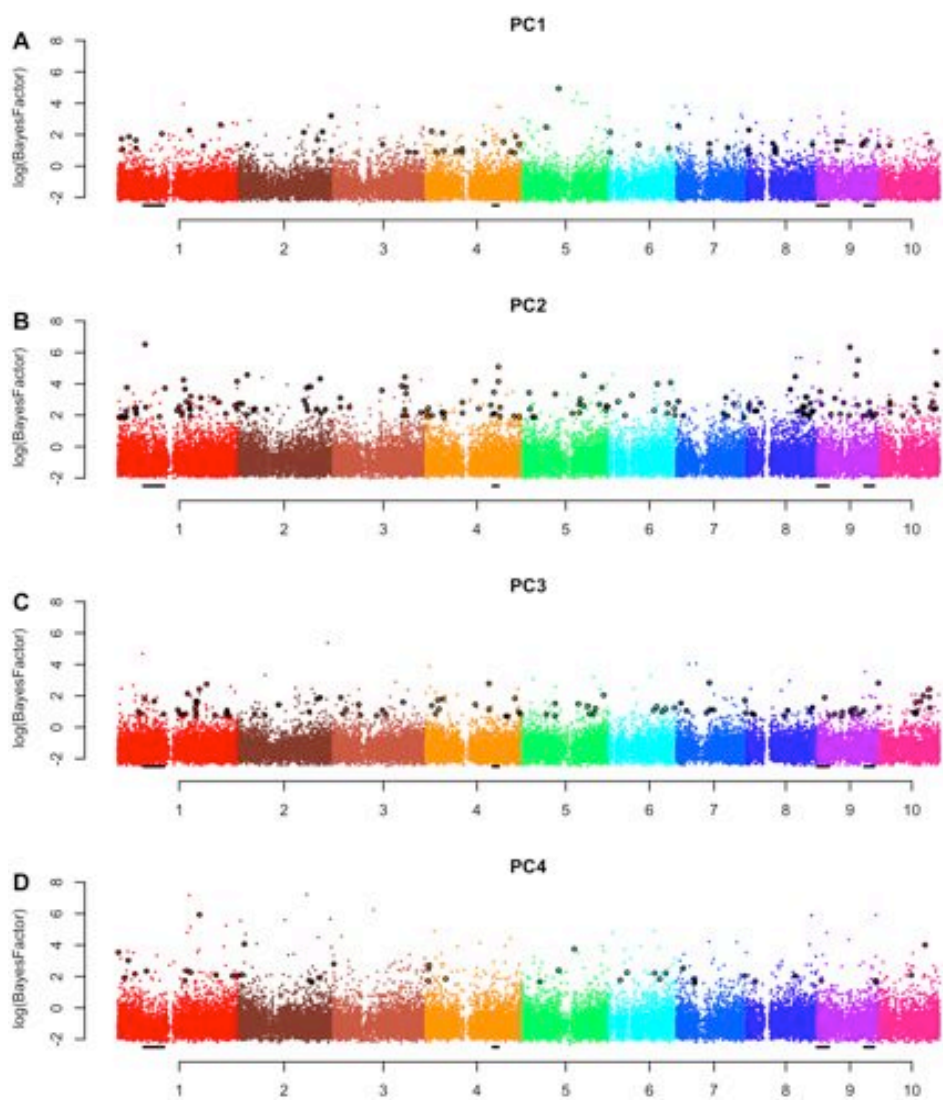

Figure S17

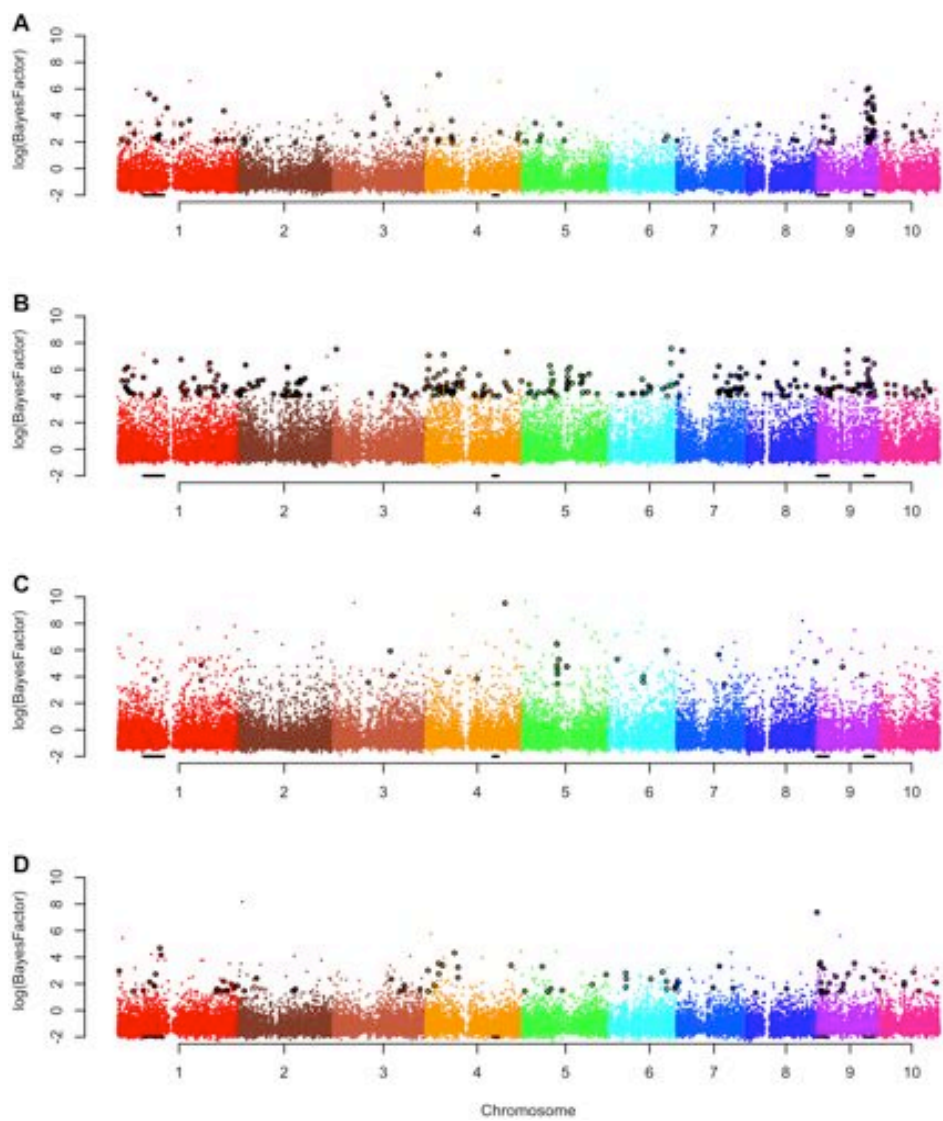

Figure S18

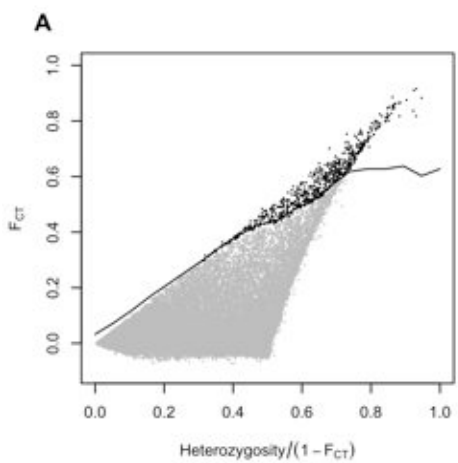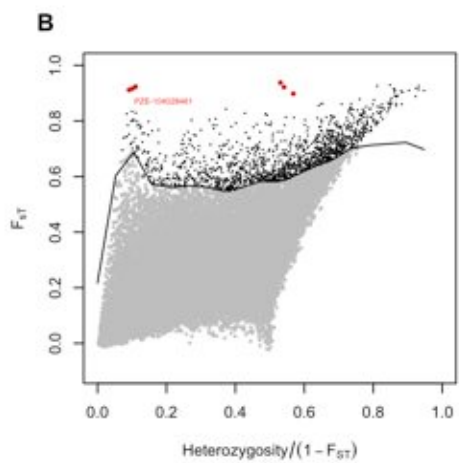

Figure S19

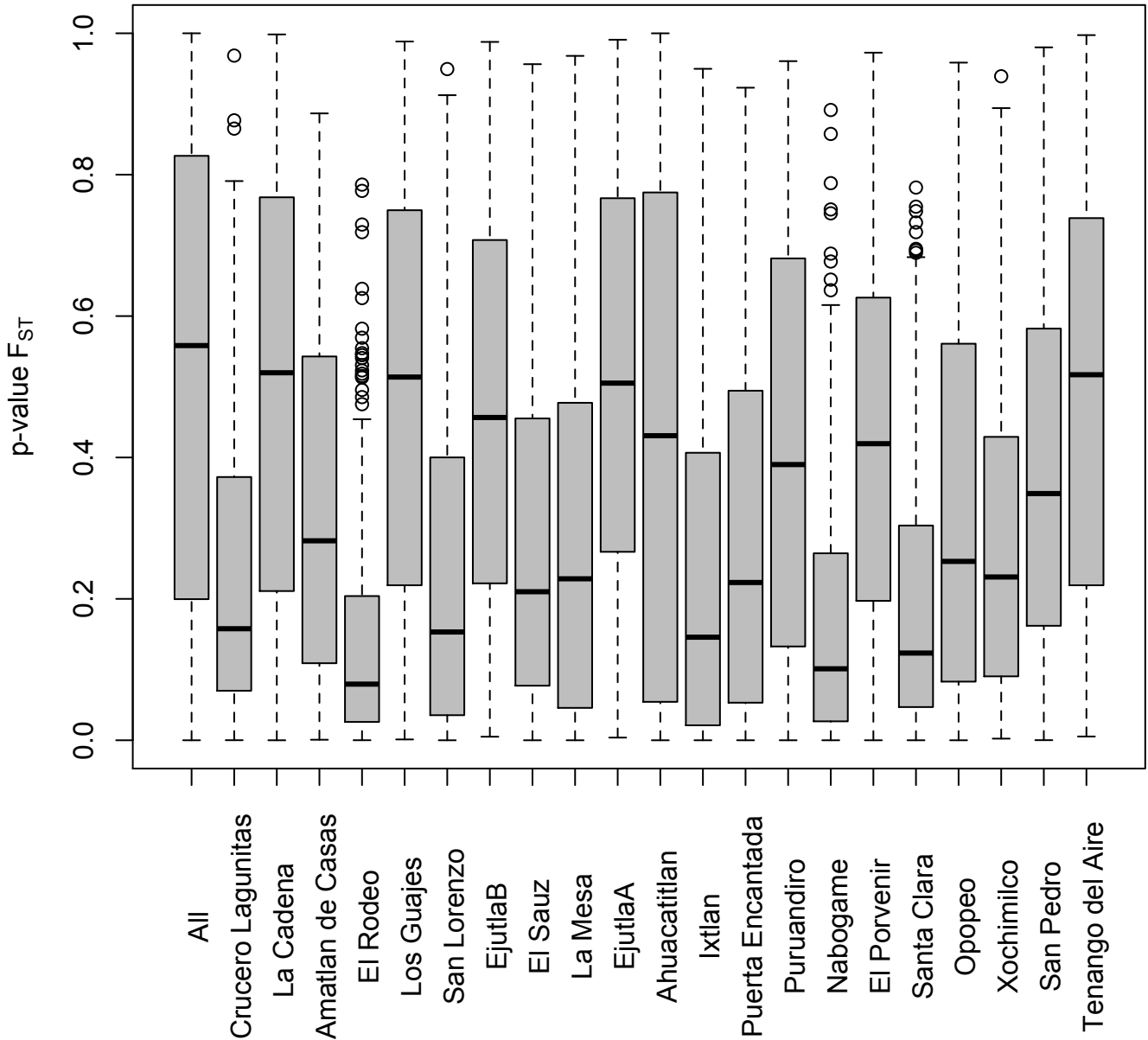

Figure S20

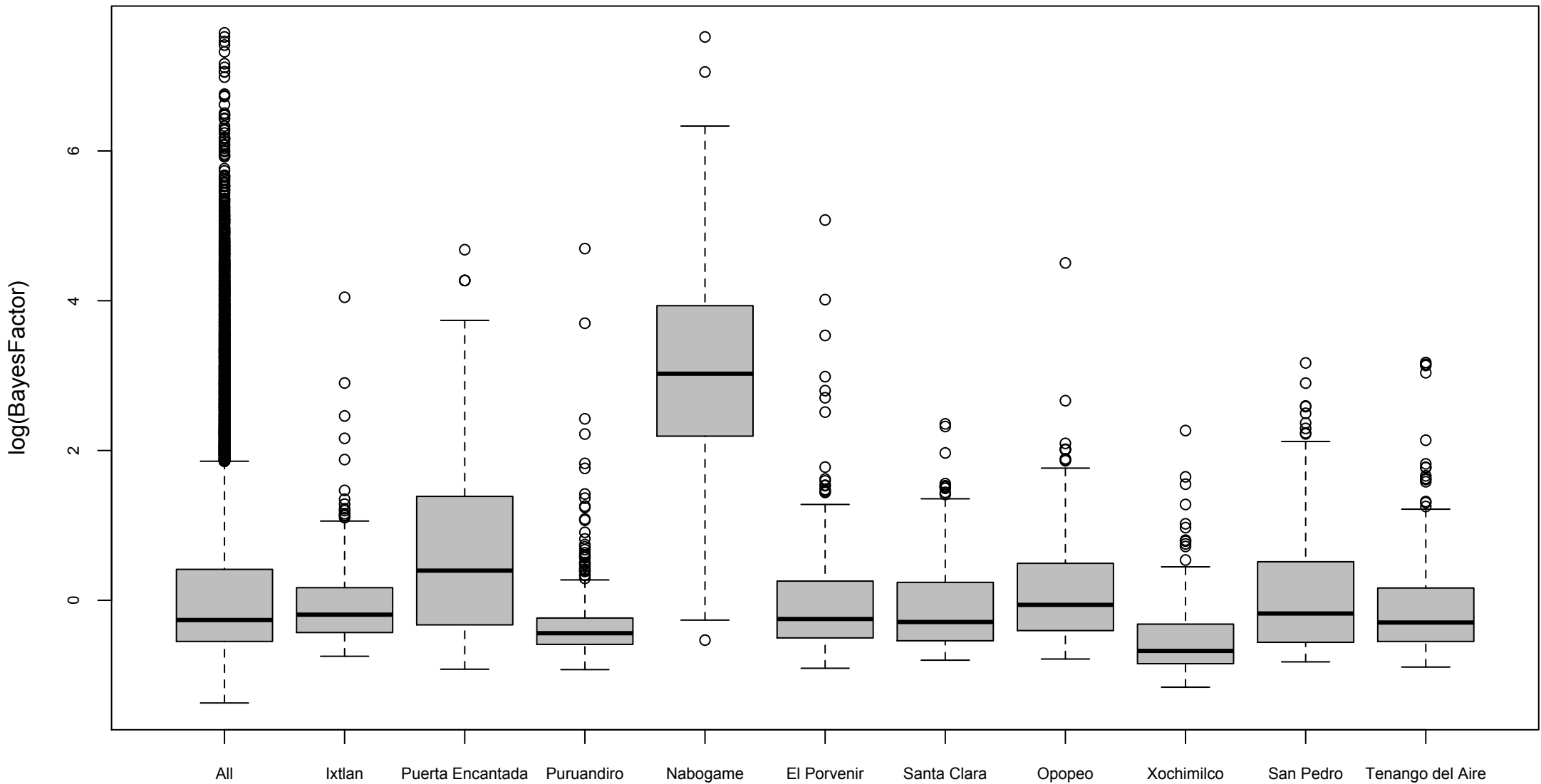

Figure 21

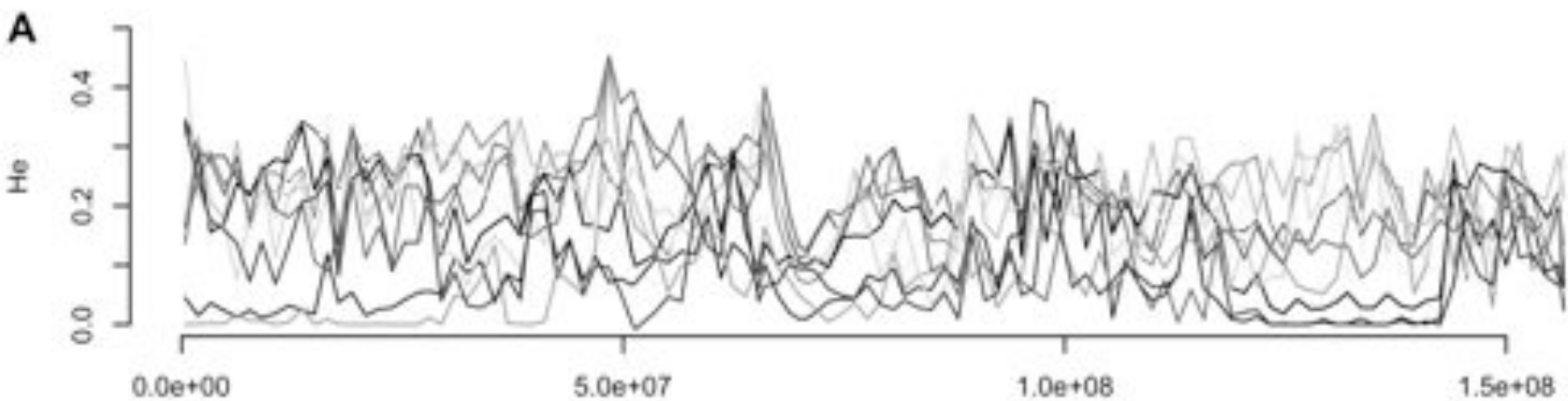
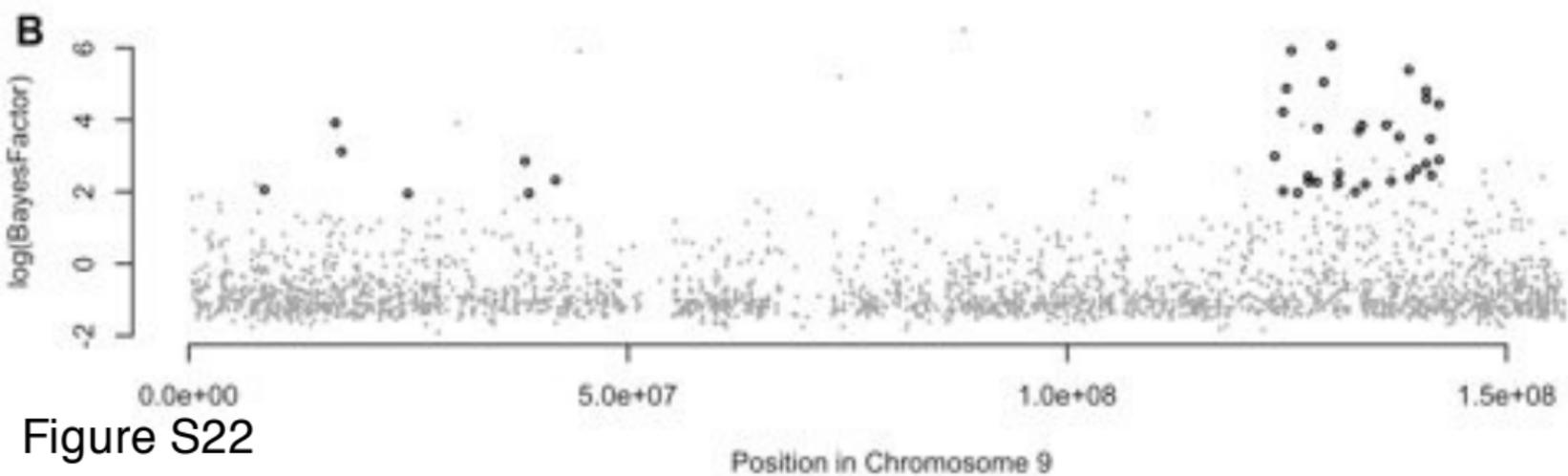

Figure S22
Position in Chromosome 9

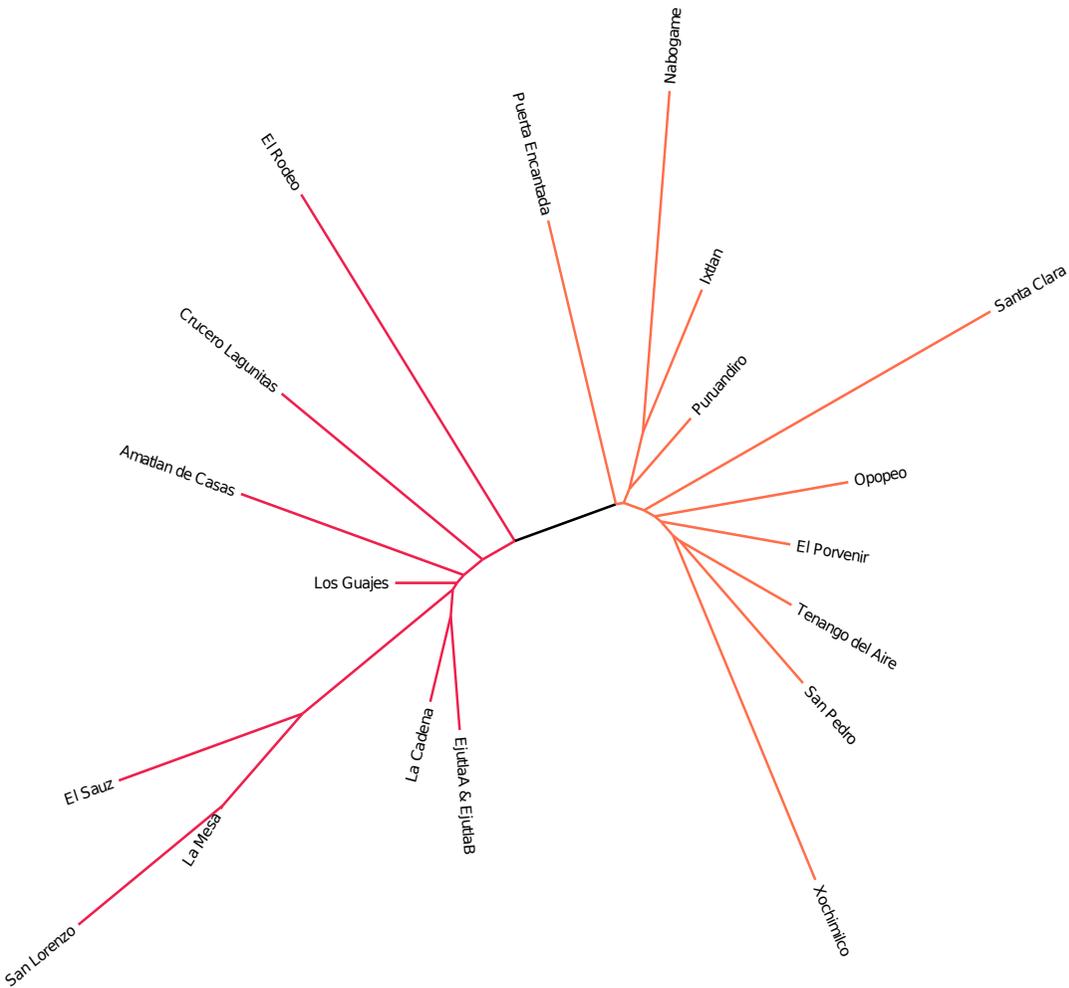

Figure S23

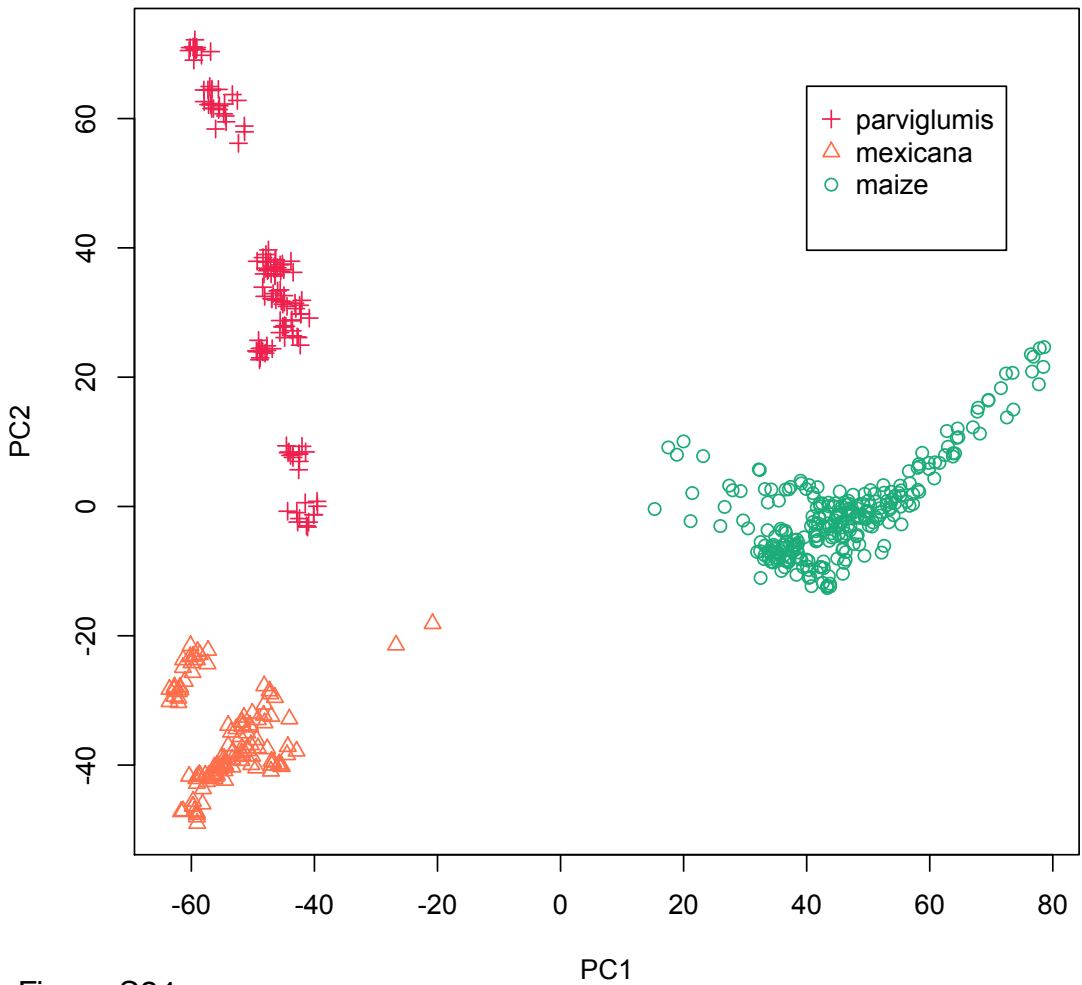

Figure S24

**Table S1** Summary statistics of genetic diversity for each population.

| Population | HWE | Polym. | $H_E$ | $F_{IS}$ | ROH length | IBS nr | IBS length |
|---|---|---|---|---|---|---|---|
| Crucero Lagunitas | 7% | 62% | 0.21 | 0.08 | 380 | 2.49 | 56 |
| La Cadena | 4% | 72% | 0.24 | 0.05 | 268 | 1.36 | 21 |
| Amatlan de Casas | 6% | 64% | 0.22 | 0.07 | 328 | 6.79 | 103 |
| El Rodeo | 6% | 54% | 0.18 | 0.03 | 282 | 6.70 | 103 |
| Los Guajes | 3% | 80% | 0.26 | -0.01 | 71 | 0.70 | 10 |
| San Lorenzo | 5% | 40% | 0.13 | -0.02 | 281 | 15.83 | 309 |
| EjutlaB | 4% | 68% | 0.23 | 0.02 | 206 | 4.22 | 54 |
| El Sauz | 6% | 54% | 0.17 | 0.03 | 333 | 9.39 | 156 |
| La Mesa | 5% | 63% | 0.19 | 0.01 | 204 | 9.52 | 134 |
| EjutlaA | 4% | 71% | 0.24 | 0.00 | 146 | 4.19 | 55 |
| Ahuacatitlan | 6% | 75% | 0.25 | 0.09 | 326 | 2.22 | 58 |
| Ixtlan | 7% | 67% | 0.20 | 0.08 | 339 | 1.84 | 29 |
| Puerta Encantada | 4% | 52% | 0.17 | -0.01 | 209 | 6.15 | 114 |
| Puruandiro | 4% | 75% | 0.24 | 0.01 | 159 | 1.63 | 26 |
| Nabogame | 4% | 62% | 0.17 | 0.02 | 401 | 10.49 | 213 |
| El Porvenir | 4% | 71% | 0.22 | 0.02 | 174 | 2.49 | 37 |
| Santa Clara | 4% | 56% | 0.18 | 0.01 | 373 | 11.55 | 226 |
| Opopeo | 4% | 69% | 0.21 | -0.01 | 207 | 5.99 | 103 |
| Xochimilco | 3% | 44% | 0.15 | -0.02 | 773 | 11.19 | 498 |
| San Pedro | 3% | 62% | 0.20 | -0.01 | 189 | 5.98 | 137 |
| Tenango del Aire | 4% | 66% | 0.20 | 0.02 | 227 | 3.80 | 74 |

HWE: proportion of SNPs deviating from HWE at 5% level, Polym.: proportion of polymorphic SNPs, $H_E$: mean expected heterozygosity, $F_{IS}$: mean inbreeding coefficient, ROH length: Average of total length of runs of homozygosity per individual (cM), IBS nr and length: Average of total number and length of identity-by-descent segments within population per individual, excluding comparisons within individuals (cM).

Table S2

Table S2 is available as an .xls file upon request from the authors.

**Table S3** Spearman's correlation between Bayes factors and p-values for $F_{CT}$ and $F_{ST}$.

| Analysis | Variable | $F_{CT}$ p-value | rho | $F_{ST}$ p-value | rho |
|---|---|---|---|---|---|
| Both | PC1 | $3.8\times10^{-69}$ | -0.09 | $1.6\times10^{-292}$ | -0.19 |
| | PC2 | 0.03 | -0.01 | 0 | -0.20 |
| | PC3 | 0.94 | 0.01 | $6.2\times10^{-122}$ | -0.12 |
| | PC4 | 0.72 | 0.00 | $1.3\times10^{-115}$ | -0.12 |
| | PC5 | 0.88 | 0.01 | $3.1\times10^{-153}$ | -0.14 |
| | PC6 | 0.26 | 0.00 | $3.6\times10^{-78}$ | -0.10 |
| parviglumis | PC1 | $2.7\times10^{-23}$ | -0.05 | $1.1\times10^{-170}$ | -0.15 |
| | PC2 | $5.5\times10^{-19}$ | -0.05 | $5.3\times10^{-167}$ | -0.15 |
| | PC3 | $4.3\times10^{-23}$ | -0.05 | $9.7\times10^{-148}$ | -0.14 |
| | PC4 | $9.4\times10^{-21}$ | -0.05 | $4.3\times10^{-132}$ | -0.13 |
| mexicana | PC1 | $1.2\times10^{-02}$ | -0.01 | $1.1\times10^{-87}$ | -0.11 |
| | PC2 | 1.00 | 0.01 | $3.2\times10^{-227}$ | -0.17 |
| | PC3 | 1.00 | 0.02 | $6.6\times10^{-80}$ | -0.10 |
| | PC4 | 0.60 | 0.00 | $1.9\times10^{-83}$ | -0.10 |

p-values are for one-sided test where alternative hypothesis is that there is a negative correlation between p-values and Bayes factors. Rho is Spearman's rank correlation coefficient.

**Table S4** Enrichment of functional properties among candidate SNPs and genes.

| Analysis | p-value gen/nongen | nonsyn/syn |
|---|---|---|
| PC1 | 0.69 | 0.37 |
| PC2 | 0.82 | 0.27 |
| PC3 | 0.55 | 0.93 |
| PC4 | 0.80 | 0.85 |
| PC5 | 0.82 | 0.87 |
| PC6 | 0.72 | 0.30 |
| PC1 parv | 0.90 | 0.20 |
| PC2 parv | 0.36 | 0.72 |
| PC3 parv | 0.54 | 0.60 |
| PC4 parv | 0.95 | 0.07 |
| PC1 mex | 0.87 | 0.04 |
| PC2 mex | 0.18 | 0.12 |
| PC3 mex | 0.92 | 0.17 |
| PC4 mex | 0.65 | 0.40 |
| $F_{CT}$ | 0.93 | 0.28 |
| $F_{ST}$ | 1.00 | 0.06 |

Analysis: candidate lists from different analyses, gen/nongen: p-value for observed ratio of genic and nongenic SNPs being higher than expected, nonsyn/syn: p-value for observed ratio of nonsynonymous and synonymous SNPs being higher than expected. All p-values are based on bootstrapping with 1000 iterations.

**Table S5** Enrichment of SNPs that are associated with a maize phenotypic trait for each list of adaptation candidates. GWAS was done using a mixed linear model that takes both population structure and relatedness into account.

| Trait | $F_{CT}$ | $F_{ST}$ | PC1 | PC2 | PC3 | PC4 | PC5 | PC6 |
|---|---|---|---|---|---|---|---|---|
| 20KernelWeight | 0.636 | 0.169 | 0.661 | 0.085 | 0.160 | 0.689 | 0.760 | 0.190 |
| CobDiameter | 0.153 | 0.238 | 0.292 | 0.026 | 0.001 | 0.231 | 0.968 | 0.336 |
| CobWeight | 0.001 | 0.105 | 0.749 | 0.849 | 0.934 | 0.760 | 0.745 | 0.148 |
| DaystoSilk | 0.601 | 0.515 | 0.171 | 0.910 | 0.933 | 0.267 | 0.042 | 0.915 |
| DaysToTassel | 0.610 | 0.563 | 0.168 | 0.502 | 0.922 | 0.112 | 0.043 | 0.790 |
| EarDiameter | 0.365 | 0.320 | 0.427 | 0.393 | 0.004 | 0.494 | 0.260 | 0.685 |
| EarLength | 0.470 | 0.914 | 0.591 | 0.688 | 0.517 | 0.961 | 0.278 | 0.452 |
| EarRankNumber | 1.000 | 0.792 | 0.357 | 0.698 | 0.228 | 0.526 | 0.652 | 0.035 |
| EarRowNumber | 0.856 | 0.908 | 0.545 | 0.174 | 0.974 | 0.597 | 0.314 | 0.860 |
| EarWeight | 0.987 | 1.000 | 0.096 | 0.093 | 0.021 | 0.096 | 1.000 | 0.344 |
| GerminationCount | 0.009 | 0.019 | 0.709 | 0.044 | 0.208 | 0.372 | 0.689 | 0.464 |
| LeafLength | 0.582 | 0.531 | 0.023 | 0.440 | 0.393 | 0.521 | 0.076 | 0.587 |
| LeafSheathLength | 0.816 | 0.649 | 0.277 | 0.440 | 0.029 | 0.619 | 0.871 | 0.692 |
| LeafWidth-BorderPlant | 0.934 | 0.989 | 0.552 | 0.416 | 0.383 | 0.781 | 0.853 | 0.760 |
| LeafWidth | 0.292 | 0.763 | 0.186 | 0.986 | 0.881 | 0.388 | 0.093 | 0.664 |
| MainSpikeLength | 0.016 | 0.024 | 0.233 | 0.658 | 0.608 | 0.334 | 0.980 | 0.031 |
| MiddleLeafAngle | 0.326 | 0.150 | 0.171 | 0.413 | 0.602 | 0.378 | 0.414 | 0.934 |
| NorthernLeafBlight | 0.399 | 0.500 | 0.020 | 1.000 | 0.657 | 0.161 | 0.031 | 0.433 |
| NumberofLeaves | 0.517 | 0.902 | 0.190 | 0.712 | 0.136 | 0.158 | 0.288 | 0.131 |
| NumberofTilleringPlants | 0.346 | 0.283 | 0.031 | 0.893 | 0.680 | 0.095 | 0.210 | 0.153 |
| Phylotaxy | 0.958 | 0.845 | 0.816 | 0.391 | 0.305 | 0.680 | 0.417 | 0.008 |
| RowQuality | 0.029 | 0.383 | 0.115 | 0.478 | 0.722 | 0.270 | 0.250 | 0.277 |
| SecondaryBranchNumber | 0.723 | 0.993 | 0.389 | 0.551 | 0.257 | 0.138 | 0.922 | 0.141 |
| SeedSetLength | 0.827 | 0.142 | 0.933 | 0.840 | 0.329 | 0.857 | 0.636 | 0.730 |
| Spikelets-MainSpike | 0.392 | 0.298 | 0.292 | 0.936 | 0.322 | 0.439 | 0.840 | 0.054 |

| Trait | PC1 parv | PC2 parv | PC3 parv | PC4 parv | PC1 mex | PC2 mex | PC3 mex | PC4 mex |
|---|---|---|---|---|---|---|---|---|
| Spikelets-PrimaryBranch | 0.109 | 0.001 | 0.203 | 0.163 | 0.206 | 0.567 | 0.127 | 0.861 |
| StandCount | 0.021 | 0.000 | 0.367 | 0.009 | 0.620 | 0.017 | 0.291 | 0.218 |
| TasselBranchLength | 0.332 | 0.081 | 0.122 | 0.620 | 0.745 | 0.927 | 0.716 | 0.179 |
| TasselLength | 0.682 | 0.919 | 0.195 | 0.460 | 0.428 | 0.920 | 0.595 | 0.106 |
| TasselPrimaryBranches | 0.016 | 0.001 | 0.236 | 0.204 | 0.419 | 0.130 | 0.174 | 0.180 |
| TasselSterility | 0.334 | 0.427 | 0.639 | 0.003 | 0.094 | 0.536 | 0.069 | 0.775 |
| TilleringIndex-BorderPlant | 0.495 | 0.302 | 0.112 | 0.407 | 0.814 | 0.315 | 0.293 | 0.453 |
| TilleringIndex | 0.834 | 0.946 | 0.859 | 0.144 | 0.830 | 0.592 | 0.082 | 0.055 |
| TotalKernelVolume | 0.870 | 0.953 | 0.270 | 0.446 | 0.169 | 0.406 | 0.996 | 0.167 |
| UpperLeafAngle-BorderPlant | 0.985 | 0.854 | 0.980 | 0.002 | 0.160 | 0.346 | 0.552 | 0.488 |
| UpperLeafAngle | 0.401 | 0.190 | 0.750 | 0.787 | 0.949 | 0.484 | 0.224 | 0.193 |
| **Trait** | **PC1 parv** | **PC2 parv** | **PC3 parv** | **PC4 parv** | **PC1 mex** | **PC2 mex** | **PC3 mex** | **PC4 mex** |
| 20KernelWeight | 0.015 | 0.262 | 0.288 | 1.000 | 0.540 | 0.964 | 0.004 | 0.765 |
| CobDiameter | 0.136 | 0.620 | 0.443 | 0.888 | 0.206 | 0.011 | 0.075 | 0.677 |
| CobWeight | 0.009 | 0.422 | 0.509 | 0.071 | 0.823 | 0.326 | 0.711 | 0.547 |
| DaystoSilk | 0.145 | 0.481 | 0.513 | 0.979 | 0.428 | 0.708 | 0.652 | 0.416 |
| DaysToTassel | 0.777 | 0.320 | 0.635 | 0.868 | 0.461 | 0.678 | 0.813 | 0.131 |
| EarDiameter | 0.002 | 0.891 | 0.140 | 0.293 | 0.032 | 0.563 | 0.901 | 0.394 |
| EarLength | 0.048 | 0.452 | 0.156 | 0.143 | 0.618 | 0.678 | 0.419 | 0.433 |
| EarRankNumber | 0.170 | 0.713 | 0.137 | 0.553 | 0.280 | 0.079 | 0.226 | 0.411 |
| EarRowNumber | 0.288 | 0.664 | 0.689 | 0.984 | 0.647 | 0.698 | 0.926 | 0.931 |
| EarWeight | 0.372 | 0.296 | 0.291 | 0.884 | 0.292 | 0.866 | 0.074 | 0.141 |
| GerminationCount | 0.058 | 0.112 | 0.608 | 0.719 | 0.978 | 0.793 | 0.713 | 0.116 |
| LeafLength | 0.000 | 0.028 | 0.407 | 0.804 | 0.165 | 0.622 | 0.100 | 0.787 |
| LeafSheathLength | 0.490 | 0.766 | 0.628 | 0.575 | 0.310 | 0.397 | 1.000 | 0.976 |
| LeafWidth-BorderPlant | 0.266 | 0.205 | 0.844 | 0.537 | 0.498 | 0.998 | 0.302 | 0.591 |
| LeafWidth | 0.154 | 0.159 | 0.367 | 0.049 | 0.121 | 0.987 | 0.554 | 0.936 |
| MainSpikeLength | 0.784 | 0.833 | 0.385 | 0.865 | 0.028 | 0.187 | 0.291 | 0.968 |
| MiddleLeafAngle | 0.170 | 0.158 | 0.769 | 0.751 | 0.761 | 0.463 | 0.086 | 0.402 |
| NorthernLeafBlight | 0.501 | 0.151 | 0.019 | 0.838 | 0.021 | 0.023 | 0.340 | 0.252 |
| NumberofLeaves | 0.389 | 0.906 | 0.190 | 0.069 | 0.771 | 0.698 | 1.000 | 0.884 |
| NumberofTilleringPlants | 0.905 | 0.538 | 0.905 | 0.793 | 0.142 | 0.504 | 0.666 | 0.522 |
| Phylotaxy | 0.945 | 0.257 | 0.123 | 0.528 | 0.525 | 0.498 | 0.691 | 0.984 |
| RowQuality | 0.612 | 0.629 | 0.514 | 0.291 | 0.211 | 0.145 | 0.677 | 0.818 |
| SecondaryBranchNumber | 0.355 | 0.243 | 0.397 | 0.094 | 0.645 | 0.390 | 0.859 | 0.303 |
| SeedSetLength | 0.634 | 0.580 | 0.298 | 0.065 | 0.782 | 0.705 | 0.008 | 0.487 |
| Spikelets-MainSpike | 0.233 | 0.510 | 0.462 | 0.460 | 0.495 | 0.637 | 0.897 | 0.097 |
| Spikelets-PrimaryBranch | 0.123 | 0.530 | 0.869 | 0.467 | 0.951 | 0.629 | 0.657 | 0.317 |
| StandCount | 0.382 | 0.421 | 0.426 | 0.389 | 0.673 | 0.072 | 0.056 | 0.259 |
| TasselBranchLength | 0.299 | 0.215 | 0.304 | 0.241 | 0.927 | 0.223 | 0.924 | 0.772 |
| TasselLength | 0.533 | 0.065 | 0.111 | 0.466 | 0.584 | 0.336 | 0.919 | 0.551 |
| TasselPrimaryBranches | 0.752 | 0.383 | 0.678 | 0.023 | 0.282 | 0.398 | 0.431 | 0.284 |
| TasselSterility | 0.780 | 0.751 | 0.906 | 1.000 | 0.120 | 0.229 | 0.498 | 0.894 |
| TilleringIndex-BorderPlant | 0.188 | 0.650 | 0.339 | 0.481 | 0.388 | 0.199 | 0.678 | 0.457 |
| TilleringIndex | 0.035 | 0.867 | 0.293 | 0.446 | 0.546 | 0.033 | 0.980 | 0.440 |
| TotalKernelVolume | 0.211 | 0.692 | 0.436 | 0.971 | 0.529 | 0.799 | 0.020 | 0.128 |
| UpperLeafAngle-BorderPlant | 0.767 | 0.659 | 0.851 | 0.833 | 0.113 | 0.171 | 1.000 | 0.253 |
| UpperLeafAngle | 0.795 | 0.044 | 0.779 | 0.984 | 0.911 | 0.962 | 0.925 | 0.416 |

**Table S6** Enrichment of SNPs that are associated with a maize phenotypic trait for each list of adaptation candidates. GWAS was done using a simple model not taking population structure into account.

| Trait | $F_{CT}$ | $F_{ST}$ | PC1 | PC2 | PC3 | PC4 | PC5 | PC6 |
|---|---|---|---|---|---|---|---|---|
| 20KernelWeight | 0.780 | 0.215 | 0.852 | 0.650 | 0.913 | 0.526 | 0.217 | 0.629 |
| CobDiameter | 0.173 | 0.046 | 0.930 | 0.058 | 0.001 | 0.274 | 0.798 | 0.067 |
| CobWeight | 0.200 | 0.031 | 0.682 | 0.555 | 0.039 | 0.649 | 0.556 | 0.415 |
| DaystoSilk | 0.003 | 0.001 | 0.288 | 0.071 | 0.343 | 0.126 | 0.256 | 0.599 |
| DaysToTassel | 0.002 | 0.000 | 0.151 | 0.023 | 0.143 | 0.166 | 0.276 | 0.254 |
| EarDiameter | 0.216 | 0.003 | 0.339 | 0.082 | 0.233 | 0.130 | 0.971 | 0.676 |
| EarLength | 0.217 | 0.015 | 0.211 | 0.860 | 0.621 | 0.868 | 0.234 | 0.394 |
| EarRankNumber | 0.997 | 0.281 | 0.480 | 0.326 | 0.105 | 0.106 | 0.871 | 0.180 |
| EarRowNumber | 0.992 | 0.765 | 0.412 | 0.302 | 0.649 | 0.050 | 0.924 | 0.936 |
| EarWeight | 0.928 | 0.373 | 0.081 | 0.478 | 0.030 | 0.002 | 0.932 | 0.730 |
| GerminationCount | 0.888 | 0.968 | 0.448 | 0.618 | 0.155 | 0.110 | 0.785 | 0.204 |
| LeafLength | 0.000 | 0.002 | 0.397 | 0.205 | 0.063 | 0.275 | 0.376 | 0.542 |
| LeafSheathLength | 0.815 | 0.727 | 0.352 | 0.256 | 0.216 | 0.565 | 0.487 | 0.139 |
| LeafWidth-BorderPlant | 0.410 | 0.845 | 0.485 | 0.617 | 0.340 | 0.969 | 0.073 | 0.330 |
| LeafWidth | 0.019 | 0.005 | 0.286 | 0.728 | 0.306 | 0.585 | 0.162 | 0.489 |
| MainSpikeLength | 0.639 | 0.824 | 0.443 | 0.443 | 0.668 | 0.319 | 0.194 | 0.124 |
| MiddleLeafAngle | 0.209 | 0.001 | 0.620 | 0.446 | 0.068 | 0.414 | 0.039 | 0.916 |
| NorthernLeafBlight | 0.239 | 0.223 | 0.177 | 0.295 | 0.243 | 0.098 | 0.041 | 0.777 |
| NumberofLeaves | 0.110 | 0.095 | 0.082 | 0.702 | 0.054 | 0.411 | 0.464 | 0.487 |
| NumberofTilleringPlants | 0.008 | 0.004 | 0.005 | 0.900 | 0.190 | 0.156 | 0.059 | 0.216 |
| Phylotaxy | 0.981 | 0.894 | 0.600 | 0.955 | 0.180 | 0.810 | 0.133 | 0.190 |
| RowQuality | 0.035 | 0.473 | 0.021 | 0.001 | 0.899 | 0.045 | 0.047 | 0.247 |
| SecondaryBranchNumber | 0.080 | 0.050 | 0.147 | 0.407 | 0.122 | 0.048 | 0.714 | 0.020 |
| SeedSetLength | 0.047 | 0.328 | 0.940 | 0.985 | 0.467 | 0.931 | 0.392 | 0.861 |
| Spikelets-MainSpike | 0.005 | 0.025 | 0.401 | 0.023 | 0.161 | 0.280 | 0.486 | 0.349 |

| Trait | PC1 parv | PC2 parv | PC3 parv | PC4 parv | PC1 mex | PC2 mex | PC3 mex | PC4 mex |
|---|---|---|---|---|---|---|---|---|
| Spikelets-PrimaryBranch | 0.000 | 0.000 | 0.180 | 0.224 | 0.295 | 0.160 | 0.085 | 0.336 |
| StandCount | 0.202 | 0.000 | 0.462 | 0.369 | 0.496 | 0.287 | 0.441 | 0.155 |
| TasselBranchLength | 0.764 | 0.775 | 0.278 | 0.068 | 0.245 | 0.513 | 0.851 | 0.485 |
| TasselLength | 0.640 | 0.795 | 0.725 | 0.307 | 0.612 | 0.366 | 0.084 | 0.565 |
| TasselPrimaryBranches | 0.095 | 0.004 | 0.087 | 0.483 | 0.235 | 0.096 | 0.432 | 0.033 |
| TasselSterility | 0.144 | 0.057 | 0.486 | 0.241 | 0.209 | 0.987 | 0.108 | 0.594 |
| TilleringIndex-BorderPlant | 0.235 | 0.059 | 0.161 | 0.432 | 0.753 | 0.809 | 0.063 | 0.411 |
| TilleringIndex | 0.938 | 0.869 | 0.553 | 0.022 | 0.054 | 0.334 | 0.472 | 0.347 |
| TotalKernelVolume | 0.577 | 0.040 | 0.030 | 0.252 | 0.239 | 0.009 | 0.964 | 0.274 |
| UpperLeafAngle-BorderPlant | 0.854 | 0.272 | 0.964 | 0.124 | 0.274 | 0.005 | 0.290 | 0.345 |
| UpperLeafAngle | 0.297 | 0.000 | 0.220 | 0.031 | 0.689 | 0.007 | 0.643 | 0.810 |
| **Trait** | **PC1 parv** | **PC2 parv** | **PC3 parv** | **PC4 parv** | **PC1 mex** | **PC2 mex** | **PC3 mex** | **PC4 mex** |
| 20KernelWeight | 0.256 | 0.762 | 0.463 | 0.998 | 0.718 | 0.484 | 0.279 | 0.405 |
| CobDiameter | 0.080 | 0.111 | 0.705 | 0.278 | 0.042 | 0.196 | 0.202 | 0.799 |
| CobWeight | 0.211 | 0.260 | 0.910 | 0.073 | 0.050 | 0.022 | 0.865 | 0.732 |
| DaystoSilk | 0.052 | 0.751 | 0.965 | 0.174 | 0.031 | 0.586 | 0.870 | 0.052 |
| DaysToTassel | 0.012 | 0.573 | 0.920 | 0.356 | 0.132 | 0.115 | 0.954 | 0.336 |
| EarDiameter | 0.124 | 0.947 | 0.971 | 0.099 | 0.011 | 0.001 | 0.565 | 0.993 |
| EarLength | 0.198 | 0.699 | 0.237 | 0.706 | 0.319 | 0.074 | 0.983 | 0.511 |
| EarRankNumber | 0.331 | 0.225 | 0.092 | 0.283 | 0.028 | 0.074 | 0.209 | 0.097 |
| EarRowNumber | 0.255 | 0.587 | 0.872 | 0.538 | 0.119 | 0.508 | 0.384 | 0.931 |
| EarWeight | 0.364 | 0.956 | 0.525 | 0.708 | 0.005 | 0.027 | 0.075 | 0.236 |
| GerminationCount | 0.022 | 0.415 | 0.444 | 0.342 | 0.856 | 0.556 | 0.643 | 0.186 |
| LeafLength | 0.058 | 0.085 | 0.923 | 0.087 | 0.789 | 0.334 | 0.889 | 0.643 |
| LeafSheathLength | 0.592 | 0.299 | 0.704 | 0.582 | 0.697 | 0.268 | 0.849 | 0.387 |
| LeafWidth-BorderPlant | 0.202 | 0.673 | 0.638 | 0.367 | 0.201 | 0.861 | 0.739 | 0.381 |
| LeafWidth | 0.016 | 0.685 | 0.551 | 0.149 | 0.304 | 0.726 | 0.668 | 0.250 |
| MainSpikeLength | 0.524 | 0.907 | 0.463 | 0.884 | 0.313 | 0.189 | 0.909 | 0.517 |
| MiddleLeafAngle | 0.014 | 0.356 | 0.171 | 0.635 | 0.473 | 0.207 | 0.650 | 0.040 |
| NorthernLeafBlight | 0.109 | 0.401 | 0.652 | 0.220 | 0.072 | 0.071 | 0.688 | 0.717 |
| NumberofLeaves | 0.115 | 0.769 | 0.270 | 0.001 | 0.334 | 0.264 | 1.000 | 0.893 |
| NumberofTilleringPlants | 0.340 | 0.167 | 0.883 | 0.878 | 0.167 | 0.458 | 0.389 | 0.301 |
| Phylotaxy | 0.930 | 0.333 | 0.037 | 0.557 | 0.705 | 0.505 | 0.704 | 0.919 |
| RowQuality | 0.154 | 0.721 | 0.543 | 0.760 | 0.087 | 0.002 | 0.774 | 0.726 |
| SecondaryBranchNumber | 0.488 | 0.718 | 0.785 | 0.222 | 0.360 | 0.423 | 0.652 | 0.355 |
| SeedSetLength | 0.743 | 0.618 | 0.279 | 0.101 | 0.493 | 0.853 | 0.106 | 0.368 |
| Spikelets-MainSpike | 0.018 | 0.302 | 0.761 | 0.408 | 0.308 | 0.502 | 0.687 | 0.385 |
| Spikelets-PrimaryBranch | 0.120 | 0.633 | 0.860 | 0.079 | 0.255 | 0.087 | 0.540 | 0.319 |
| StandCount | 0.765 | 0.095 | 0.744 | 0.664 | 0.775 | 0.029 | 0.619 | 0.308 |
| TasselBranchLength | 0.255 | 0.792 | 0.665 | 0.784 | 0.635 | 0.558 | 0.650 | 0.631 |
| TasselLength | 0.652 | 0.914 | 0.582 | 0.123 | 0.763 | 0.195 | 0.839 | 0.689 |
| TasselPrimaryBranches | 0.403 | 0.901 | 0.980 | 0.054 | 0.376 | 0.111 | 0.121 | 0.421 |
| TasselSterility | 0.306 | 0.826 | 0.848 | 1.000 | 0.318 | 0.157 | 1.000 | 0.821 |
| TilleringIndex-BorderPlant | 0.075 | 0.501 | 0.243 | 0.626 | 0.905 | 0.111 | 0.752 | 0.310 |
| TilleringIndex | 0.342 | 0.463 | 0.604 | 0.365 | 0.696 | 0.019 | 0.177 | 0.427 |
| TotalKernelVolume | 0.297 | 0.430 | 0.859 | 0.836 | 0.007 | 0.009 | 0.142 | 0.063 |
| UpperLeafAngle-BorderPlant | 0.528 | 0.055 | 0.082 | 0.559 | 0.035 | 0.416 | 0.080 | 0.282 |
| UpperLeafAngle | 0.608 | 0.526 | 0.203 | 0.364 | 0.104 | 0.261 | 0.536 | 0.188 |

**Table S7** Information about sampled populations ordered by ascending altitude.

| Population | Subspecies | N | Lat | Long | Alt | T | P |
|---|---|---|---|---|---|---|---|
| Crucero Lagunitas | *parviglumis* | 12 | 16.98 | -99.28 | 590 | 26.2 | 1625 |
| La Cadena | *parviglumis* | 10 | 19.06 | -101.21 | 759 | 26.6 | 1005 |
| Amatlan de Casas | *parviglumis* | 12 | 20.82 | -104.41 | 880 | 24.4 | 849 |
| El Rodeo | *parviglumis* | 12 | 16.35 | -97.02 | 982 | 22.3 | 1033 |
| Los Guajes | *parviglumis* | 12 | 19.23 | -100.49 | 985 | 22.9 | 1049 |
| San Lorenzo | *parviglumis* | 12 | 19.95 | -104.00 | 989 | 23.2 | 845 |
| EjutlaB | *parviglumis* | 12 | 19.91 | -104.11 | 1013 | 23.2 | 891 |
| El Sauz | *parviglumis* | 12 | 19.44 | -103.98 | 1103 | 22.4 | 1048 |
| La Mesa | *parviglumis* | 12 | 19.96 | -104.05 | 1225 | 22.5 | 849 |
| EjutlaA | *parviglumis* | 12 | 19.90 | -104.18 | 1246 | 21.3 | 809 |
| Ahuacatitlan | *parviglumis* | 12 | 18.36 | -99.81 | 1528 | 22.3 | 1081 |
| Ixtlan | *mexicana* | 12 | 20.17 | -102.37 | 1547 | 20.1 | 793 |
| Puerta Encantada | *mexicana* | 12 | 18.97 | -99.03 | 1658 | 20.0 | 957 |
| Puruandiro | *mexicana* | 12 | 20.11 | -101.49 | 1915 | 18.9 | 811 |
| Nabogame | *mexicana* | 12 | 26.25 | -106.92 | 2020 | 15.8 | 978 |
| El Porvenir | *mexicana* | 12 | 19.68 | -100.64 | 2094 | 16.4 | 918 |
| Santa Clara | *mexicana* | 12 | 19.42 | -101.64 | 2173 | 15.5 | 1143 |
| Opopeo | *mexicana* | 12 | 19.42 | -101.61 | 2213 | 15.6 | 1142 |
| Xochimilco | *mexicana* | 12 | 19.29 | -99.08 | 2237 | 16.0 | 727 |
| San Pedro | *mexicana* | 12 | 19.09 | -98.49 | 2459 | 14.3 | 1017 |
| Tenango del Aire | *mexicana* | 12 | 19.12 | -99.59 | 2609 | 13.7 | 893 |

N: sample size, Lat: latitude, Long: longitude, Alt(m): altitude in meters above sea level, T: mean annual temperature in Celsius degrees, P: mean annual precipitation in millimeters

**Table S8** Environmental variables and abbreviations used in the study.

| Environmental variable | Abbreviation |
|---|---|
| Annual Mean Temperature | bio1 |
| Mean Diurnal Range (Mean of monthly (max temp - min temp)) | bio2 |
| Isothermality (bio2/bio7) (* 100) | bio3 |
| Temperature Seasonality (standard deviation *100) | bio4 |
| Max Temperature of Warmest Month | bio5 |
| Min Temperature of Coldest Month | bio6 |
| Temperature Annual Range (bio5-bio6) | bio7 |
| Mean Temperature of Wettest Quarter | bio8 |
| Mean Temperature of Driest Quarter | bio9 |
| Mean Temperature of Warmest Quarter | bio10 |
| Mean Temperature of Coldest Quarter | bio11 |
| Annual Precipitation | bio12 |
| Precipitation of Wettest Month | bio13 |
| Precipitation of Driest Month | bio14 |
| Precipitation Seasonality (Coefficient of Variation) | bio15 |
| Precipitation of Wettest Quarter | bio16 |
| Precipitation of Driest Quarter | bio17 |
| Precipitation of Warmest Quarter | bio18 |
| Precipitation of Coldest Quarter | bio19 |
| Monthly mean, minimum and maximum temperature | tmean#, tmin#, tmax# |
| Monthly total precipitation | prec# |
| Rooting conditions[1] | sq3 |
| Oxygen availability to roots[2] | sq4 |
| Workability[3] | sq7 |
| Sand in topsoil | ts_sand |
| Loam in topsoil | ts_loam |
| Clay in topsoil | ts_clay |
| Cracking clays | v_mod |
| Volcanic | |

[1] Soil textures, bulk density, coarse fragments, vertical soil properties and soil phases affecting root penetration and soil depth and soil volume
[2] Soil drainage and soil phases affecting soil drainage
[3] Soil texture, effective soil depth/volume, and soil phases constraining soil management (soil depth, rock outcrop, stoniness, gravel/concretions and hardpans)

**Table S9** Variable loadings/rotations for each of 6 PCs that were used in BAYENV for the joint dataset of 20 *parviglumis* and *mexicana* populations

| PC1 | | PC2 | | PC3 | | PC4 | | PC5 | | PC6 | |
|---|---|---|---|---|---|---|---|---|---|---|---|
| Var | Rot | Var | Rot | Var | Rot | Var | Rot | Var | Rot | Var | Rot |
| bio1 | 0.146 | bio4 | 0.244 | prec7 | 0.287 | ts_clay | 0.410 | bio2 | 0.380 | bio2 | 0.412 |
| tmean11 | 0.146 | bio3 | 0.241 | prec8 | 0.276 | v_mod | 0.359 | sq4 | 0.328 | x_mod | 0.365 |
| tmean12 | 0.145 | bio7 | 0.241 | prec11 | 0.262 | ts_sand | 0.329 | ts_loam | 0.289 | sq7 | 0.332 |
| bio11 | 0.145 | prec6 | 0.237 | bio13 | 0.247 | bio15 | 0.272 | ts_sand | 0.266 | bio7 | 0.312 |
| tmax12 | 0.145 | sq7 | 0.218 | prec1 | 0.246 | prec4 | 0.259 | sq7 | 0.231 | v_mod | 0.307 |
| tmin5 | 0.145 | prec9 | 0.217 | bio16 | 0.242 | x_mod | 0.244 | bio18 | 0.213 | prec11 | 0.230 |
| tmean1 | 0.145 | sq3 | 0.207 | prec12 | 0.240 | prec3 | 0.226 | bio13 | 0.207 | bio18 | 0.220 |
| tmean2 | 0.145 | prec12 | 0.207 | bio19 | 0.238 | sq3 | 0.210 | prec11 | 0.183 | sq3 | 0.176 |
| tmin4 | 0.145 | bio12 | 0.204 | bio12 | 0.231 | prec5 | 0.210 | bio7 | 0.170 | sq4 | 0.153 |
| tmax1 | 0.145 | bio19 | 0.196 | prec2 | 0.222 | prec7 | 0.190 | bio16 | 0.163 | ts_sand | 0.153 |
| tmean4 | 0.145 | prec2 | 0.188 | bio18 | 0.221 | sq4 | 0.186 | bio4 | 0.157 | bio4 | 0.148 |
| tmin11 | 0.144 | prec1 | 0.185 | sq4 | 0.200 | bio3 | 0.185 | bio12 | 0.156 | prec7 | 0.139 |
| tmax11 | 0.144 | prec10 | 0.184 | prec9 | 0.180 | bio18 | 0.178 | bio3 | 0.155 | tmax3 | 0.127 |
| tmin12 | 0.144 | bio16 | 0.183 | prec10 | 0.171 | sq7 | 0.132 | prec6 | 0.154 | tmax4 | 0.126 |
| tmin2 | 0.144 | prec8 | 0.170 | prec5 | 0.161 | bio14 | 0.116 | x_mod | 0.152 | bio5 | 0.118 |
| tmean5 | 0.144 | prec5 | 0.165 | prec4 | 0.154 | bio13 | 0.099 | prec9 | 0.144 | tmax5 | 0.115 |
| tmean10 | 0.144 | bio14 | 0.158 | sq3 | 0.147 | bio16 | 0.095 | prec8 | 0.143 | bio3 | 0.107 |
| bio6 | 0.144 | bio13 | 0.151 | bio2 | 0.143 | prec8 | 0.090 | v_mod | 0.142 | ts_loam | 0.095 |
| tmax2 | 0.144 | bio17 | 0.149 | bio17 | 0.129 | bio7 | 0.077 | bio15 | 0.136 | ts_clay | 0.078 |
| tmean3 | 0.144 | prec3 | 0.144 | ts_loam | 0.127 | bio4 | 0.075 | prec7 | 0.112 | tmin9 | 0.072 |
| tmin1 | 0.143 | ts_clay | 0.141 | v_mod | 0.123 | bio2 | 0.074 | prec4 | 0.108 | tmin8 | 0.071 |
| tmin10 | 0.143 | bio2 | 0.129 | prec3 | 0.113 | prec2 | 0.074 | bio14 | 0.096 | prec9 | 0.070 |
| Altitude | 0.143 | prec7 | 0.108 | x_mod | 0.111 | bio19 | 0.068 | tmax7 | 0.093 | tmin10 | 0.069 |
| bio9 | 0.143 | tmax6 | 0.107 | bio14 | 0.099 | prec12 | 0.056 | tmax8 | 0.092 | tmin12 | 0.069 |
| tmin3 | 0.143 | x_mod | 0.106 | bio4 | 0.070 | ts_loam | 0.053 | prec1 | 0.091 | tmin7 | 0.066 |
| bio10 | 0.142 | bio15 | 0.098 | tmax3 | 0.067 | tmax12 | 0.047 | prec2 | 0.086 | tmean4 | 0.066 |
| tmax10 | 0.142 | ts_loam | 0.088 | ts_clay | 0.065 | bio17 | 0.047 | tmin11 | 0.086 | tmax2 | 0.066 |
| tmax3 | 0.142 | tmean6 | 0.085 | bio15 | 0.056 | bio9 | 0.043 | prec5 | 0.082 | tmax6 | 0.064 |
| tmax4 | 0.142 | tmin7 | 0.082 | tmax2 | 0.055 | tmax8 | 0.042 | bio17 | 0.082 | tmean3 | 0.064 |

| | | | | | | | | | |
|---|---|---|---|---|---|---|---|---|---|
| tmin6 | 0.142 | bio5 | 0.082 | tmean3 | 0.052 | tmax1 | 0.041 | tmin12 | 0.080 | bio9 | 0.058 |
| tmean9 | 0.141 | tmean7 | 0.081 | ts_sand | 0.050 | tmax5 | 0.039 | prec3 | 0.078 | tmin11 | 0.058 |
| tmin9 | 0.141 | prec4 | 0.080 | prec6 | 0.048 | tmax7 | 0.039 | tmax9 | 0.078 | prec3 | 0.056 |
| tmean8 | 0.141 | tmax7 | 0.079 | sq7 | 0.048 | prec10 | 0.038 | tmin1 | 0.077 | tmean5 | 0.056 |
| bio8 | 0.140 | bio8 | 0.079 | tmin7 | 0.046 | Altitude | 0.037 | tmin10 | 0.074 | bio10 | 0.051 |
| tmean6 | 0.140 | tmax9 | 0.077 | bio3 | 0.044 | tmax10 | 0.035 | bio6 | 0.071 | bio15 | 0.047 |
| tmean7 | 0.140 | tmean8 | 0.076 | tmax4 | 0.043 | tmax2 | 0.033 | prec12 | 0.067 | tmin1 | 0.045 |
| tmin8 | 0.140 | tmin8 | 0.076 | bio7 | 0.042 | tmax9 | 0.030 | tmin2 | 0.061 | prec12 | 0.045 |
| tmax5 | 0.140 | tmax5 | 0.074 | tmax1 | 0.036 | tmean12 | 0.029 | tmin6 | 0.061 | bio17 | 0.045 |
| tmax9 | 0.139 | tmax8 | 0.074 | tmin3 | 0.035 | tmax11 | 0.027 | tmax6 | 0.059 | tmean10 | 0.043 |
| tmax8 | 0.139 | tmean9 | 0.072 | bio9 | 0.035 | tmean1 | 0.027 | tmin3 | 0.052 | prec4 | 0.043 |
| bio5 | 0.139 | bio18 | 0.070 | tmin8 | 0.034 | bio5 | 0.026 | bio8 | 0.046 | Altitude | 0.041 |
| tmax7 | 0.139 | v_mod | 0.066 | tmean4 | 0.031 | tmean5 | 0.026 | tmean11 | 0.041 | bio6 | 0.039 |
| tmin7 | 0.138 | tmin9 | 0.066 | tmean2 | 0.031 | bio11 | 0.025 | tmean1 | 0.040 | tmin6 | 0.038 |
| tmax6 | 0.135 | bio10 | 0.065 | tmax12 | 0.030 | tmean2 | 0.022 | tmin9 | 0.040 | tmean12 | 0.035 |
| bio17 | 0.107 | tmax10 | 0.063 | tmean7 | 0.028 | tmin6 | 0.022 | tmean12 | 0.040 | tmean9 | 0.035 |
| bio14 | 0.095 | tmin6 | 0.061 | tmin9 | 0.023 | prec1 | 0.022 | tmax10 | 0.038 | prec5 | 0.033 |
| prec3 | 0.094 | prec11 | 0.056 | Altitude | 0.021 | tmean7 | 0.020 | tmin5 | 0.038 | tmean8 | 0.030 |
| bio15 | 0.092 | ts_sand | 0.054 | tmax11 | 0.019 | tmean8 | 0.020 | bio11 | 0.037 | tmin2 | 0.030 |
| prec4 | 0.088 | sq4 | 0.049 | tmin4 | 0.018 | tmean9 | 0.019 | tmean8 | 0.035 | tmax1 | 0.030 |
| prec10 | 0.084 | tmin3 | 0.048 | tmax5 | 0.017 | prec11 | 0.019 | tmean7 | 0.033 | prec1 | 0.025 |
| ts_sand | 0.073 | tmean10 | 0.047 | tmin10 | 0.017 | tmean11 | 0.018 | tmean2 | 0.030 | bio13 | 0.024 |
| ts_loam | 0.072 | tmean5 | 0.044 | tmin6 | 0.017 | tmax4 | 0.016 | bio19 | 0.029 | tmean11 | 0.023 |
| prec6 | 0.069 | tmax4 | 0.042 | tmean6 | 0.017 | tmean10 | 0.016 | bio9 | 0.027 | tmax7 | 0.023 |
| prec9 | 0.063 | tmax11 | 0.041 | tmax6 | 0.017 | tmin5 | 0.015 | tmin7 | 0.026 | tmean7 | 0.023 |
| prec2 | 0.063 | bio6 | 0.038 | tmin12 | 0.017 | tmean6 | 0.015 | tmin4 | 0.025 | prec10 | 0.022 |
| bio4 | 0.060 | tmin1 | 0.038 | bio11 | 0.016 | prec6 | 0.014 | tmean5 | 0.023 | prec2 | 0.020 |
| prec5 | 0.058 | tmin2 | 0.037 | tmin11 | 0.015 | bio6 | 0.012 | tmean3 | 0.022 | bio8 | 0.018 |
| bio13 | 0.055 | Altitude | 0.036 | tmin1 | 0.013 | tmin1 | 0.012 | tmin8 | 0.022 | tmean2 | 0.017 |
| x_mod | 0.053 | tmin10 | 0.034 | tmean5 | 0.013 | tmin12 | 0.011 | tmean10 | 0.021 | bio16 | 0.016 |
| prec1 | 0.050 | bio9 | 0.032 | tmean8 | 0.013 | tmin2 | 0.011 | tmean9 | 0.019 | tmean6 | 0.014 |
| bio16 | 0.049 | tmean1 | 0.029 | bio6 | 0.012 | bio1 | 0.011 | tmean4 | 0.019 | bio19 | 0.012 |
| bio7 | 0.043 | bio1 | 0.028 | tmean9 | 0.010 | tmin4 | 0.009 | ts_clay | 0.018 | tmax11 | 0.012 |

| | | | | | | | | | | | | |
|---|---|---|---|---|---|---|---|---|---|---|---|---|
| bio19 | 0.042 | tmean3 | 0.026 | tmean1 | 0.010 | tmax6 | 0.008 | prec10 | 0.015 | tmax10 | 0.011 |
| bio12 | 0.042 | tmean2 | 0.024 | bio10 | 0.010 | tmin11 | 0.007 | bio1 | 0.014 | tmax8 | 0.010 |
| bio3 | 0.040 | bio11 | 0.022 | tmax8 | 0.009 | bio8 | 0.007 | tmax4 | 0.014 | tmean1 | 0.009 |
| prec7 | 0.034 | tmean11 | 0.019 | tmax7 | 0.009 | bio10 | 0.007 | bio10 | 0.011 | bio14 | 0.009 |
| prec12 | 0.033 | tmin12 | 0.019 | bio5 | 0.008 | tmin9 | 0.006 | tmax5 | 0.009 | prec8 | 0.007 |
| bio2 | 0.033 | tmax1 | 0.017 | tmin5 | 0.008 | prec9 | 0.005 | bio5 | 0.008 | bio11 | 0.007 |
| prec8 | 0.032 | tmean4 | 0.016 | tmin2 | 0.007 | bio12 | 0.004 | tmax11 | 0.008 | tmin5 | 0.006 |
| sq4 | 0.026 | tmin5 | 0.013 | bio1 | 0.007 | tmax3 | 0.003 | tmax12 | 0.008 | tmax9 | 0.005 |
| sq3 | 0.024 | tmean12 | 0.012 | tmax10 | 0.006 | tmean4 | 0.003 | tmax3 | 0.007 | bio12 | 0.004 |
| v_mod | 0.015 | tmax2 | 0.011 | tmean10 | 0.006 | tmin3 | 0.002 | sq3 | 0.007 | tmin3 | 0.004 |
| ts_clay | 0.012 | tmin4 | 0.010 | tmean12 | 0.005 | tmin7 | 0.002 | tmax2 | 0.006 | prec6 | 0.004 |
| sq7 | 0.004 | tmax3 | 0.003 | bio8 | 0.004 | tmean3 | 0.002 | Altitude | 0.004 | tmin4 | 0.003 |
| prec11 | 0.004 | tmax12 | 0.002 | tmax9 | 0.003 | tmin10 | 0.002 | tmean6 | 0.002 | bio1 | 0.002 |
| bio18 | 0.001 | tmin11 | 0.001 | tmean11 | 0.001 | tmin8 | 0.000 | tmax1 | 0.001 | tmax12 | 0.002 |

**Table S10** Variable loadings/rotations for each of 4 PCs that were used in BAYENV for 10 *parviglumis* populations

| PC1 | | PC2 | | PC3 | | PC4 | |
|---|---|---|---|---|---|---|---|
| Var | Rot | Var | Rot | Var | Rot | Var | Rot |
| tmax12 | 0.145 | bio4 | 0.258 | x_mod | 0.300 | prec4 | 0.315 |
| bio1 | 0.144 | bio7 | 0.248 | ts_loam | 0.300 | prec11 | 0.278 |
| bio11 | 0.144 | bio3 | 0.232 | prec5 | 0.241 | prec1 | 0.276 |
| tmean2 | 0.143 | ts_clay | 0.205 | prec11 | 0.214 | bio2 | 0.256 |
| tmean12 | 0.143 | prec9 | 0.185 | ts_sand | 0.214 | prec5 | 0.233 |
| tmax1 | 0.143 | bio12 | 0.185 | prec4 | 0.196 | prec12 | 0.222 |
| tmean11 | 0.142 | prec10 | 0.184 | bio15 | 0.194 | ts_clay | 0.213 |
| tmean1 | 0.142 | tmax6 | 0.177 | bio2 | 0.186 | bio19 | 0.208 |
| tmin5 | 0.142 | prec6 | 0.173 | bio18 | 0.176 | x_mod | 0.183 |
| tmean3 | 0.141 | tmax5 | 0.169 | prec8 | 0.165 | ts_loam | 0.183 |
| tmax2 | 0.140 | bio2 | 0.167 | prec12 | 0.161 | tmax8 | 0.170 |
| tmin4 | 0.139 | bio5 | 0.165 | prec7 | 0.159 | sq3 | 0.166 |
| Altitude | 0.139 | tmean6 | 0.160 | tmin10 | 0.157 | sq7 | 0.166 |
| bio9 | 0.139 | prec8 | 0.157 | sq3 | 0.155 | bio18 | 0.164 |
| tmax11 | 0.138 | bio16 | 0.157 | sq7 | 0.155 | tmax9 | 0.149 |
| tmin2 | 0.138 | ts_sand | 0.156 | bio13 | 0.151 | tmin12 | 0.131 |
| tmin3 | 0.138 | prec5 | 0.155 | tmax3 | 0.146 | tmax7 | 0.130 |
| tmean4 | 0.137 | bio13 | 0.139 | tmin7 | 0.145 | tmin11 | 0.125 |
| tmin11 | 0.136 | prec1 | 0.138 | tmin9 | 0.142 | bio17 | 0.124 |
| bio6 | 0.135 | bio17 | 0.136 | tmin8 | 0.142 | bio9 | 0.119 |
| bio10 | 0.134 | tmin6 | 0.127 | bio16 | 0.141 | bio3 | 0.118 |
| tmean5 | 0.133 | tmin7 | 0.126 | bio7 | 0.129 | prec2 | 0.116 |
| tmax3 | 0.133 | tmax4 | 0.125 | prec2 | 0.129 | tmin1 | 0.113 |
| tmin1 | 0.133 | tmin9 | 0.120 | tmax4 | 0.128 | tmean8 | 0.111 |
| tmin12 | 0.132 | bio8 | 0.120 | tmax2 | 0.115 | bio8 | 0.108 |
| tmean10 | 0.130 | tmean7 | 0.119 | prec9 | 0.112 | tmean9 | 0.106 |
| tmax10 | 0.128 | tmean9 | 0.117 | tmean4 | 0.111 | prec10 | 0.106 |
| tmax7 | 0.128 | tmin8 | 0.116 | tmean10 | 0.110 | tmax10 | 0.102 |
| tmean7 | 0.127 | bio14 | 0.116 | tmean3 | 0.106 | bio6 | 0.102 |
| tmean8 | 0.126 | tmax9 | 0.113 | bio5 | 0.096 | bio7 | 0.096 |
| tmin10 | 0.125 | tmax10 | 0.112 | tmin12 | 0.090 | tmin2 | 0.088 |
| bio8 | 0.125 | tmean8 | 0.110 | tmin11 | 0.088 | prec7 | 0.082 |
| tmax8 | 0.125 | prec4 | 0.110 | tmax5 | 0.085 | Altitude | 0.081 |
| tmin6 | 0.125 | bio10 | 0.109 | tmean9 | 0.085 | tmean7 | 0.073 |
| tmean9 | 0.125 | tmax7 | 0.107 | tmean7 | 0.083 | tmin4 | 0.071 |
| tmax9 | 0.123 | prec2 | 0.105 | tmean5 | 0.082 | bio4 | 0.070 |
| tmax4 | 0.122 | tmean10 | 0.098 | bio12 | 0.080 | tmin3 | 0.068 |
| tmin8 | 0.121 | tmean5 | 0.096 | tmin4 | 0.079 | tmin6 | 0.068 |
| tmean6 | 0.120 | tmax8 | 0.096 | prec1 | 0.079 | ts_sand | 0.066 |
| tmin9 | 0.119 | prec12 | 0.095 | tmean8 | 0.078 | tmean12 | 0.063 |
| bio19 | 0.118 | tmax11 | 0.094 | tmin6 | 0.077 | tmin5 | 0.058 |
| tmin7 | 0.118 | tmin1 | 0.086 | prec10 | 0.069 | prec3 | 0.055 |
| prec7 | 0.117 | tmin10 | 0.081 | tmax1 | 0.068 | tmean11 | 0.053 |
| prec2 | 0.117 | bio6 | 0.080 | tmean2 | 0.066 | tmax6 | 0.050 |
| prec3 | 0.116 | bio18 | 0.077 | tmin5 | 0.066 | prec6 | 0.049 |
| bio13 | 0.113 | prec3 | 0.072 | bio3 | 0.062 | bio11 | 0.049 |
| bio5 | 0.113 | prec7 | 0.071 | bio8 | 0.058 | tmean4 | 0.048 |
| tmax6 | 0.112 | tmin2 | 0.070 | tmin3 | 0.057 | tmean5 | 0.046 |
| tmax5 | 0.112 | sq7 | 0.069 | tmax10 | 0.055 | tmin9 | 0.045 |
| sq3 | 0.112 | sq3 | 0.069 | tmax12 | 0.053 | tmean10 | 0.042 |
| sq7 | 0.112 | tmin3 | 0.068 | bio17 | 0.050 | tmean1 | 0.041 |
| bio17 | 0.108 | bio19 | 0.067 | bio9 | 0.049 | tmax2 | 0.039 |
| prec6 | 0.108 | tmean1 | 0.062 | bio10 | 0.047 | tmax1 | 0.038 |
| bio15 | 0.108 | tmin12 | 0.054 | tmean11 | 0.046 | bio16 | 0.032 |
| bio16 | 0.106 | bio1 | 0.053 | bio4 | 0.040 | prec8 | 0.032 |
| bio14 | 0.106 | tmax3 | 0.052 | tmin1 | 0.039 | bio14 | 0.031 |
| bio12 | 0.104 | tmean11 | 0.049 | tmean6 | 0.039 | tmin8 | 0.028 |
| prec9 | 0.100 | tmean4 | 0.049 | prec3 | 0.038 | tmean2 | 0.028 |
| prec8 | 0.098 | bio11 | 0.041 | tmax9 | 0.037 | tmin10 | 0.028 |
| prec10 | 0.098 | tmean12 | 0.039 | bio6 | 0.032 | tmax5 | 0.027 |
| prec12 | 0.096 | bio9 | 0.037 | ts_clay | 0.031 | tmean3 | 0.025 |
| ts_sand | 0.089 | tmin4 | 0.035 | tmax7 | 0.031 | bio10 | 0.025 |
| prec1 | 0.078 | tmean2 | 0.031 | bio14 | 0.024 | tmax12 | 0.022 |
| bio3 | 0.068 | x_mod | 0.030 | tmax8 | 0.023 | tmax11 | 0.022 |
| ts_clay | 0.064 | ts_loam | 0.030 | tmean12 | 0.022 | prec9 | 0.019 |
| x_mod | 0.055 | tmax1 | 0.029 | bio11 | 0.021 | tmax3 | 0.018 |
| ts_loam | 0.055 | Altitude | 0.023 | tmin2 | 0.014 | tmax4 | 0.016 |
| bio4 | 0.043 | tmax12 | 0.020 | tmean1 | 0.013 | bio12 | 0.013 |
| prec4 | 0.034 | tmax2 | 0.008 | tmax6 | 0.010 | bio15 | 0.011 |
| bio7 | 0.026 | bio15 | 0.008 | prec6 | 0.009 | bio5 | 0.009 |
| prec11 | 0.017 | prec11 | 0.006 | bio19 | 0.005 | bio1 | 0.006 |
| bio2 | 0.014 | tmean3 | 0.006 | tmax11 | 0.004 | bio13 | 0.005 |
| bio18 | 0.005 | tmin5 | 0.003 | bio1 | 0.002 | tmin7 | 0.003 |
| prec5 | 0.003 | tmin11 | 0.001 | Altitude | 0.002 | tmean6 | 0.003 |

**Table S11** Variable loadings/rotations for each of 4 PCs that were used in BAYENV for 10 *mexicana* populations

| PC1 | | PC2 | | PC3 | | PC4 | |
|---|---|---|---|---|---|---|---|
| Var | Rot | Var | Rot | Var | Rot | Var | Rot |
| bio1 | 0.152 | bio4 | 0.254 | bio13 | 0.308 | ts_sand | 0.456 |
| tmean4 | 0.150 | prec12 | 0.235 | prec7 | 0.308 | v_mod | 0.338 |
| tmean10 | 0.150 | bio7 | 0.230 | bio16 | 0.303 | bio3 | 0.300 |
| tmean5 | 0.150 | bio19 | 0.228 | prec8 | 0.295 | ts_clay | 0.269 |
| tmean11 | 0.150 | prec2 | 0.226 | bio12 | 0.270 | bio2 | 0.241 |
| tmax11 | 0.149 | prec6 | 0.219 | x_mod | 0.246 | bio18 | 0.232 |
| tmin5 | 0.149 | prec1 | 0.215 | prec9 | 0.238 | sq4 | 0.181 |
| tmax4 | 0.149 | bio3 | 0.206 | prec10 | 0.228 | ts_loam | 0.170 |
| tmin10 | 0.148 | bio17 | 0.191 | prec11 | 0.227 | prec2 | 0.163 |
| tmin4 | 0.147 | prec5 | 0.187 | prec4 | 0.227 | prec9 | 0.153 |
| tmax5 | 0.146 | prec3 | 0.183 | bio2 | 0.184 | tmax7 | 0.148 |
| tmax12 | 0.146 | bio14 | 0.179 | sq4 | 0.179 | prec12 | 0.147 |
| Altitude | 0.144 | bio18 | 0.177 | bio15 | 0.164 | sq7 | 0.142 |
| tmean12 | 0.144 | sq7 | 0.175 | ts_sand | 0.152 | prec3 | 0.138 |
| tmax10 | 0.144 | prec9 | 0.135 | prec5 | 0.142 | bio19 | 0.129 |
| bio10 | 0.144 | prec11 | 0.135 | prec6 | 0.125 | bio12 | 0.120 |
| bio5 | 0.143 | tmin7 | 0.129 | prec1 | 0.121 | tmax8 | 0.120 |
| tmin6 | 0.143 | tmax6 | 0.128 | bio9 | 0.103 | prec10 | 0.112 |
| tmin11 | 0.142 | prec10 | 0.125 | bio19 | 0.095 | tmin6 | 0.108 |
| tmean9 | 0.142 | tmin3 | 0.121 | ts_loam | 0.091 | x_mod | 0.107 |
| bio11 | 0.141 | tmean7 | 0.119 | tmin1 | 0.084 | tmax9 | 0.097 |
| bio8 | 0.141 | tmin8 | 0.115 | bio6 | 0.080 | tmax12 | 0.095 |
| bio9 | 0.141 | bio2 | 0.107 | prec2 | 0.075 | tmax10 | 0.094 |
| tmean2 | 0.141 | tmin1 | 0.106 | prec12 | 0.074 | tmin9 | 0.093 |
| tmax2 | 0.140 | bio6 | 0.105 | sq7 | 0.071 | bio14 | 0.088 |
| tmax1 | 0.140 | tmax7 | 0.104 | tmin12 | 0.070 | tmax11 | 0.076 |
| tmax3 | 0.140 | tmean8 | 0.104 | prec3 | 0.067 | prec5 | 0.073 |
| tmin9 | 0.139 | tmean3 | 0.103 | tmin11 | 0.066 | tmin8 | 0.071 |
| tmean6 | 0.139 | tmean6 | 0.102 | bio7 | 0.065 | tmax1 | 0.071 |
| tmean8 | 0.139 | tmin2 | 0.102 | bio3 | 0.064 | tmax6 | 0.069 |
| tmax9 | 0.139 | tmean1 | 0.099 | Altitude | 0.056 | tmin10 | 0.066 |
| tmean1 | 0.139 | bio8 | 0.095 | tmin10 | 0.055 | tmean12 | 0.064 |
| tmean3 | 0.138 | tmean2 | 0.095 | ts_clay | 0.055 | tmin5 | 0.059 |
| tmin12 | 0.138 | bio15 | 0.093 | tmax2 | 0.051 | tmin7 | 0.059 |
| tmax8 | 0.137 | tmax9 | 0.092 | tmin7 | 0.050 | tmean7 | 0.058 |
| tmin2 | 0.135 | tmean9 | 0.092 | tmin6 | 0.046 | tmean1 | 0.058 |
| tmean7 | 0.133 | bio11 | 0.090 | bio18 | 0.045 | bio15 | 0.054 |
| tmin8 | 0.133 | tmax8 | 0.089 | tmax8 | 0.044 | bio11 | 0.053 |
| tmax7 | 0.133 | tmax1 | 0.087 | tmin2 | 0.043 | bio16 | 0.047 |
| tmin3 | 0.132 | tmin9 | 0.086 | tmax12 | 0.040 | Altitude | 0.047 |
| bio6 | 0.131 | tmax2 | 0.083 | tmax3 | 0.039 | prec8 | 0.047 |
| tmin1 | 0.131 | tmin12 | 0.080 | tmin5 | 0.038 | tmax2 | 0.045 |
| tmax6 | 0.130 | tmax3 | 0.079 | tmax11 | 0.034 | bio6 | 0.042 |
| tmin7 | 0.127 | bio10 | 0.078 | tmin8 | 0.033 | bio7 | 0.041 |
| ts_loam | 0.112 | prec4 | 0.077 | v_mod | 0.032 | prec11 | 0.040 |
| ts_clay | 0.111 | bio5 | 0.075 | tmax10 | 0.032 | tmean11 | 0.040 |
| bio15 | 0.106 | ts_clay | 0.074 | tmin9 | 0.031 | tmin1 | 0.040 |
| v_mod | 0.101 | tmean12 | 0.069 | tmean1 | 0.030 | tmean5 | 0.038 |
| prec4 | 0.094 | x_mod | 0.069 | tmean5 | 0.030 | tmean2 | 0.038 |
| bio14 | 0.091 | tmax10 | 0.063 | tmax9 | 0.029 | bio17 | 0.036 |
| prec3 | 0.090 | tmin6 | 0.062 | tmax1 | 0.027 | tmean8 | 0.033 |
| bio17 | 0.089 | tmin11 | 0.056 | bio5 | 0.027 | tmin12 | 0.031 |
| sq7 | 0.077 | tmax5 | 0.052 | tmean6 | 0.026 | bio10 | 0.029 |
| bio12 | 0.065 | tmax12 | 0.052 | bio10 | 0.026 | bio8 | 0.029 |
| prec11 | 0.064 | Altitude | 0.049 | tmax5 | 0.025 | bio4 | 0.028 |
| prec5 | 0.062 | tmin4 | 0.048 | tmax7 | 0.022 | tmin2 | 0.027 |
| sq4 | 0.058 | v_mod | 0.048 | tmean12 | 0.018 | bio1 | 0.026 |
| prec1 | 0.055 | ts_loam | 0.047 | tmean11 | 0.017 | tmax5 | 0.024 |
| prec8 | 0.052 | bio12 | 0.046 | bio11 | 0.014 | bio13 | 0.021 |
| prec10 | 0.051 | bio13 | 0.040 | tmean3 | 0.012 | prec7 | 0.021 |
| bio18 | 0.048 | prec7 | 0.040 | bio14 | 0.012 | prec4 | 0.017 |
| prec9 | 0.043 | bio9 | 0.040 | tmin3 | 0.012 | tmax3 | 0.017 |
| prec2 | 0.042 | tmean10 | 0.035 | tmean10 | 0.011 | tmean9 | 0.017 |
| bio19 | 0.035 | bio16 | 0.030 | tmean7 | 0.011 | tmin4 | 0.016 |
| bio2 | 0.032 | ts_sand | 0.028 | tmax6 | 0.010 | tmean10 | 0.016 |
| bio16 | 0.032 | tmean4 | 0.028 | bio8 | 0.010 | prec6 | 0.016 |
| x_mod | 0.029 | sq4 | 0.025 | bio1 | 0.010 | bio9 | 0.012 |
| prec12 | 0.025 | tmean11 | 0.023 | bio17 | 0.008 | tmean3 | 0.010 |
| bio7 | 0.022 | tmean5 | 0.021 | tmax4 | 0.008 | tmax4 | 0.010 |
| bio13 | 0.019 | tmax11 | 0.013 | tmean8 | 0.007 | bio5 | 0.008 |
| prec7 | 0.019 | tmin5 | 0.013 | tmin4 | 0.007 | tmean6 | 0.008 |
| prec6 | 0.017 | tmax4 | 0.009 | bio4 | 0.004 | tmin11 | 0.007 |
| bio3 | 0.003 | tmin10 | 0.004 | tmean9 | 0.004 | tmean4 | 0.006 |
| bio4 | 0.003 | bio1 | 0.002 | tmean2 | 0.003 | prec1 | 0.002 |
| ts_sand | 0.000 | prec8 | 0.001 | tmean4 | 0.003 | tmin3 | 0.001 |

**Text S1** List of 278 maize inbred lines used in the association analysis

33.16, 38.11, 4226, 4722, A188, A214N, A239, A272, A441.5, A554, A556, A6, A619, A632, A634, A635, A641, A654, A659, A661, A679, A680, A682, AB28A, B10, B103, B104, B105, B109, B115, B14A, B164, B2, B37, B46, B52, B57, B64, B68, B73, B73HTRHM, B75, B76, B77, B79, B84, B97, C103, C123, C49A, CH701.30, CH9, CI.7, CI187.2, CI21E, CI28A, CI31A, CI3A, CI64, CI66, CI90C, CI91B, CM105, CM174, CM37, CM7, CML10, CML103, CML108, CML11, CML14, CML154Q, CML157Q, CML158Q, CML218, CML220, CML228, CML238, CML247, CML254, CML258, CML261, CML264, CML277, CML281, CML287, CML311, CML314, CML321, CML322, CML323, CML328, CML331, CML332, CML333, CML341, CML38, CML45, CML5, CML52, CML61, CML69, CML77, CML91, CML92, CMV3, CO106, CO125, CO255, D940Y, DE.2, DE.3, DE1, DE811, E2558W, EP1, F2834T, F44, F6, F7, GA209, GT112, H105W, H49, H84, H91, H95, H99, HI27, HP301, HY, I137TN, I205, I29, IA2132, IA5125, IDS28, IDS69, IDS91, IL101, IL14H, IL677A, K148, K4, K55, K64, KI11, KI14, KI2021, KI21, KI3, KI43, KI44, KY21, KY226, KY228, L317, L578, M14, M162W, M37W, MEF156.55.2, MO17, MO18W, MO1W, MO24W, MO44, MO45, MO46, MO47, MOG, MP339, MS1334, MS153, MS71, MT42, N192, N28HT, N6, N7A, NC222, NC230, NC232, NC236, NC238, NC250, NC258, NC260, NC262, NC264, NC290A, NC294, NC296, NC296A, NC298, NC300, NC302, NC304, NC306, NC310, NC314, NC318, NC320, NC324, NC326, NC328, NC33, NC336, NC338, NC340, NC342, NC344, NC346, NC348, NC350, NC352, NC354, NC356, NC358, NC360, NC362, NC364, NC366, NC368, ND246, OH40B, OH43, OH43E, OH603, OH7B, OS420, P39, PA762, PA875, PA880, PA91, R109B, R168, R177, R229, R4, SA24, SC213R, SC357, SC55, SD40, SD44, SG1533, SG18, T232, T234, T8, TX303, TX601, TZI10, TZI11, TZI16, TZI18, TZI25, TZI9, U267Y, VA102, VA14, VA17, VA22, VA26, VA35, VA59, VA85, VA99, VAW6, W117HT, W153R, W182B, W22, WD, WF9, YU796.NS